\pdfoutput=1
\documentclass[12pt]{iopart}

\usepackage{iopams}
\usepackage{setspace}
\usepackage[left=3.5cm, right=3.5cm, top=2cm]{geometry}
\usepackage{amssymb}
\usepackage{amsbsy}
\usepackage{rotating}      
\usepackage{float}
\usepackage{mwe}
\usepackage[color=yellow]{todonotes}
\usepackage{graphicx}
\usepackage{caption,subcaption}
\usepackage{natbib}
\def\newblock{\ }%
\setcitestyle{authoryear,open={(},close={)}}
\usepackage{siunitx}
\usepackage{csquotes}
\usepackage{multirow}
\usepackage{varwidth}
\usepackage{hyperref}
\hypersetup{
	colorlinks   = true, 
	urlcolor     = black, 
	linkcolor    = blue, 
	citecolor   = blue 
}

\begin{document}

\title[Efficient UQ for MC dose calculations using importance (re-)weighting]{Efficient uncertainty quantification for Monte Carlo dose calculations using importance (re-)weighting}

\author{P. Stammer$^{\hyperlink{1}{1},\hyperlink{2}{2},\hyperlink{3}{3}}$, L. Burigo$^{\hyperlink{2}{2}, \hyperlink{4}{4}}$, O. Jäkel$^{\hyperlink{2}{2},\hyperlink{3}{3}, \hyperlink{4}{4},\hyperlink{5}{5}}$, M. Frank$^{\hyperlink{1}{1},\hyperlink{3}{3}}$, N. Wahl$^{\hyperlink{2}{2}, \hyperlink{4}{4}}$}

\address{ \hypertarget{1}{$^1$} Karlsruhe Institute of Technology, Steinbuch Centre for Computing, Karlsruhe, Germany}
\address{ \hypertarget{2}{$^2$} German Cancer Research Center - DKFZ, Department of Medical Physics in Radiation Oncology, Heidelberg, Germany}
\address{ \hypertarget{3}{$^3$} HIDSS4Health - Helmholtz Information and Data Science School for Health, Karlsruhe/Heidelberg, Germany}
\address{\hypertarget{4}{$^4$} Heidelberg Institute for Radiation Oncology (HIRO), Heidelberg, Germany}
\address{\hypertarget{5}{$^5$} Heidelberg Ion Beam Therapy Center - HIT, Department of Medical Physics in Radiation Oncology, Heidelberg, Germany}

\ead{pia.stammer@kit.edu}

\begin{abstract}

The high precision and conformity of intensity-modulated particle therapy (IMPT) comes at the cost of susceptibility to treatment uncertainties in particle range and patient set-up. Dose uncertainty quantification and mitigation, which is usually based on sampled error scenarios, however becomes challenging when computing the dose with computationally expensive but accurate Monte Carlo (MC) simulations. This paper introduces an importance (re-)weighting method in MC history scoring to concurrently construct estimates for error scenarios, the expected dose and its variance from a single set of MC simulated particle histories.
The approach relies on a multivariate Gaussian input and uncertainty model, which assigns probabilities to the initial phase space sample, enabling the use of different correlation models. Exploring and adapting bivariate emittance parametrizations for the beam shape, accuracy can be traded between that of the uncertainty or the nominal dose estimate.

The method was implemented using the MC code TOPAS and tested for proton IMPT plan delivery in comparison to a reference scenario estimate. We achieve accurate results for set-up uncertainties ($\gamma_{\SI{3}{\mm}/\SI{3}{\percent}} \geq  \SI{99.99}{\percent}$) and expectedly lower but still sufficient agreement for range uncertainties, which are approximated with uncertainty over the energy distribution ($\gamma_{\SI{3}{\mm}/\SI{3}{\percent}} \geq  \SI{99.50}{\percent}$ ($E[\boldsymbol{d}]$), $\gamma_{\SI{3}{\mm}/\SI{3}{\percent}} \geq  \SI{91.69}{\percent}$ ($\sigma(\boldsymbol{d})$)\,). Initial experiments on a water phantom, a prostate and a liver case show that the re-weighting approach lowers the CPU time by more than an order of magnitude. Further, we show that uncertainty induced by interplay and other dynamic influences may be approximated using suitable error correlation models.

\end{abstract}
\vspace{0.5pc}

\section{Introduction}
Monte Carlo methods are considered the gold standard for dose calculation in radiotherapy treatment planning due to their accuracy \citep{paganettiRangeUncertaintiesProton2012, wengVectorizedMonteCarlo2003}. However, the accuracy of a simulated compared to a delivered dose is not only determined by the chosen dose engine, but also compromised by treatment uncertainties in water-equivalent path length (WEPL), patient set-up and anatomy. Especially in proton and carbon-ion therapy, the high dose localization in the Bragg-peak usually does not allow for uncertainty quantification and mitigation using approximations known for photon therapy, such as the static dose cloud \citep{lomaxIntensityModulatedProton2008, lomaxIntensityModulatedProton2008a}. 

Consequently, particle therapy demands personalized robustness analyses and mitigation. Such techniques may be based on explicit propagation of input uncertainties using probabilistic methods and statistical analysis \citep{bangertAnalyticalProbabilisticModeling2013,wahlEfficiencyAnalyticalSamplingbased2017,wahlAnalyticalProbabilisticModeling2020,kraanDoseUncertaintiesIMPT2013,parkStatisticalAssessmentProton2013,perkoFastAccurateSensitivity2016} or worst-case estimates \citep{mcgowanDefiningRobustnessProtocols2015, casiraghiAdvantagesLimitationsWorst2013, loweIncorporatingEffectFractionation2016}. Most of these methods then further translate to robust and probabilistic optimization to extend the conventional, generic margin approach to uncertainty mitigation \citep{sobottaRobustOptimizationBased2010, liuRobustOptimizationIntensity2012,fredrikssonCharacterizationRobustRadiation2012,unkelbachRobustRadiotherapyPlanning2018}.

The additional computational effort of robustness analyses and robust optimization techniques, however, clashes with the long computation times of Monte Carlo dose calculation. The use of faster, less accurate deterministic pencil-beam dose calculation algorithms instead is not always feasible, because their accuracy is low in particularly heterogeneous anatomies like lung \citep{taylorPencilBeamAlgorithms2017}, which at the same time show high sensitivity to uncertainties in range and set-up.

More efficient uncertainty quantification approaches for Monte Carlo methods, developed for example by the radiative transport community \citep[e.g.][]{poetteGPCintrusiveMonteCarloScheme2018,huStochasticGalerkinMethod2016}, often do not demonstrate an application to realistic patient data and it is not clear how well the results transfer. Also in many cases, more sophisticated methods are intrusive, which limits the applicability when using proprietary MC simulation engines. 

In this paper, we introduce a simple, minimally-intrusive method for uncertainty quantification in Monte Carlo dose computations. It is based on (re-)weighting a single set of MC simulated particle histories. In contrast to the conventional approach of simulating different scenarios separately, our method significantly reduces the required computational effort. The weighting step, which can be represented as multiplications of a weight vector with a history dose matrix, replaces simulations of different dose scenarios. The method enables uncertainty propagation during the simulation, making it possible to estimate the dose uncertainty induced by range and setup errors from nominal dose calculations. We demonstrate the application of this method to specifically approximate expected value and variance of dose, given a respective uncertainty model for set-up and range errors, which includes the choice of different beam and pencil beam correlation scenarios. 

The remainder of this paper is organized as follows: In section \ref{sec:methods}, we introduce basic definitions and notation, derive a direct computation of the expected value before introducing the concept of importance (re-)weighting for set-up and range uncertainty models. Section \ref{sec:results} then compares estimated expected doses and corresponding standard deviations to reference computations based on scenario sampling. Discussion and Conclusion follow in sections \ref{sec:discussion} and \ref{sec:conclusion}, respectively.

\section{Materials and methods}
\label{sec:methods}
\subsection{The Monte Carlo method for dose computation}
\label{sec:MCmethod}
First, we briefly recapitulate the basic principles of the Monte Carlo method for radiotherapy. This serves the purpose of establishing notation and parameters used to introduce our method and simplifying the illustration of later adaptations. For a more detailed description we refer to other sources, such as \citet{paganettiRangeUncertaintiesProton2012,fippelMonteCarloDose2004, maMonteCarloDose2002, bielajewMonteCarloModeling1994b, mackieApplicationsMonteCarlo1990}, among many others. 

The Monte Carlo method is a numerical integration technique, based on random sampling. When used for dose calculations, a set of particles is created with properties including position, momentum and energy, which evolve dynamically over the course of a simulation. The initial values of these properties constitute the random input parameters of the MC simulation and are sampled from a known probability distribution function. On this basis, the trajectories of each primary particle and its secondaries are simulated and the deposited dose is aggregated, by sampling interactions such as scattering and energy loss according to physical laws and material properties. While this appears to be an intuitive simulation of the actual physical process, it is essentially a statistical method to solve the linear Boltzmann transport equation and therefore compute the expected value of a model with random input. 

Let $\boldsymbol{\xi}$ be the vector of random input parameters of the dose simulation. $\Phi_0(\boldsymbol{\xi})$ is the joint density of these parameters and is assumed to be known. For our purposes, which will not interfere with the simulation itself, we assume that the trajectory of a primary particle is given by the "black box" simulation engine, yielding the dose deposited in voxel $i$ within an individual particle's history $h_i(\boldsymbol{\xi})$. 

The nominal dose $d_i$ in voxel $i$ can now be estimated with the expected value $\mathbb{E}_{\Phi_0}$ of all histories via the sample mean

\begin{equation}
\label{eq:nomDose}
d_i = \mathbb{E}_{\Phi_0}[h_{i}(\boldsymbol{\xi})] \approx \frac{1}{H}\sum_{p=1}^H h_{i}(\boldsymbol{\Xi}_p) \;,
\end{equation}
where $H$ is the sample size (number of computed primary particle histories) and $\boldsymbol{\Xi}_p$, $p=1,...,H$ are realizations of primary particle properties. 

Here we omit the dependence on random factors within the simulation, such as particle scattering, as well as their probability distribution. Particle histories $h_{i}(\boldsymbol{\Xi}_p)$, for input realizations $\boldsymbol{\Xi}_p$, implicitly also include realizations of these random parameters. For a large number of histories, their effect on the dose estimates can however be assumed to be constant.

\subsection{Beam model}
\label{sec:beamModel}
The initial state of each particle is represented by a point in the seven-dimensional phase space, which encompasses the particle position $\boldsymbol{r}=(r_x,r_y,r_z)$, momentum $\boldsymbol{p}=(p_x, p_y, p_z)$ and energy $E$. We assume a Gaussian emittance model, i.e. the parameters within each pencil beam are multivariate normal distributed with $\Phi_0^b(\boldsymbol{\xi})=\Phi_0^b(\boldsymbol{r},\boldsymbol{\varphi},E)=\mathcal{N}(\boldsymbol{\mu}_{\xi}^b,\boldsymbol{\Sigma}_{\xi}^b)$, for pencil beams $b=1,..,B$. Here, $\boldsymbol{\varphi}=(\varphi_x,\varphi_y)=(\frac{dp_x}{dp_z},\frac{dp_y}{dp_z})$ describes the transverse divergence of the momentum direction from the axial beam direction.

The joint density over all pencil beams is then defined by a Gaussian mixture model 
\begin{equation}
\Phi_0(\boldsymbol{\xi})=\sum_{b=1}^B w_b \Phi_0^b(\boldsymbol{\xi})\;,
\end{equation} where $w_b$ are the pencil beam weights. 

To introduce our method, we will initially assume a simplified phase space where $\varphi_x = \varphi_y = 0$. Results including a distribution in the momentum direction can be found in the Appendix. 

\subsection{Uncertainties}
\label{sec:uncertainties}
Among the most important sources of uncertainty in proton therapy are errors in the patient set-up $\boldsymbol{\delta}_r=(\delta_{r_x},\delta_{r_y},\delta_{r_z}
)$ and the proton range $\delta_{\rho}$ \citep[comp.][]{parkStatisticalAssessmentProton2013, liuRobustOptimizationIntensity2012,perkoFastAccurateSensitivity2016,lomaxIntensityModulatedProton2008, lomaxIntensityModulatedProton2008a}. While these errors are random variables with, in principle, unknown probability distributions, we follow the common approach of assuming normally distributed errors \citep{wieserImpactGaussianUncertainty2020,unkelbachAccountingRangeUncertainties2007,perkoFastAccurateSensitivity2016, bangertAnalyticalProbabilisticModeling2013, fredrikssonMinimaxOptimizationHandling2011}.

 Set-up errors directly affect the primary particle positions in an additive way, such that the actual position $\boldsymbol{r}_{\delta}$ of a primary particle under uncertainty is given by its position $\boldsymbol{r}$ according to the emmitance model, plus the error $\boldsymbol{\delta}_r$.
 
 Uncertainties in the particle range are caused by a variety of factors, ranging from the conversion of Hounsfield units to stopping powers and imaging artifacts, over changes in the patient geometry to biological effects and inaccuracies in physics models \citep{unkelbachAccountingRangeUncertainties2007,paganettiRangeUncertaintiesProton2012, lomaxIntensityModulatedProton2008,mcgowanTreatmentPlanningOptimisation2013}. Here, we focus on calculational uncertainties, such as conversion errors, and model these by scaling the complete tissue density with the random factor $\delta_{\rho}$ \citep[comp.][]{lomaxIntensityModulatedProton2008, sourisTechnicalNoteMonte2019,malyapaEvaluationRobustnessSetup2016}. Since the density is assumed to be deterministic,  the error is not directly linked to a random input parameter. In section \ref{sec:rangeError}, we however present an approximation which models range errors using the initial energy distribution.
 
 Sampling-based uncertainty quantification approaches, similar to \citet{parkStatisticalAssessmentProton2013} or \citet{kraanDoseUncertaintiesIMPT2013}, rely on repeated dose calculations for different realizations $\boldsymbol{\Delta}_k, k=1,...,K$ of the error vector. For an individual error scenario $\boldsymbol{\Delta}_k$, the dose is computed as
 \begin{equation}
 \label{eq:errorScenarioDose}
 d^{\Delta_k}_i = \mathbb{E}_{\Phi(\boldsymbol{\xi},\boldsymbol{\Delta}_k)}[h_{i}(\boldsymbol{\xi},\boldsymbol{\Delta}_k)] \approx \frac{1}{H}\sum_{p=1}^H h_{i}(\boldsymbol{\Xi}_p) \;,\;\; \boldsymbol{\Xi}_p \sim \Phi(\boldsymbol{\xi},\boldsymbol{\Delta}_k)
 \end{equation}
 
In the case of set-up uncertainties $\Phi(\boldsymbol{\xi},\boldsymbol{\Delta}_k)$ for example corresponds to the nominal parameter density $\Phi_0(\boldsymbol{\xi})$, where all particle positions are shifted by $\boldsymbol{\Delta}_{r;k}$. Due to its accuracy, this procedure is later used to obtain reference values to validate our results. It is however extremely computationally expensive, since it requires numerous runs of the complete Monte Carlo dose simulation.

\subsection{Direct computation of the expected value}
\label{sec:directExpDose}
When the distribution $\Psi(\boldsymbol{\xi}_{\delta})$ of the initial parameters under uncertainty can be explicitly defined, it is possible to compute the expected dose directly by replacing the nominal parameter distribution $\Phi_0$ with $\Psi$ in the Monte Carlo dose simulation as follows:

\begin{equation}
\label{eq:expectedDose}
E(d_i) = \mathbb{E}_{\Psi(\boldsymbol{\xi}_{\delta})}[h_{i}(\boldsymbol{\xi}_{\delta})] \approx \frac{1}{H}\sum_{p=1}^H h_{i}(\boldsymbol{\Xi}_p) \;,\;\; \boldsymbol{\Xi}_p \sim \Psi(\boldsymbol{\xi}_{\delta})
\end{equation}
For example when the error is additive, i.e. 
\begin{equation}
\boldsymbol{\xi}_{\delta}=\boldsymbol{\xi} + \boldsymbol{\delta}
\end{equation}
and $\boldsymbol{\xi}\sim\mathcal{N}(\boldsymbol{\mu}_{\xi},\boldsymbol{\Sigma}_{\xi})$, as well as $\boldsymbol{\delta}\sim\mathcal{N}(\boldsymbol{\mu}_{\delta},\boldsymbol{\Sigma}_{\delta})$, the distribution of $\boldsymbol{\xi}_{\delta}$ is the convolution $\Psi=\mathcal{N}(\boldsymbol{\mu}_{\xi}+\boldsymbol{\mu}_{\delta},\boldsymbol{\Sigma}_{\xi}+\boldsymbol{\Sigma}_{\delta})$. For $\boldsymbol{\mu}_{\delta}=0$, this is just a wider Gaussian distribution.

\subsection{Importance (re-)weighting}
\label{sec:importanceRew}

We now consider the dose deposited by histories $h(\boldsymbol{\xi},\boldsymbol{\delta})$, which are a function of the random input parameters $\boldsymbol{\xi}\sim \Phi_0(\boldsymbol{\xi})$ and random error vector $\boldsymbol{\delta}\sim p_{\delta}$. In the following we focus on computing estimates for the dose expected value and standard deviation, the method can however be analogously applied to the computation of several worst case scenarios.

We propose a replacement of the dose calculations for different error scenarios by a more efficient weighting of particle histories $h$. For this, we adopt the concept of importance sampling \citep{kahnRandomSamplingMonte1950, hastingsMonteCarloSampling1970}. Instead of sampling primary particles from $\Phi(\boldsymbol{\xi},\boldsymbol{\Delta}_k)$ for different error scenarios, we sample from a different density function - e.g. the nominal parameter distribution $\Phi_0(\boldsymbol{\xi})$. Then, the dose for all scenarios can be estimated using histories from the nominal dose calculation:

\begin{eqnarray}
\label{eq:shiftedDoseIR}
d^{\Delta_k}_i &= \mathbb{E}_{\Phi(\boldsymbol{\xi},\boldsymbol{\Delta}_k)}[h_{i}(\boldsymbol{\xi},\boldsymbol{\Delta}_k)]\nonumber\\ 
 &\approx \frac{1}{H}\sum_{p=1}^H h_{i}(\boldsymbol{\Xi}_p) \frac{\Phi(\boldsymbol{\Xi}_p,\boldsymbol{\Delta}_k)}{\Phi_0(\boldsymbol{\Xi}_p)} \;,\;\; \boldsymbol{\Xi}_p \sim \Phi_0(\boldsymbol{\xi})
\end{eqnarray}

Thus, scenario computation reduces to a scoring problem. Dose expectation and variance can now be computed through sample mean and variance over the respectively obtained scenarios.

\begin{equation}
\label{eq:varianceIR}
Var(d_i) \approx \frac{1}{K-1}\sum_{k=1}^K (d^{\Delta_k}_{i}- E[d_i])^2\;.
\end{equation}

\begin{equation}
\label{eq:expDoseIR}
E[d_i] \approx \frac{1}{K}\sum_{k=1}^K d^{\Delta_k}_{i}\;.
\end{equation}

 When it is possible to compute the expected value directly as discussed in \ref{sec:directExpDose}, only one (re-)weighting step is necessary:

\begin{equation}
\label{eq:expDoseIRDirect}
E[d_i] \approx \frac{1}{H}\sum_{p=1}^H h_{i}(\boldsymbol{\Xi}_p) \frac{\Psi(\boldsymbol{\Xi}_p)}{\Phi_0(\boldsymbol{\Xi}_p)} \;,\;\; \boldsymbol{\Xi}_p \sim \Phi_0(\boldsymbol{\xi})
\end{equation}

\subsection{Modeling set-up uncertainties}
\label{sec:setupUncert}

Set-up uncertainties correspond to a shift of the patient position or equivalently the positions of primary particles relative to the patient. While errors occur in three dimensional space, shifts along the beam axis do not affect the dose distribution. In the Gaussian model, set-up errors can hence be assumed to follow a bivariate normal distribution for each pencil beam $b=1,..,B$:
\begin{equation}
\boldsymbol{\delta}_r^b=(\delta^b_{r_x},\delta^b_{r_y}) \sim \mathcal{N}(\boldsymbol{\mu}_{\delta_r}^b,\boldsymbol{\Sigma}_{\delta_r}^b) \;,
\end{equation} with $\boldsymbol{\mu}_{\delta_r}^b \in \mathbb{R}^2$ and $\boldsymbol{\Sigma}_{\delta_r}^b \in \mathbb{R}^{2\times 2}$.

Particles are initialized in a 2D plane, thus the primary particle positions follow a bivariate Gaussian mixture (\ref{sec:beamModel})
\begin{equation} 
\label{eq:nomDistSetup}
\Phi_{0;r}(\boldsymbol{r})= \sum_{b=1}^B w_b \Phi^b_{0;r}(\boldsymbol{r}) \,,\;\; \Phi^b_{0;r}=\mathcal{N}(\boldsymbol{r};\boldsymbol{\mu}_r^b,\boldsymbol{\Sigma}_r^b) \; .
\end{equation}

Here $\boldsymbol{\mu}_r^b$ is the mean lateral position of initial particles in pencil beam $b$ in beam's eye view i.e., in the 2D plane perpendicular to the central beam axis. Then, according to \ref{sec:uncertainties} the initial position $\boldsymbol{r}_\delta$ of a particle under uncertainty is determined by

\begin{equation}
\boldsymbol{r}_\delta=\boldsymbol{r}+\boldsymbol{\delta}_r\;
\end{equation}
and $\boldsymbol{r}_\delta$ is distributed with the convolution function
\begin{equation} 
\label{eq:expDistSetup}
\Psi_r = \sum_{b=1}^B w_b \Psi_r^b \,,\;\;\Psi_r^b=\mathcal{N}(\boldsymbol{\mu}_r^b+\boldsymbol{\mu}_{\delta_r}^b,\boldsymbol{\Sigma}_r^b + \boldsymbol{\Sigma}_{\delta_r}^b) \;.
\end{equation}

An individual error realization $\boldsymbol{\Delta}_{r;k}\sim p_{\delta_r}$ then formally just corresponds to a shift of the original primary positions, which now follow the distribution
\begin{equation}
\label{eq:shiftedDistErrorScen}
\Phi_r(\boldsymbol{r},\boldsymbol{\Delta}_k) = \sum_{b=1}^B w_b \cdot \mathcal{N}(\boldsymbol{r};\boldsymbol{\mu}_r^b+\boldsymbol{\Delta}^b_{r;k},\boldsymbol{\Sigma}_r^b) \;,
\end{equation} 
corresponding to the nominal distribution shifted by $\boldsymbol{\Delta}_{r;k}$.

The above distributions can be directly used with (\ref{eq:shiftedDoseIR}), (\ref{eq:varianceIR}) and (\ref{eq:expDoseIRDirect}) to obtain the expected dose and variance for set-up uncertainties.

\subsection{Modeling range uncertainties}
\label{sec:rangeError}

The proposed approach could be analogously applied to any type of uncertainty directly affecting input parameters of the simulation, which have an a-priori probability distribution. Range uncertainties, however, modify the density values, which are deterministic and can thus not be directly modeled within the proposed framework.

To still approximate our quantities of interest, we exploit that the largest dose uncertainty is induced near the range of a beam \citep{bortfeldAnalyticalApproximationBragg1997}, although the uncertain density variation affects the whole trajectory. Range can be expressed in terms of the initial energy of particles, using the Bragg-Kleemann rule 
\begin{equation}
\label{eq:braggKleemann}
R=\alpha \cdot E_0^{p} \;, 
\end{equation}

where $R$ is the range, $E_0$ is the initial energy and $\alpha$ and $p$ are application-specific parameters. For the case of the slow-down of therapeutic protons in water, values of $\alpha=0.022$  mm/MeV$^p$ and $p=1.77$ can be chosen \citep{ulmerTheoreticalMethodsCalculation2011}. 

The initial energy spectrum of a scanned pencil beam at the exit of the nozzle can be approximately represented by a Gaussian \citep{bortfeldAnalyticalApproximationBragg1997,kimstrandBeamSourceModel2007,tourovskyMonteCarloDose2005,soukupPencilBeamAlgorithm2005}. We can use this to model range uncertainties through random variations of the initial energy \citep[compare treatment of range straggling in][]{pedroniExperimentalCharacterizationPhysical2005}. 

Let's assume range uncertainties are normally distributed, i.e. $\delta_{\rho} \sim \mathcal{N}(0,\sigma_R^2)$ \citep[comp.][]{lomaxIntensityModulatedProton2008, yangComprehensiveAnalysisProton2012}. With a Taylor approximation (order 1 for the mean and 2 for the variance) around $X=E[X]$, we can determine the parameters $\mu_{E_0},\sigma^2_{E_0}$ of the energy distribution due to range uncertainties.
\begin{eqnarray}
E_0&=(\frac{1}{\alpha}R)^{\frac{1}{p}}=:g(E[R])) \\
\mu_{E_0}&=E[g(E[R])] \approx g(E[R]) = (\frac{1}{\alpha}E[R])^{\frac{1}{p}} \\
\sigma^2_{E_0}&=Var(g(E[R])) \approx g'(E[R])^2 Var(E[R]) \\
 &= \left( \frac{1}{p \alpha}(E[R]\frac{1}{\alpha})^{\frac{1}{p}-1} \right)^2\sigma^2_R \nonumber \end{eqnarray}

Thus the randomness in range is approximated through an energy distribution $E_0 \sim \mathcal{N}(\mu_{E_0},\sigma^2_{E_0})$ and the expected dose and variance can be computed by (re-)weighting histories analogously to \ref{sec:setupUncert}. This can again be extended to multiple pencil beams using Gaussian mixtures.

Note that, again, the expected value can be directly computed from simulations with an energy spectrum convolved with the Gaussian uncertainty kernel. Alternatively to the nominal energy distribution, this convolved distribution can also be used to obtain the required histories. In this case, the nominal distribution is replaced by the convolved distribution in (\ref{eq:shiftedDoseIR})-(\ref{eq:expDoseIRDirect}).

\subsection{Correlation Models}
\label{sec:corrModels}
In the previous sections, the distributions of different types of phase space parameters were considered independently. Note that the derived distributions are however all marginals of the joint multivariate Gaussian mixture spanning the complete phase space and all pencil beams (see also \ref{sec:beamModel}). 

Similarly, the univariate normal distributions of errors of different types and in different pencil beams can be connected using a joint multivariate Gaussian distribution. This framework in principle allows for the definition of arbitrary correlation models for uncertainties between pencil beams. For the dose variance computation, these correlations can be easily implemented using the covariance matrix of this joint distribution, since the samples for the weighted scenarios are directly drawn from the respective multivariate normal distribution. The expected dose is constant under varying correlation assumptions. 

Using the example of set-up uncertainties, if we define the errors in each beam by
$\boldsymbol{\delta}_r = (\delta^1_{r},...,\delta^B_{r})^T$, the multivariate Gaussian $\mathcal{N}(\boldsymbol{\mu},\boldsymbol{C})$ would be parametrized with

\begin{equation*}
     \boldsymbol{\mu} =\left(\matrix{ \boldsymbol{\mu}_{\delta_r}^1 \cr\boldsymbol{\mu}_{\delta_r}^2 \cr\vdots\cr\boldsymbol{\mu}_{\delta_r}^B }\right),
\end{equation*}
\begin{equation} 
\boldsymbol{C} = \left( \begin{array}{ccccc}
\boldsymbol{\Sigma}_r^1 & \begin{array}{cc}
\rho_{xx}^{12} & \rho_{xy}^{12} \\
\rho_{yx}^{12} & \rho_{yy}^{12}\\
\end{array}  & \cdots & \begin{array}{cc}
\rho_{xx}^{1B} & \rho_{xy}^{1B}\\
\rho_{yx}^{1B} & \rho_{yy}^{1B}\\
\end{array}\\

\begin{array}{cc}
\rho_{xx}^{21} & \rho_{xy}^{21}\\
\rho_{yx}^{21} & \rho_{yy}^{21} 
\end{array}& \boldsymbol{\Sigma}_r^2 & &\\

\vdots & &\ddots & \vdots\\

\begin{array}{cc}\rho_{xx}^{B1} & \rho_{xy}^{B1}\\

\rho_{yx}^{B1} & \rho_{yy}^{B1}\\
\end{array} & \cdots & &\boldsymbol{\Sigma}_r^B \\\end{array} \right) \;,
\end{equation}
where $\rho_{xy}^{ab}$ is the covariance between set-up errors in the x-direction in pencil beam $a$ and errors in the y-direction in pencil beam $b$.

\begin{figure}[ht!]
	\begin{tabular}{c c c c c}
		\multicolumn{5}{l}{\includegraphics[width=0.8\textwidth]{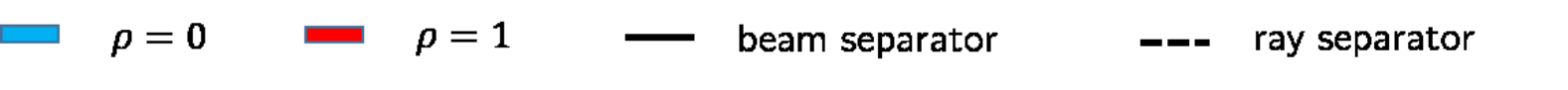}}\\[-0.5ex]
		\includegraphics[width=0.17\textwidth]{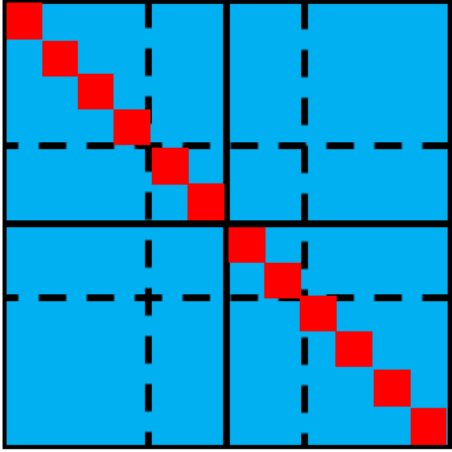}&\includegraphics[width=0.17\textwidth]{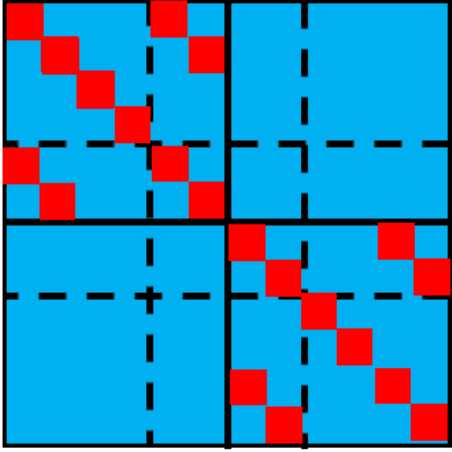}&\includegraphics[width=0.17\textwidth]{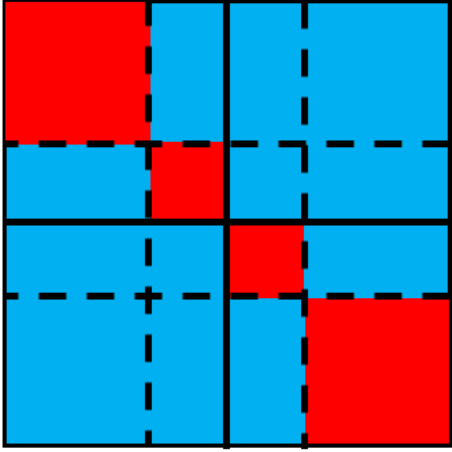}&\includegraphics[width=0.17\textwidth]{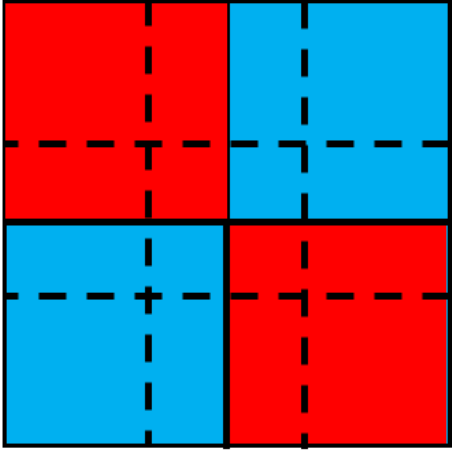}&\includegraphics[width=0.17\textwidth]{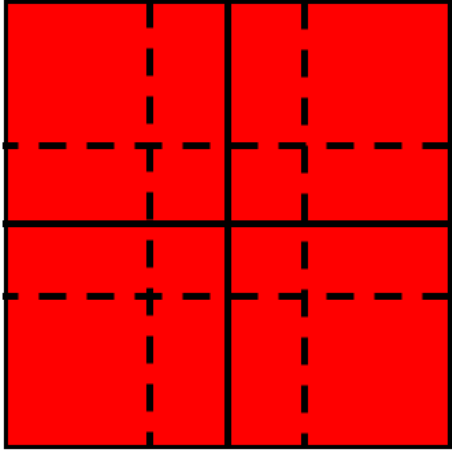} \\
		& & & & \\[-1.5ex]
		(a)&(b)&(c)&(d)&(e)\\
	\end{tabular}
	\caption{Adapted from \citet{wahlAnalyticalModelsProbabilistic2018a}. Covariance matrices for different correlation assumptions. Rows and columns of the matrices correspond to the individual pencil beams, beam and ray separators indicate sections of pencil beams with the same irradiation angle and lateral position, respectively. (a) No correlation between pencil beams,  (b) correlation of energy levels within one beam, (c) ray-wise correlation, all pencil beams with the same lateral position, i.e.\ hitting the same material are fully correlated (d) beam-wise correlation, pencil beams with the same irradiation angle are fully correlated and (e) errors in all pencil beams are fully correlated.}
	\label{fig:corrModels}
\end{figure}

A few simple examples for correlation models are shown in figure \ref{fig:corrModels}, more can be found in literature  \citep{bangertAnalyticalProbabilisticModeling2013, pflugfelderWorstCaseOptimization2008, unkelbachReducingSensitivityIMPT2009}.

In case the correlation matrix is singular (perfect correlation between some pencil beams), the dimension of the uncertain vector can be reduced and one joint error can be sampled for the respective perfectly correlated pencil beams. 

More complex correlation models are possible.

\subsection{Implementation}
\label{sec:implementation}
For the proof-of-concept in this work, the weighting method was implemented as a post-processing routine in Matlab. Radiation plans were generated with matRad \citep{wieserDevelopmentOpensourceDose2017b} and exported to the Monte Carlo simulation engine TOPAS  \citep{perlTOPASInnovativeProton2012} for dose calculations. The required particle histories $h(\boldsymbol{\xi})$ are stored during the simulation using a custom extension for TOPAS.

All reference computations rely on Monte Carlo dose calculations with TOPAS as well.

To reduce the number of required realizations, both for the reference computation and the (re-)weighting steps, a quasi-Monte Carlo approach was used to sample the random parameters \citep[see e.g.][]{caflischMonteCarloQuasiMonte1998}.

\subsection{Investigated patients and uncertainties}

In the following, we consider range and set-up uncertainties, as well as a combination of both. For the set-up uncertainties, we assume a symmetric, bivariate normal distribution with zero mean (no systematic errors) and a standard deviation of \SI{3}{\milli\meter} \citep[comp.][]{wahlAnalyticalModelsProbabilistic2018a, perkoFastAccurateSensitivity2016}. For range uncertainties,  in the reference computations we scale the density with a normally distributed factor, where the mean is equal to the nominal density and the standard deviation is \SI{3}{\percent} \citep[as recommended in][]{lomaxIntensityModulatedProton2008a, yangComprehensiveAnalysisProton2012}. The corresponding parameters of the energy distribution, used to approximate range errors in the importance (re-)weighting estimate, are determined based on this distribution as detailed in \ref{sec:rangeError}. Table \ref{table:overviewPatientsUnc} provides an overview of which uncertainty models were computed for which patient, as well as the irradiation angles used for different patients.

\begin{table}[htb!]
	\centering
		\caption{Overview of uncertainties investigated for each patient/test case. Beam irradiation angles are given as (couch angle, gantry angle) in degrees and error values refer to the standard deviations of the corresponding normal distributions.}
	\begin{tabular}{l l l l l l }
		\br
		 & &Patient & Water phantom & Prostate & Liver \\
		&&Angles& (\ang{0}, \ang{0}) & (\ang{0}, \ang{90}/\ang{270}) &(\ang{0}, \ang{315})  \\
		Correlation & Type &  &&&\\
		\mr
		\multirow{ 3}{*}{Full}& Set-up && \SI{3}{\milli\meter}& \SI{3}{\milli\meter}& \SI{3}{\milli\meter}\\
		&Range &&\SI{3}{\percent}&-&\SI{3}{\percent}\\
		&Both& & \SI{3}{\milli\meter}/\SI{3}{\percent}&-& \SI{3}{\milli\meter}/\SI{3}{\percent}\\
		Beam& Set-up& &-& \SI{3}{\milli\meter}&-\\
		Ray& Set-up& &-& \SI{3}{\milli\meter}&-\\
		Energy& Both &&-& \SI{3}{\milli\meter}/\SI{3}{\percent}&-\\
		None& Set-up &&-& \SI{3}{\milli\meter}& -\\
		\br
	\end{tabular}
	\label{table:overviewPatientsUnc}
\end{table}

The number of histories and pencil beams for each considered patient, as well as the number of error scenarios computed for the importance (re-)weighting estimates can be found in table \ref{table:patientStatistics}. Note, that the number of histories per pencil beam were determined based on (non-normalized) weights from the optimized radiation plan (see \ref{sec:implementation}), where around $10^5$ histories are computed for the pencil beams with the highest weights. 

\begin{table}[htb!]
	\centering
	\caption{Overview of simulated plans and error scenarios per patient.}
	\lineup
	\begin{tabular}{l l l l l }
		\br
		Patient & Water phantom & Liver & \multicolumn{2}{c}{Prostate} \\
		\mr
		Irradiation angles &\0\0 (\ang{0}, \ang{0}) & \0(\ang{0}, \ang{315}) &\0\0(\ang{0}, \ang{90}) &\0(\ang{0}, \ang{270}) \\
		Number of pencil beams & \,\0\0\0\0\0147 & \0\0\0\0\01\,378 & \0\0\0\0\01\,375 & \0\0\0\0\01\,383 \\
		Number of histories &2\,566\,453 &13\,528\,430 & 16\,992\,193 & 16\,748\,034\\
		Number of error scenarios & \,\0\0\0\0\0100 & \,\0\0\0\0\0\0100 & \multicolumn{2}{c}{100} \\
		\br
	\end{tabular}
	\label{table:patientStatistics}
\end{table}

\subsection{Evaluation criteria}

To compare our results to the respective reference computations, we plot two-dimensional slices of the dose cubes as well as a difference map and employ a global three-dimensional $\gamma$-analysis. For the difference maps, we compute 
\begin{equation}
 \text{diff}_i(d^\text{ref}_i,d_i^\text{est}) = d_i^\text{ref} - d_i^\text{est} \; ,
\end{equation}
for each voxel $i$ in the reference result $d^\text{ref}$ and (re-)weighting estimate $d^\text{est}$. For the  $\gamma$-analysis, we use the matRad implementation based on \citet{lowTechniqueQuantitativeEvaluation1998}, with a distance to agreement of $\SI{3}{\milli\meter}$ and a dose difference criterion of $\SI{3}{\percent}$.

\section{Results}
\label{sec:results}

In the following we present results for the cases given in table \ref{table:overviewPatientsUnc}. Unless specified otherwise, results were computed on the basis of histories from nominal dose calculations, i.e. with phase space parameters sampled from $\Phi_0$ (see \ref{sec:beamModel}). The references computed for nominal and expected dose stem from Monte Carlo dose calculations with the respective phase space distributions $\Phi_0$ and $\Psi$ (see \ref{sec:directExpDose}), the reference standard deviation is derived using numerous such Monte Carlo simulations for different error scenarios sampled from the joint error distribution. Therefore, the importance (re-)weighting estimate for the nominal dose only differs from the reference by round-off errors introduced in post-processing, as can be seen in figure \ref{fig:WB_Setup}. It is omitted thereafter, as is the reference. 

\subsection{Set-up errors}
\begin{figure}[H]
	\centering 
	\begin{tabular}{c c c c}
		& \textbf{Estimate}\hspace*{0.5cm} & \textbf{Reference}\hspace*{0.5cm}&\textbf{Difference}\hspace*{0.5cm} \\
		$\boldsymbol{d}$  & 	\raisebox{-.5\height}{\includegraphics[width=0.25\textwidth]{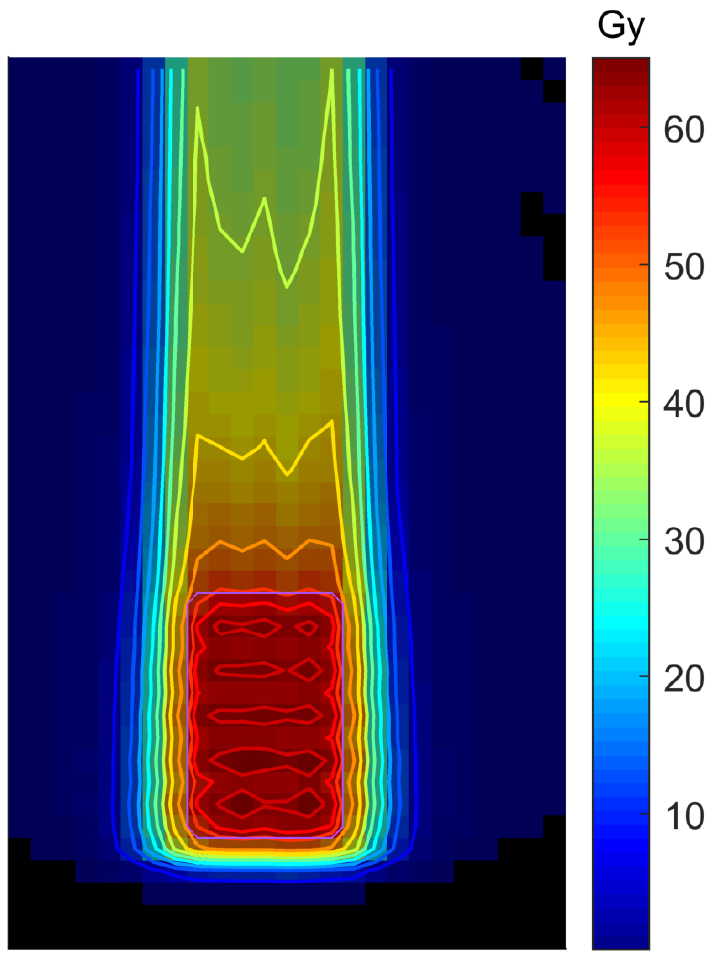}}&\raisebox{-.5\height}{\includegraphics[width=0.25\textwidth]{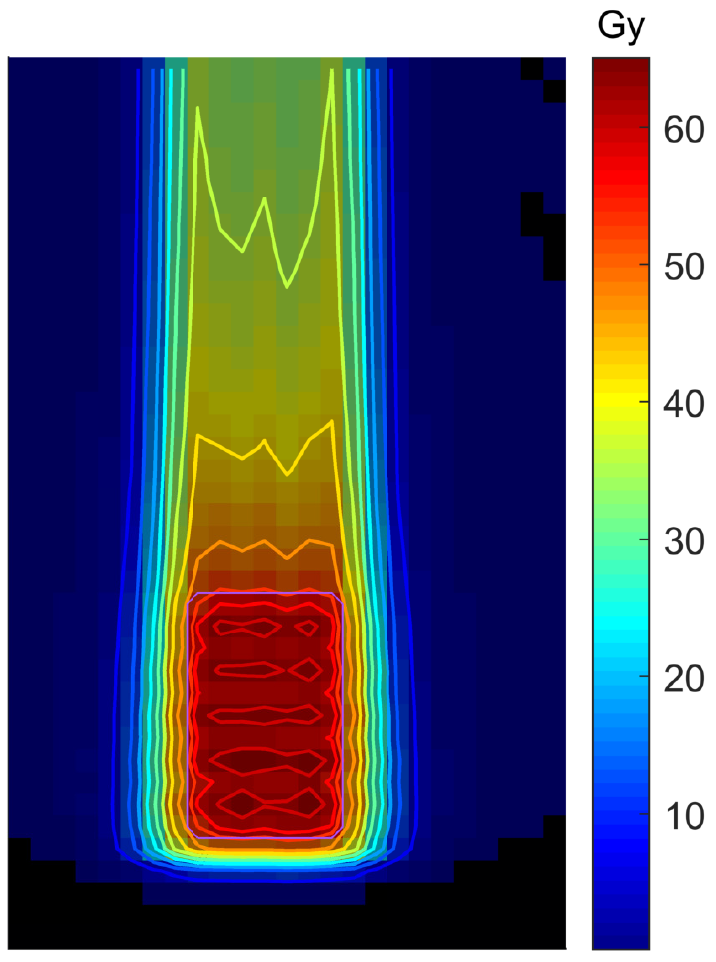}}&\raisebox{-.5\height}{\includegraphics[width=0.255\textwidth]{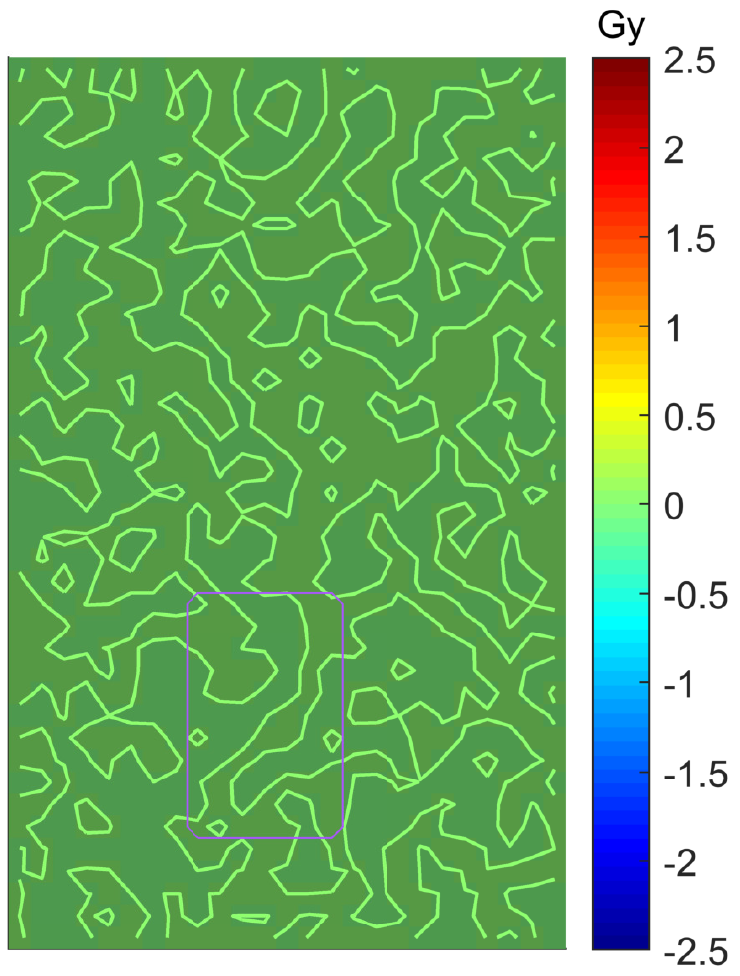}} \\
		
		$E[\boldsymbol{d}]$&\raisebox{-.5\height}{\includegraphics[width=0.25\textwidth]{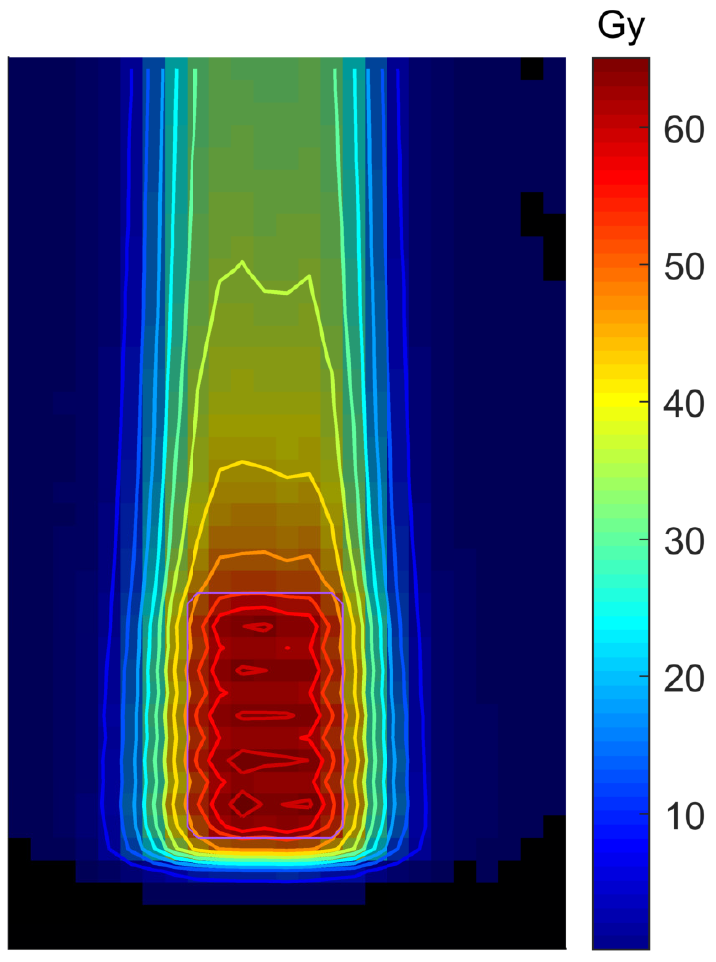}}&\raisebox{-.5\height}{\includegraphics[width=0.25\textwidth]{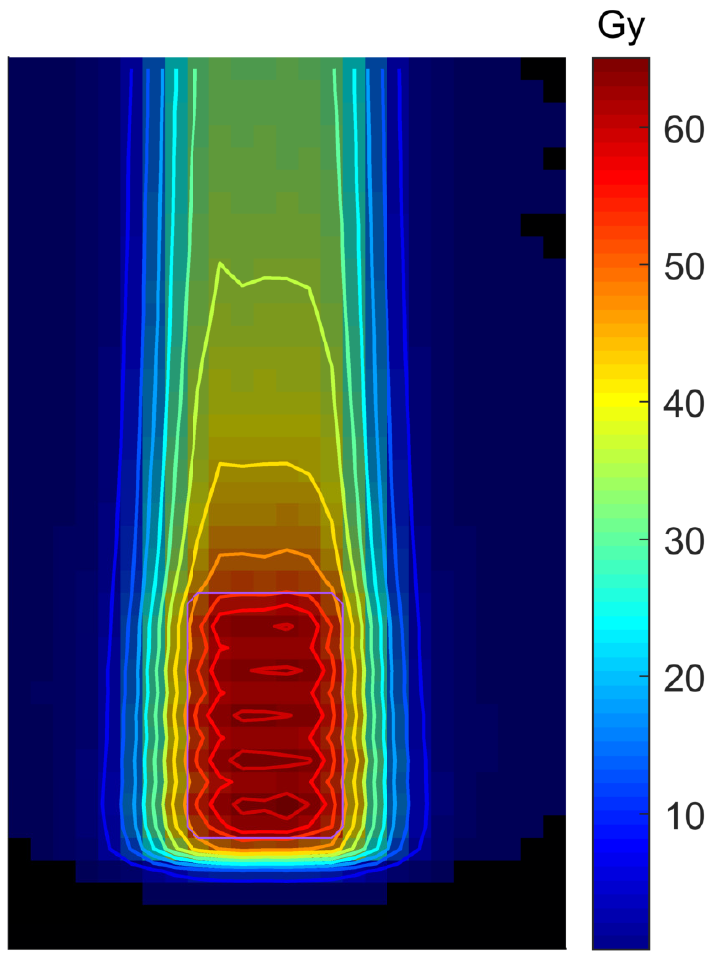}}&\raisebox{-.5\height}{\includegraphics[width=0.255\textwidth]{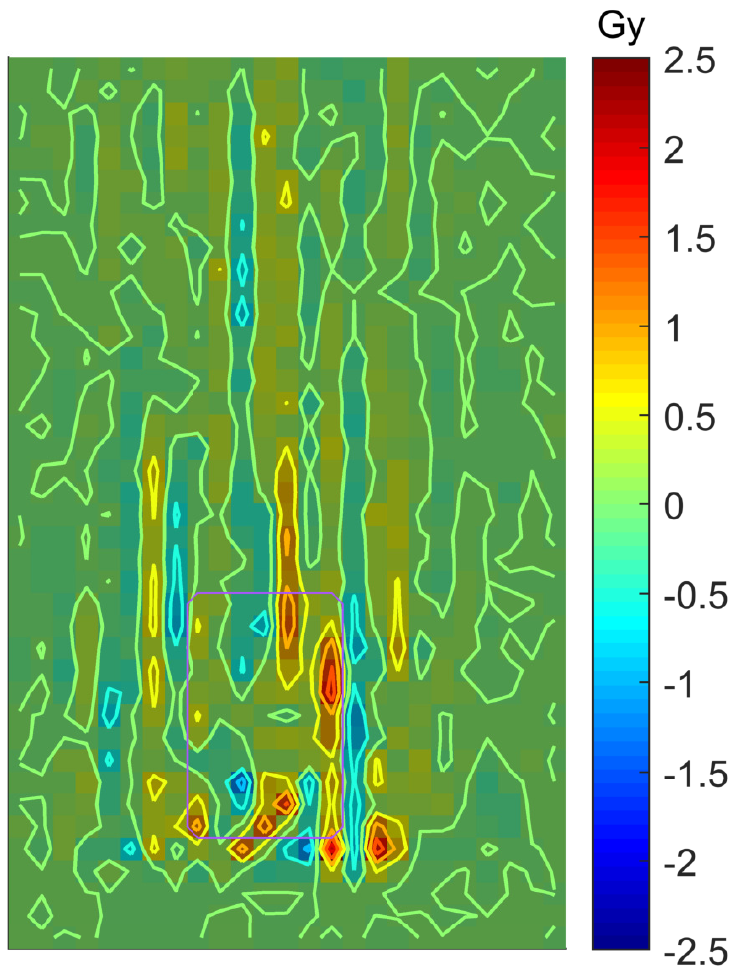}}\\
		
		$\boldsymbol{\sigma(d)}$&\raisebox{-.5\height}{\includegraphics[width=0.25\textwidth]{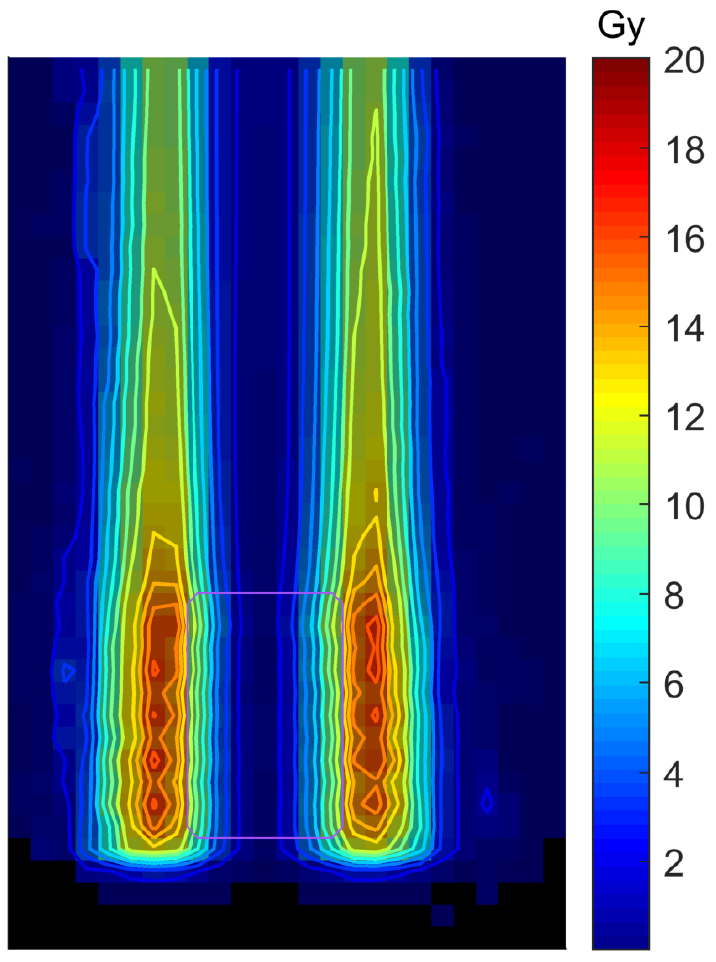}}&\raisebox{-.5\height}{\includegraphics[width=0.25\textwidth]{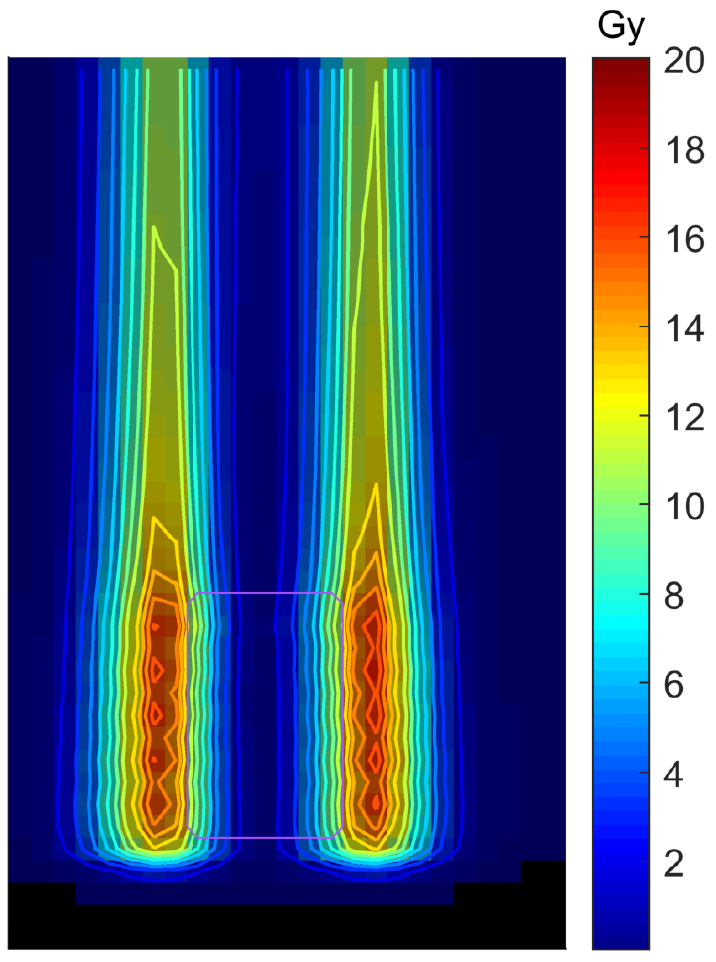}}&\raisebox{-.5\height}{\includegraphics[width=0.255\textwidth]{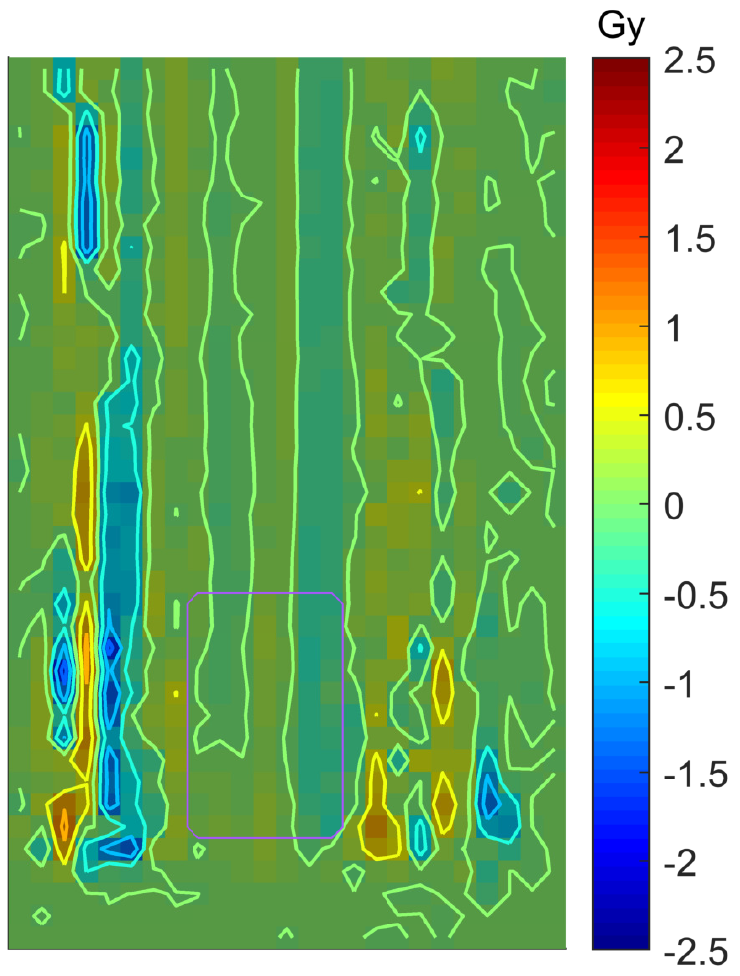}}\\
		
	\end{tabular}
	\caption{Nominal dose $\boldsymbol{d}$, expected dose $E[\boldsymbol{d}]$ and standard deviation $\boldsymbol{\sigma(d)}$ w.r.t.\ set-up uncertainties with $\SI{3}{\milli\meter}$ standard deviation for a spread out Bragg peak in a water phantom. The left column shows the estimate computed with the proposed (re-)weighting approach, the middle column the respective reference and the right column the difference between both simulations.}
	\label{fig:WB_Setup}
\end{figure}

Figure \ref{fig:WB_Setup} displays the nominal dose, expected value and standard deviation estimates for a water phantom, computed using the (re-)weighting approach in comparison to the respective references. While we see some minor deviations in difference maps for the expected dose and standard deviations, they do not appear systematic. 

\begin{figure}[H]
	\centering 
	\begin{tabular}{c@{\hspace{-0.1ex}} c @{\hspace{-0.1ex}} c c@{\hspace{-0.1ex}} c}
	& \textbf{Estimate}\hspace*{0.5cm} & \textbf{Difference}\hspace*{0.5cm}&  \textbf{Estimate}\hspace*{0.5cm} & \textbf{Difference}\hspace*{0.5cm} \\
	
	$E[\boldsymbol{d}]$&\raisebox{-.5\height}{\includegraphics[width=0.22\textwidth]{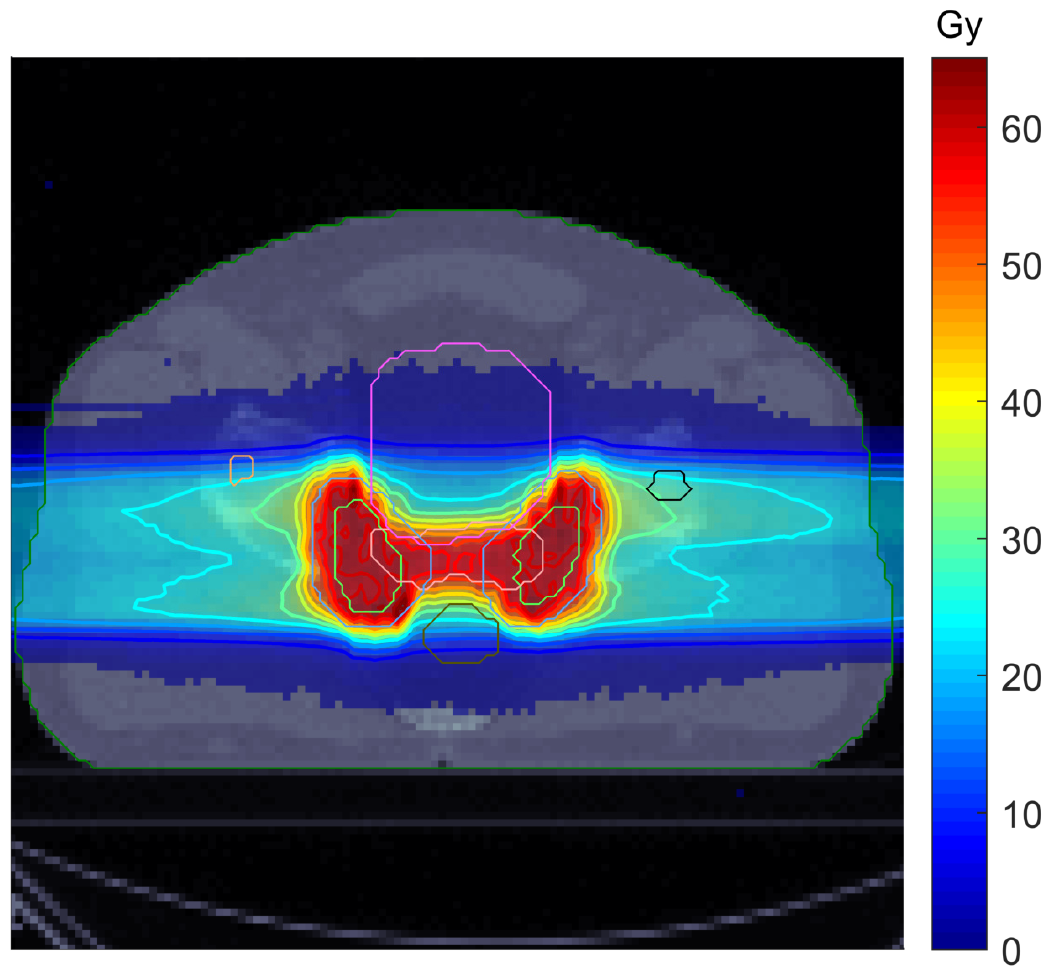}}&\raisebox{-.5\height}{\includegraphics[width=0.225\textwidth]{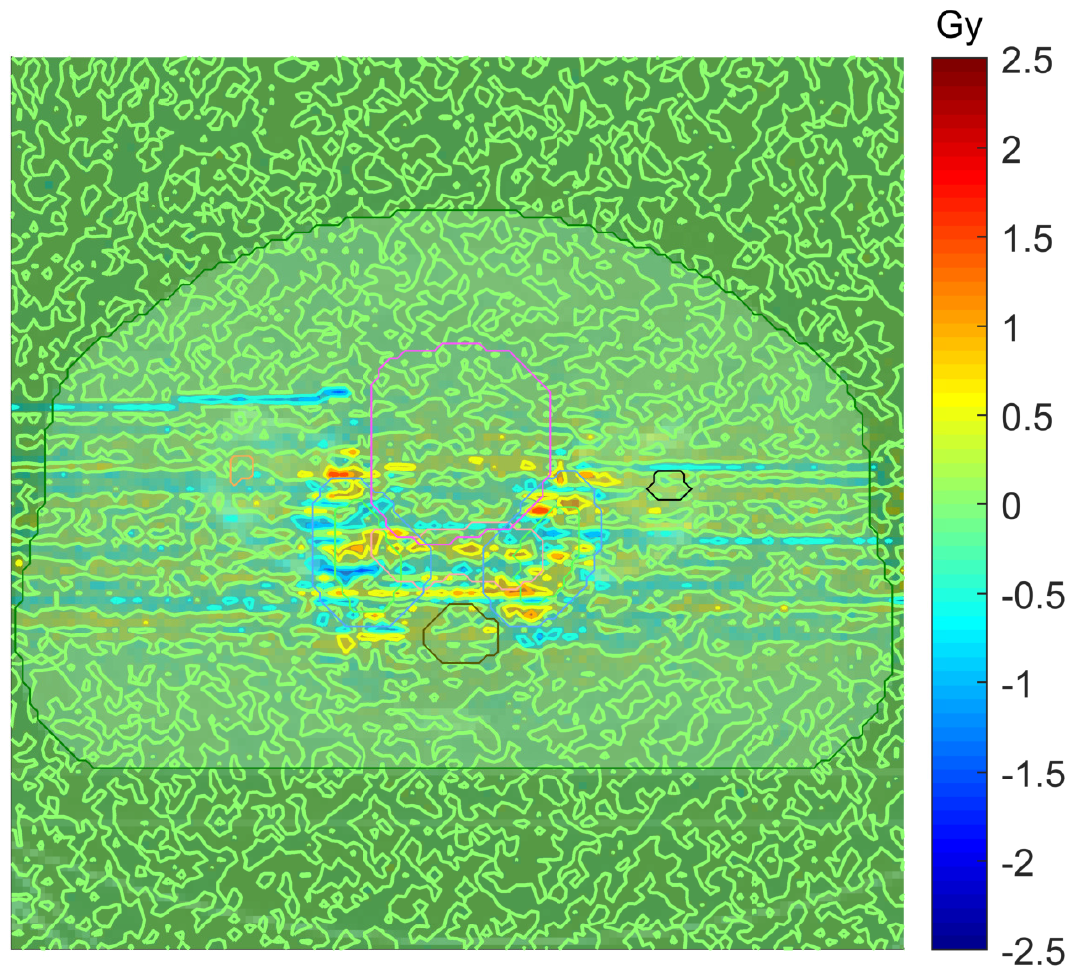}}&\raisebox{-.5\height}{\includegraphics[width=0.22\textwidth]{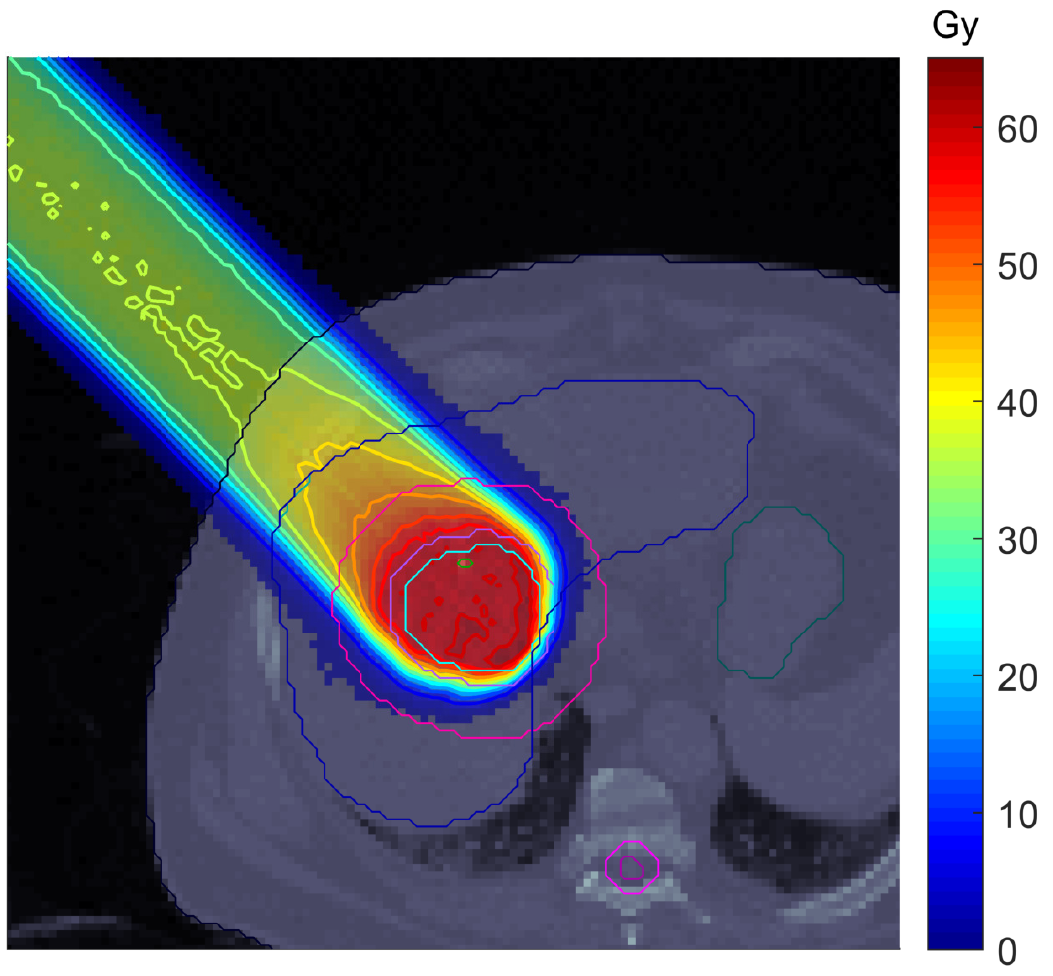}}&\raisebox{-.5\height}{\includegraphics[width=0.225\textwidth]{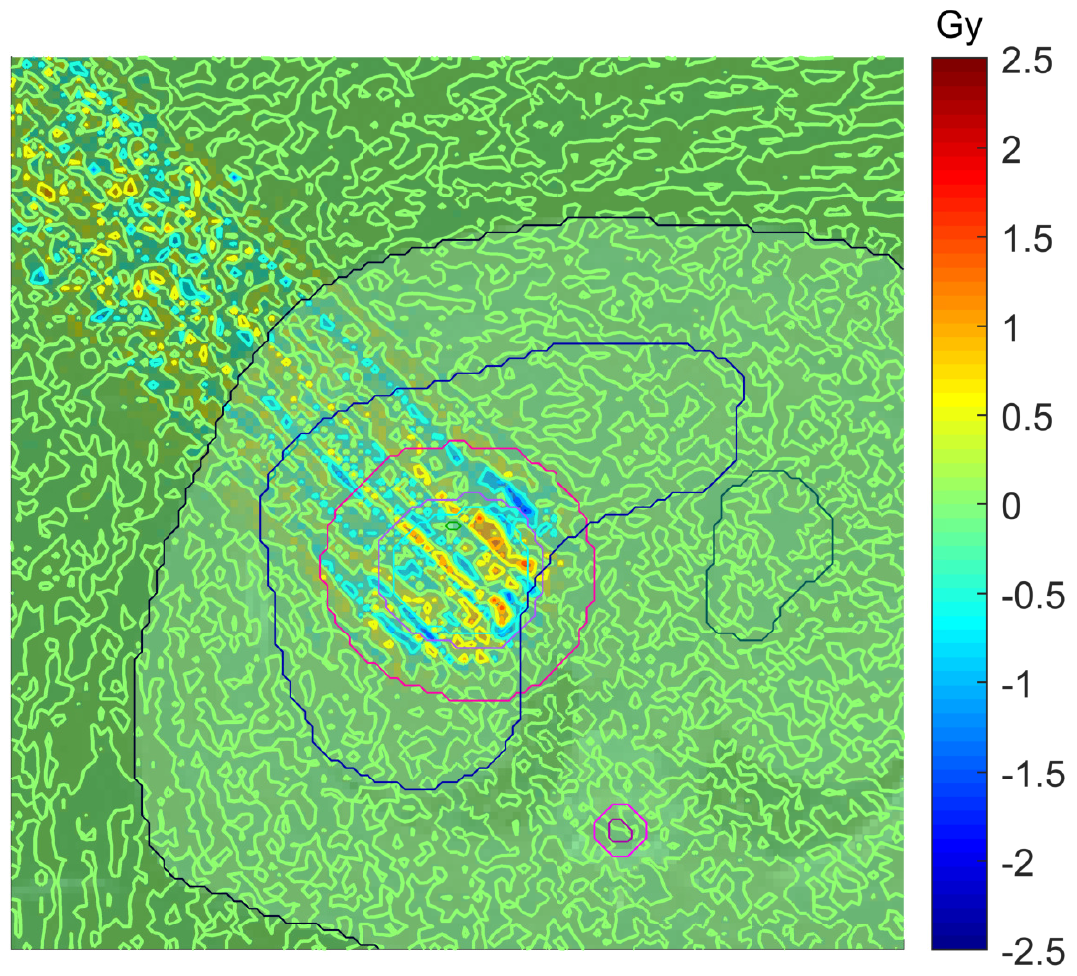}}\\
	
	$\boldsymbol{\sigma(d)}$&\raisebox{-.5\height}{\includegraphics[width=0.22\textwidth]{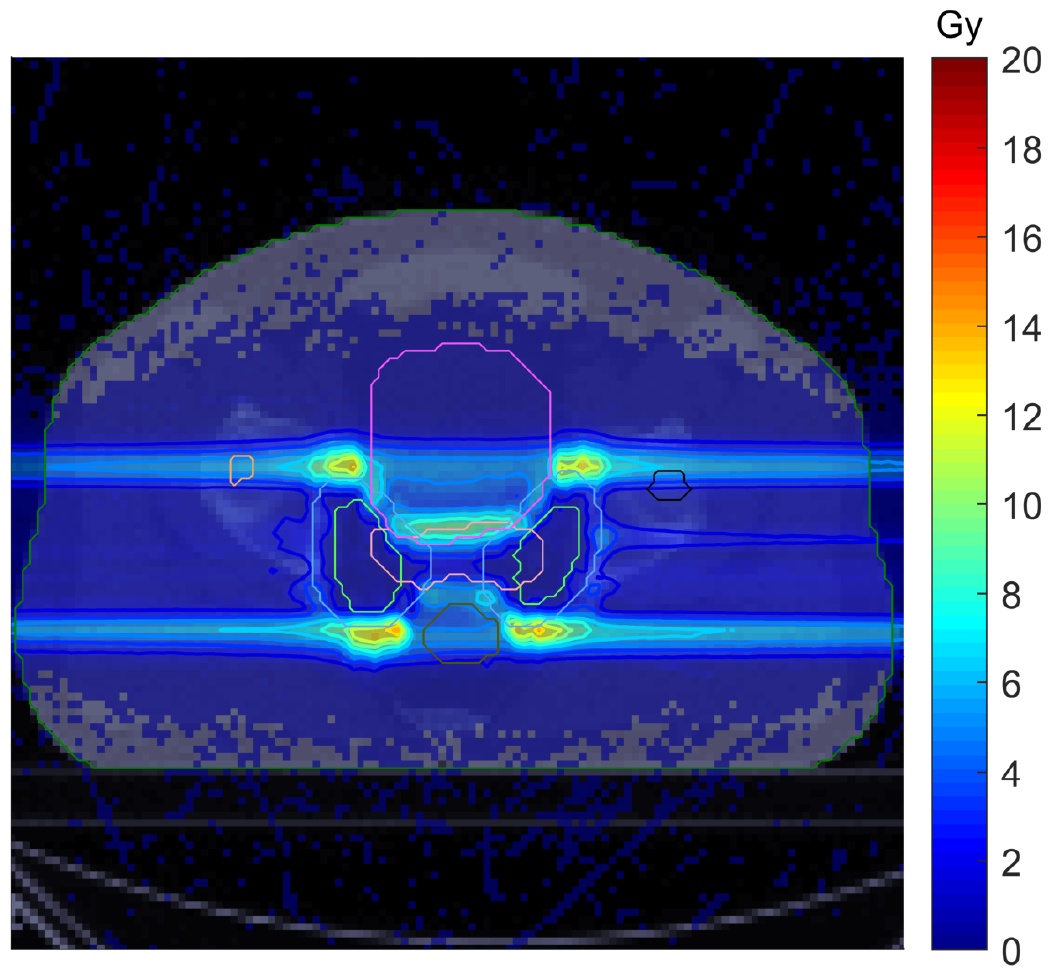}}&\raisebox{-.5\height}{\includegraphics[width=0.225\textwidth]{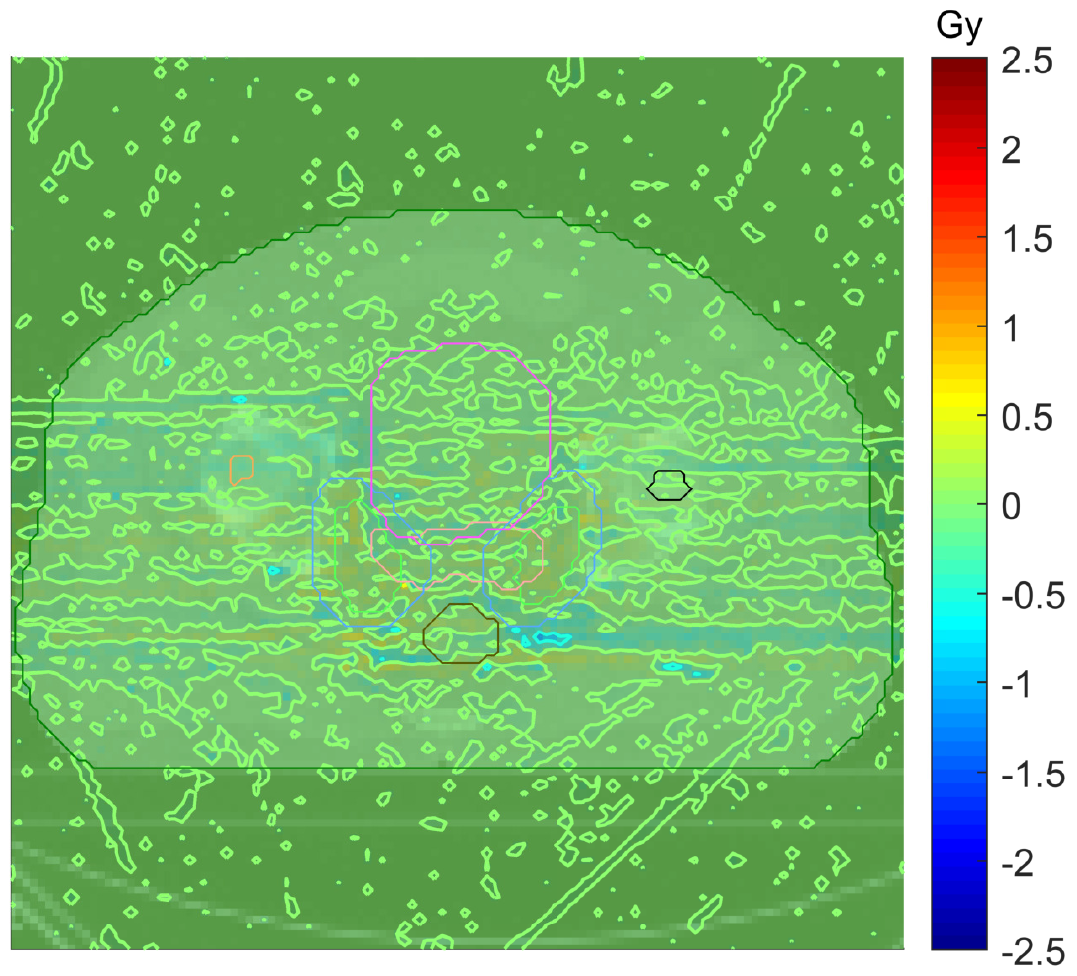}}&\raisebox{-.5\height}{\includegraphics[width=0.22\textwidth]{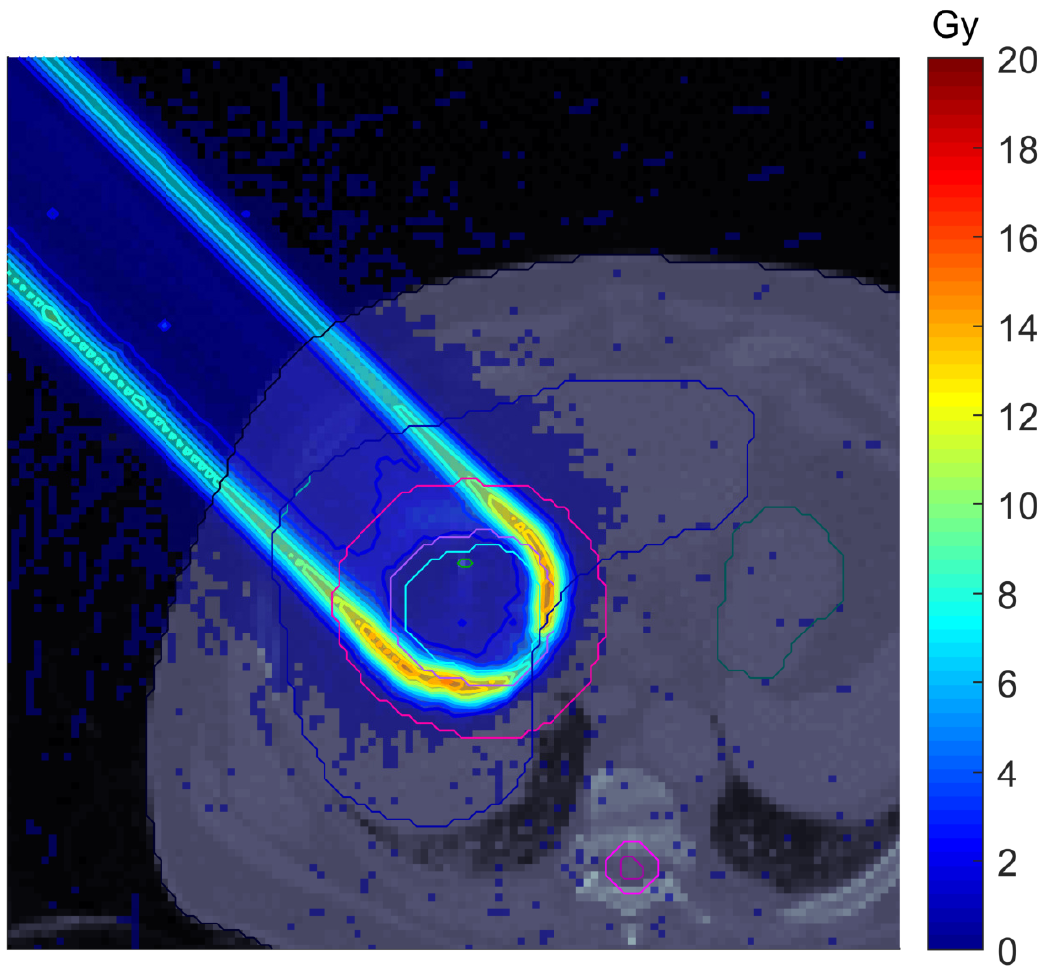}}&\raisebox{-.5\height}{\includegraphics[width=0.225\textwidth]{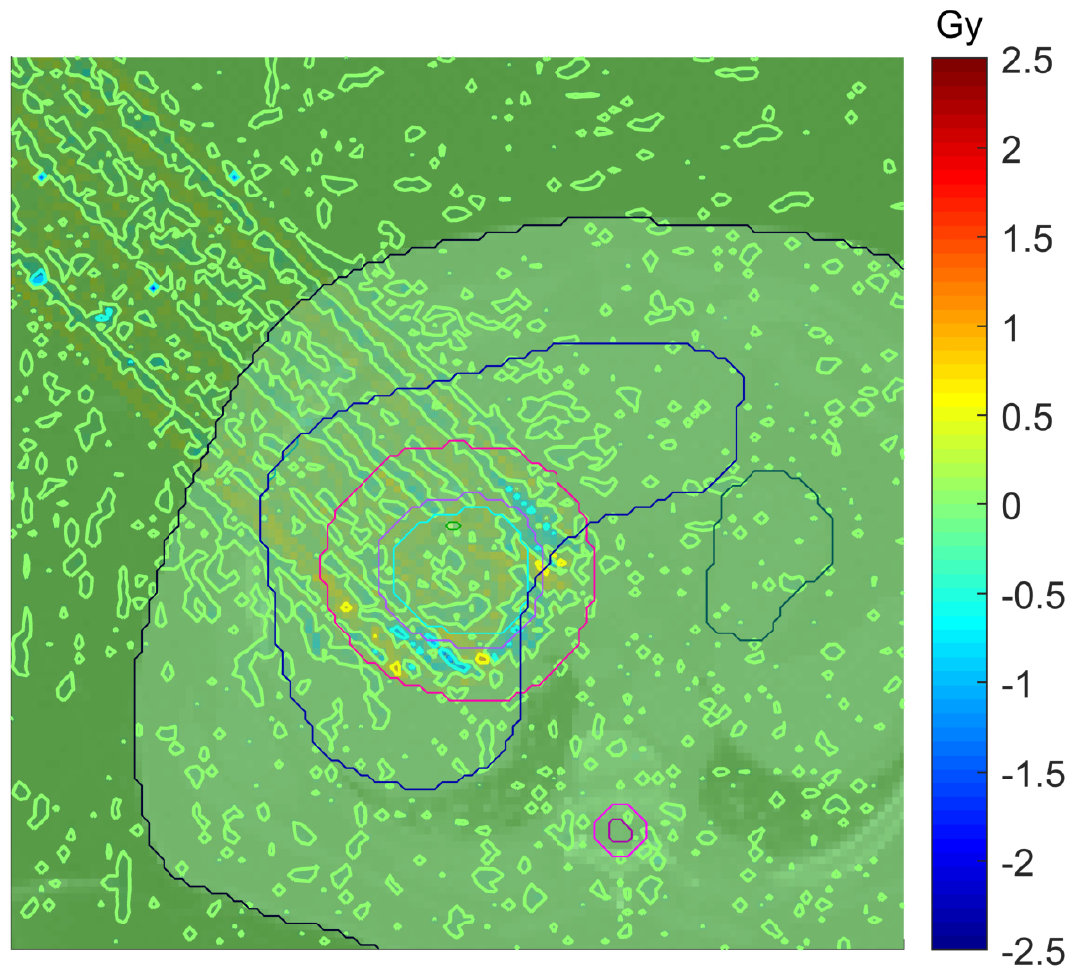}}\\
	& & & & \\[-1.5ex]
	& \multicolumn{2}{c}{(a) Prostate}& \multicolumn{2}{c}{(b) Liver} \\
	
	\end{tabular}
	\caption{Expected dose $E[\boldsymbol{d}]$ and standard deviation $\boldsymbol{\sigma(d)}$ w.r.t.\ set-up uncertainties with $\SI{3}{\milli\meter}$ standard deviation for (a) a prostate patient (couch angle \ang{0}, gantry angles \ang{90} and \ang{270}) and (b) a liver patient (couch angle \ang{0}, gantry angle \ang{315}). The left columns show the estimates computed with the proposed (re-)weighting approach and the right columns the difference to the corresponding references.}
	\label{fig:Setup_LiverProstate}
\end{figure}

The distance-to-agreement analysis using the $\gamma$-criterion supports this quantitatively (table \ref{table:gammaWaterBox}), with $\gamma^{\SI{3}{\milli\meter}}_{\SI{3}{\percent}}$-pass rate of \SI{100}{\percent} for the nominal and expected dose and  \SI{99.97}{\percent} for the standard deviation. 
Figure \ref{fig:Setup_LiverProstate} demonstrates that this is transferable to  the more complex patient cases. With overall $\gamma^{\SI{3}{\milli\meter}}_{\SI{3}{\percent}}$-pass rates of \SI{99.99}{\percent} (prostate patient) and \SI{100}{\percent} (liver patient), the standard deviation agrees as well with the reference computations as the expected value, with \SI{100}{\percent} and \SI{99.99}{\percent}, respectively (table \ref{table:gammaProstate}, \ref{table:gammaLiver}).
\begin{table}[h]
	\centering
	\caption{$\gamma^{\SI{3}{\milli\meter}}_{\SI{3}{\percent}}$-pass rates in volumes of interest (VOI) of the water phantom.}
	\lineup
	\begin{tabular}{l  l  l  l }
		\br
		Water Phantom &  Nominal dose & Expected value & Standard deviation\\
		\mr
		Error type	& \multicolumn{3}{c}{Set-up} \\
		\mr
		\textbf{Overall} &100  &100&\099.97\\
		Body &100 &100 & \099.97\\
		Target &100 & 100 & 100 \\
		\br
	\end{tabular}
	\label{table:gammaWaterBox}
\end{table}

\begin{table}[h]
	\centering
		\caption{$\gamma^{\SI{3}{\milli\meter}}_{\SI{3}{\percent}}$-pass rates in volumes of interest (VOI) of the prostate patient.}
	\label{table:gammaProstate}
	\lineup
	\begin{tabular}{l l l }
		\br
		Prostate& Expected value & Standard deviation \\
		\mr
		Error type	& \multicolumn{2}{c}{Set-up}  \\
		\mr
		\textbf{Overall}  &100&\099.99\\
		Rectum &100&100\\
		Penile bulb &100&100\\
		Lymph nodes &100&100\\
		Rt femoral head  &100&100\\
		Prostate bed &100&100\\
		PTV 68 & 100&100\\
		PTV 56 & 100&\099.99\\
		Bladder & 100&\099.99\\
		Body & 100&\099.99\\
		Lt femoral head  &100 &100\\
		\br
	\end{tabular}
\end{table}

\begin{table}[htb!]
	\centering
		\caption{$\gamma^{\SI{3}{\milli\meter}}_{\SI{3}{\percent}}$-pass rates in volumes of interest (VOI) of the liver patient (initial particles sampled from $\Phi_0$).}
	\label{table:gammaLiver}
	\lineup
	\begin{tabular}{l | l l l | l l l}
		Liver & \multicolumn{3}{l|}{Expected Value} &\multicolumn{3}{l}{Standard Deviation}\\
		\hline
		Error type	& Set-up & Range & both & Set-up & Range & both\\
		\hline
		\textbf{Overall} &\099.99&100&99.50 &100 & 91.69&93.12\\
		GTV &100&100&97.59&100&99.01&87.10\\
		Liver &100&100&99.75&\099.99&92.78&84.04\\
		Heart &100&100&96.67&100&91.26&98.84\\
		CTV &100&100&98.90&100&93.01&90.99\\
		Contour &100&100&98.69&100&91.48&90.75\\
		PTV &100&100&99.28&\099.99&83.60&90.69\\
		
	\end{tabular}
\end{table}

\subsection{Range errors}
\label{sec:resultsRangeError}
\begin{figure}[H]
	\centering 
	\begin{tabular}{c@{\hspace{-0.1ex}} c@{\hspace{1ex}} c@{\hspace{1ex}} c@{\hspace{1ex}} c@{\hspace{1ex}}}
		& \textbf{Estimate ($\Phi_0$)}\hspace*{0.25cm} & \textbf{Difference}\hspace*{0.25cm}&  \textbf{Estimate ($\Psi$)}\hspace*{0.25cm} & \textbf{Difference}\hspace*{0.25cm} \\
		
		$E[\boldsymbol{d}]$&\raisebox{-.5\height}{\includegraphics[width=0.22\textwidth]{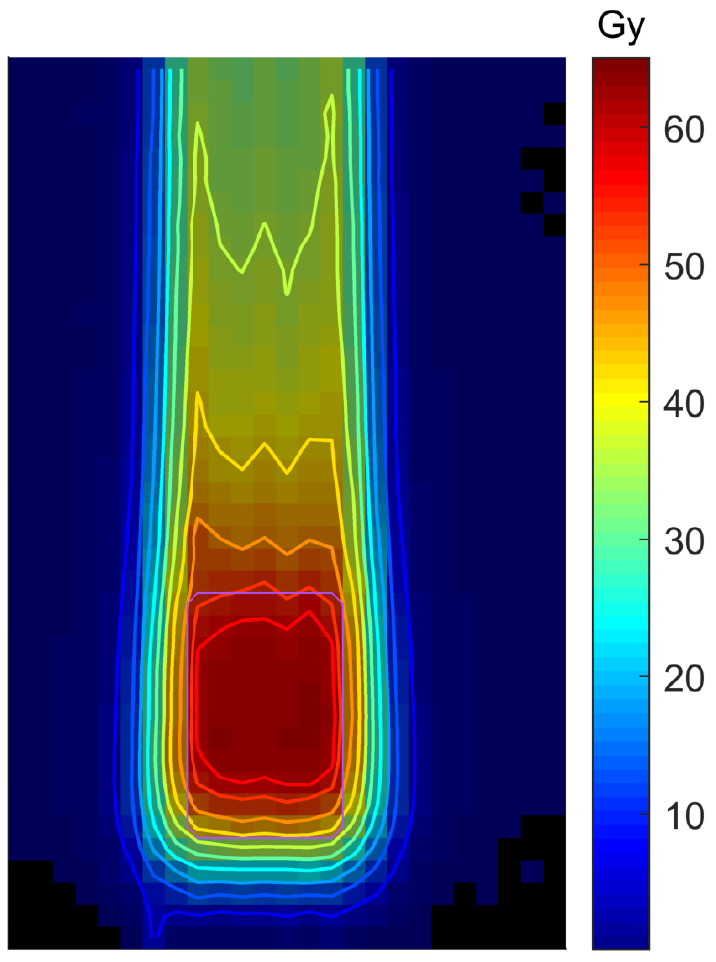}}&\raisebox{-.5\height}{\includegraphics[width=0.225\textwidth]{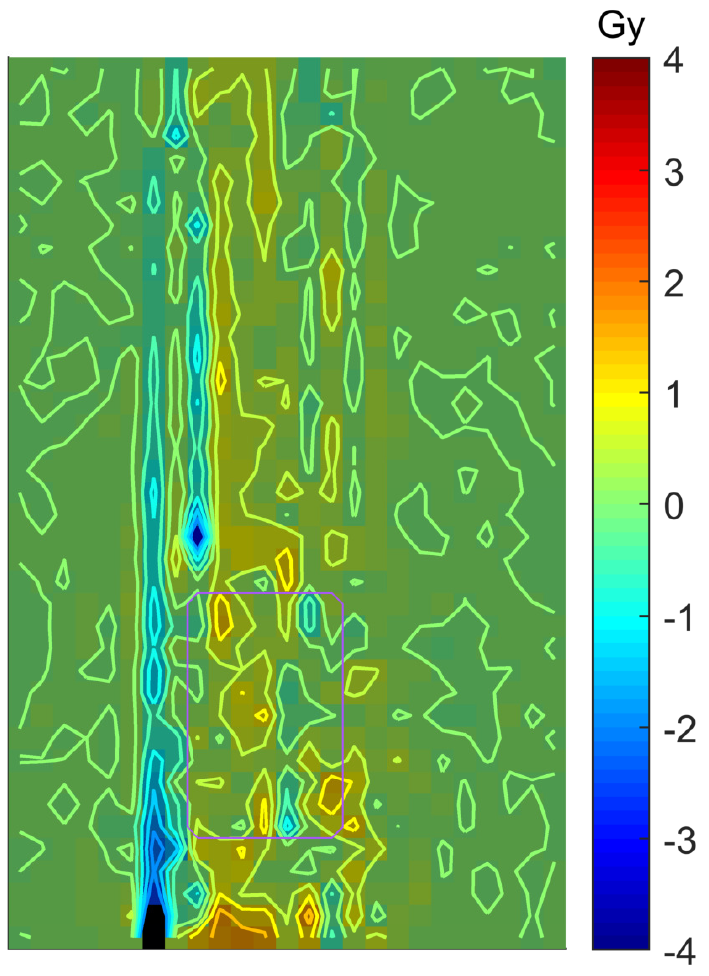}}&\raisebox{-.5\height}{\includegraphics[width=0.22\textwidth]{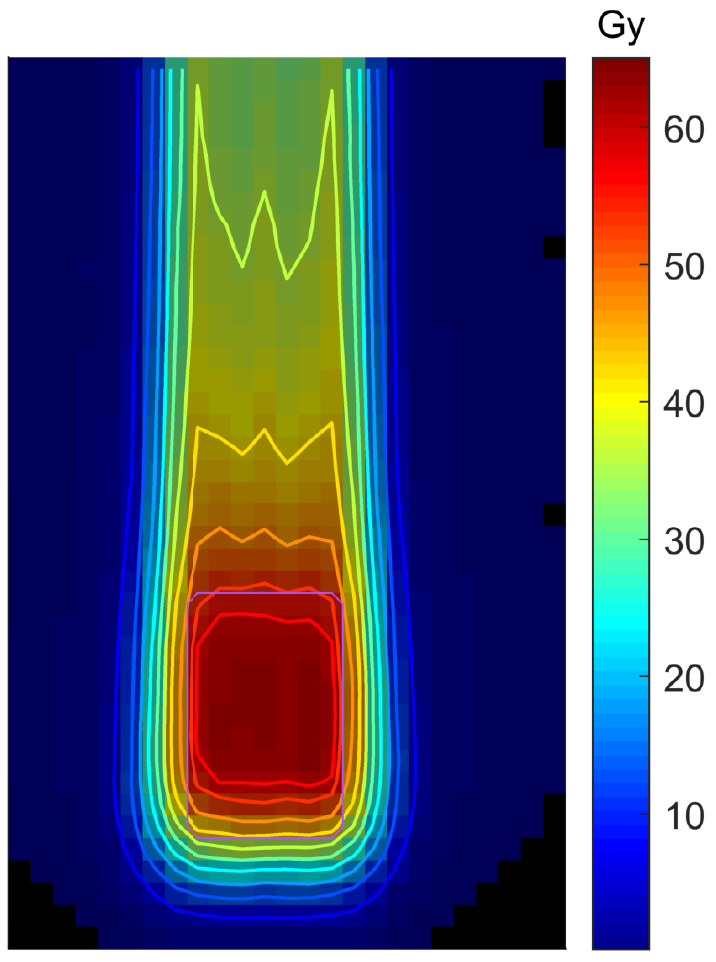}}&\raisebox{-.5\height}{\includegraphics[width=0.225\textwidth]{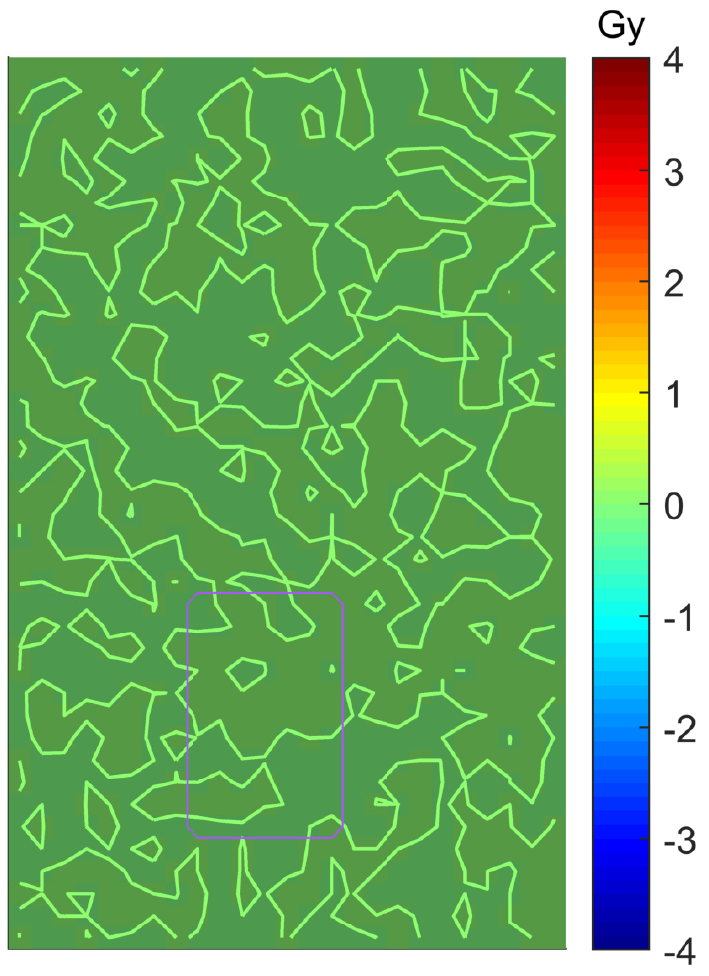}}\\
		
		$\boldsymbol{\sigma(d)}$&\raisebox{-.5\height}{\includegraphics[width=0.22\textwidth]{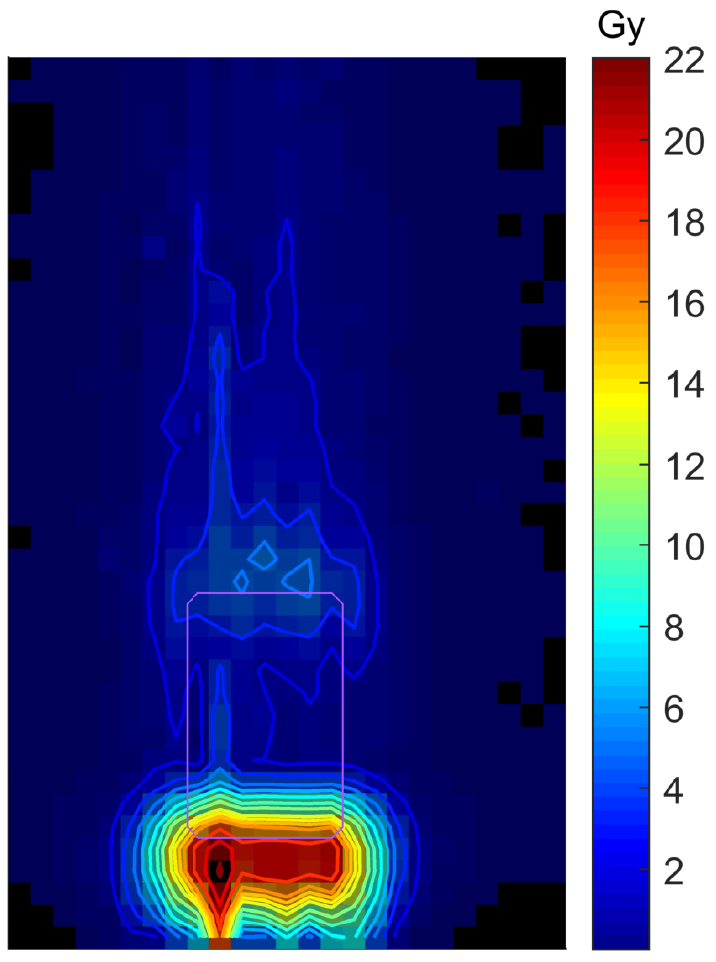}}&\raisebox{-.5\height}{\includegraphics[width=0.225\textwidth]{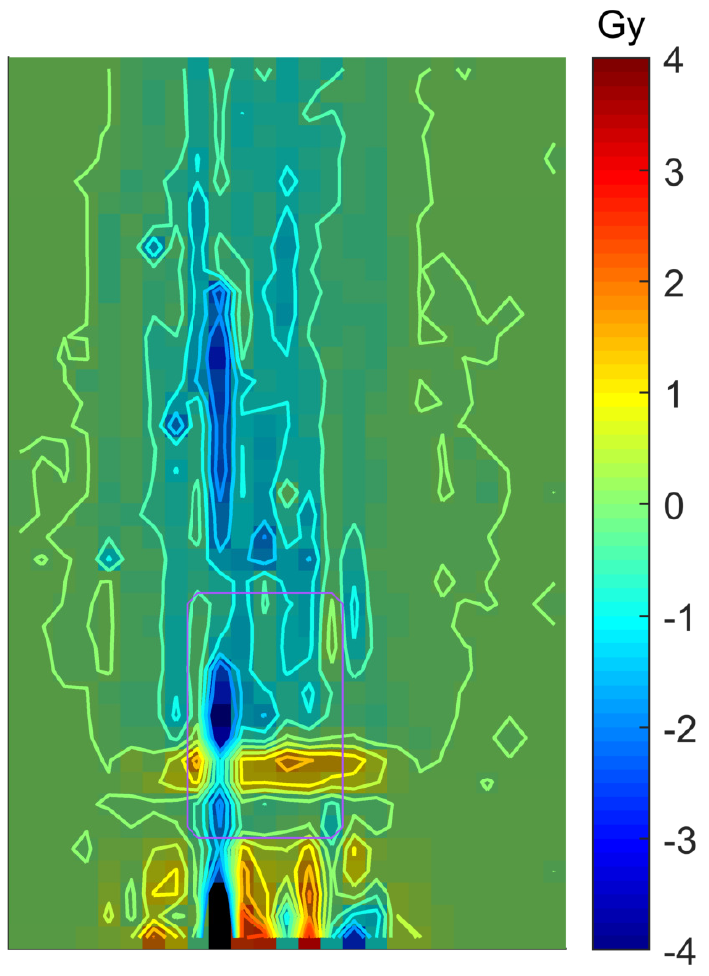}}&\raisebox{-.5\height}{\includegraphics[width=0.22\textwidth]{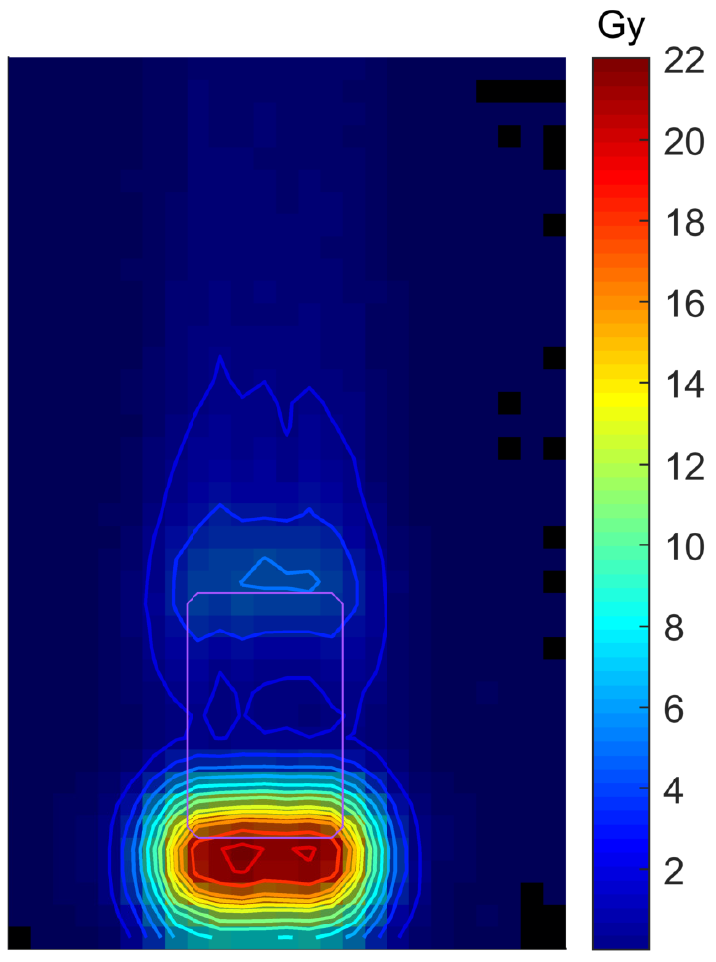}}&\raisebox{-.5\height}{\includegraphics[width=0.225\textwidth]{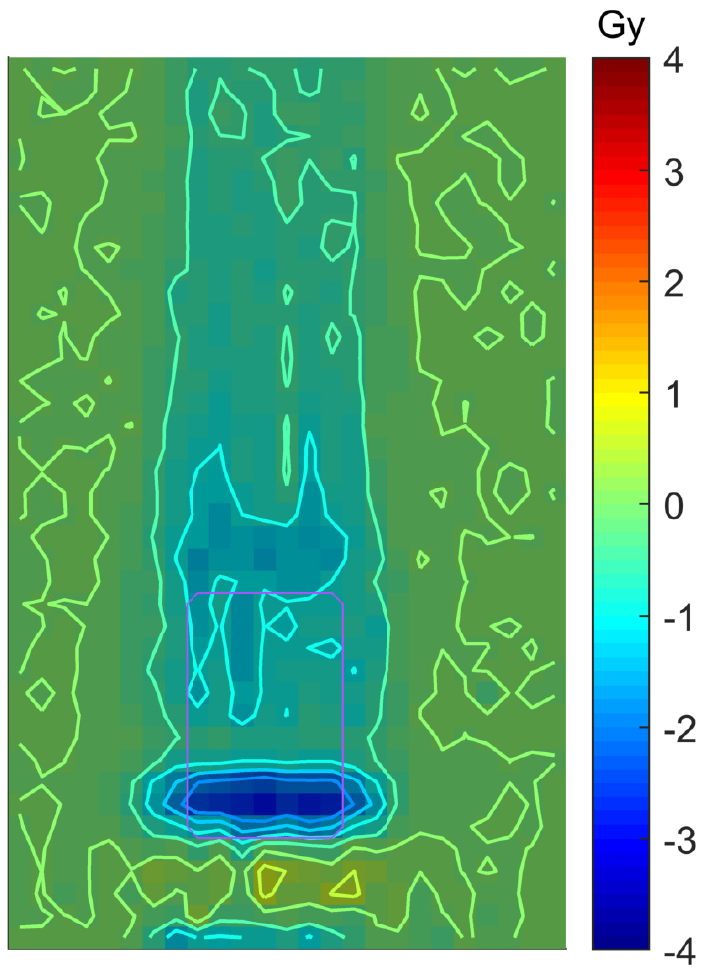}}\\
		& & & &\\[-1.5ex]
		& \multicolumn{4}{c}{(a) Range errors}\\
		
		& \textbf{Estimate ($\Phi_0$)}\hspace*{0.5cm} & \textbf{Difference}\hspace*{0.5cm}&  \textbf{Estimate ($\Psi$)}\hspace*{0.5cm} & \textbf{Difference}\hspace*{0.5cm} \\
		
		$E[\boldsymbol{d}]$&\raisebox{-.5\height}{\includegraphics[width=0.22\textwidth]{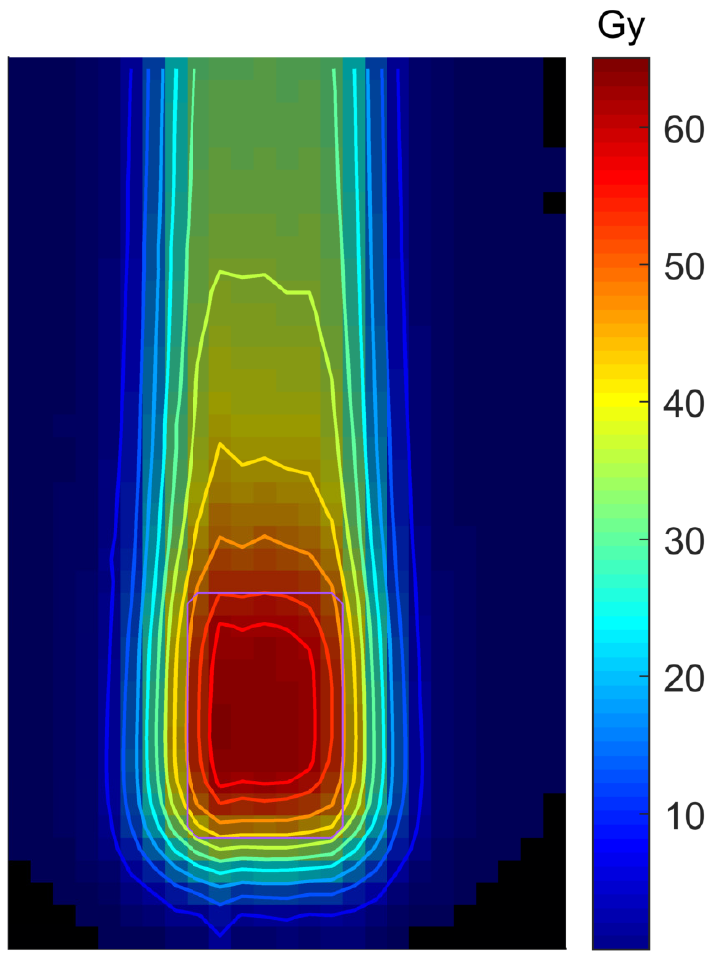}}&\raisebox{-.5\height}{\includegraphics[width=0.225\textwidth]{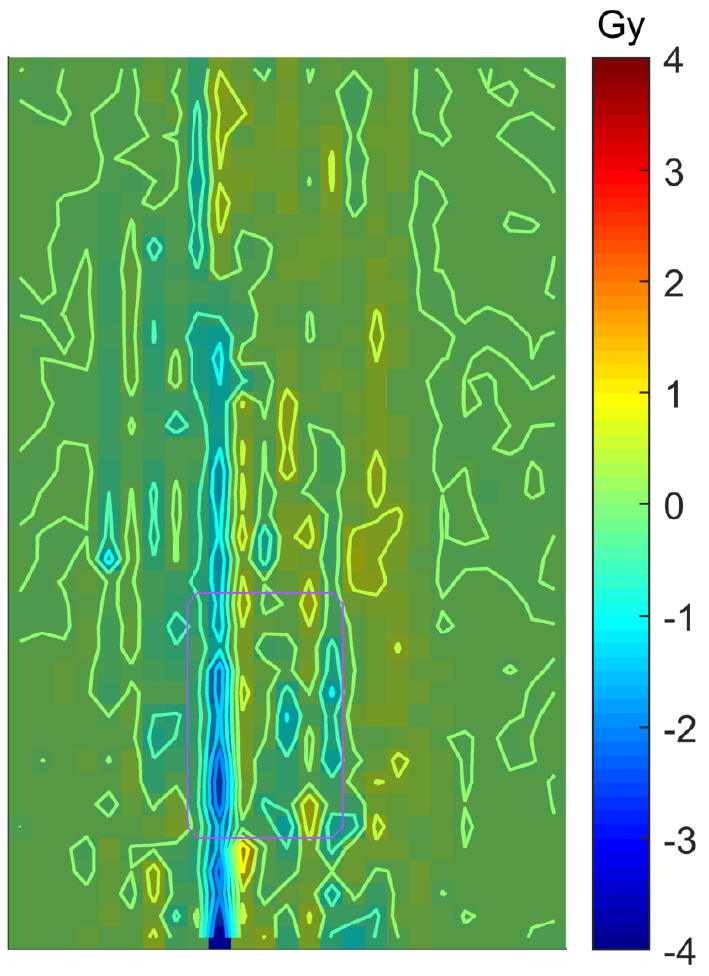}}&\raisebox{-.5\height}{\includegraphics[width=0.22\textwidth]{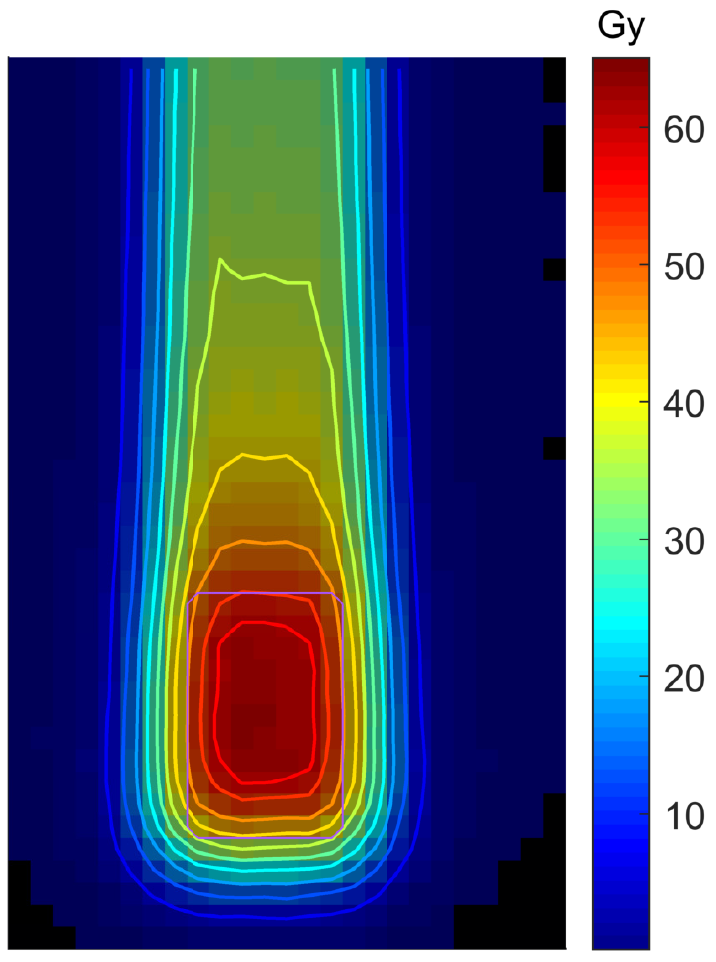}}&\raisebox{-.5\height}{\includegraphics[width=0.225\textwidth]{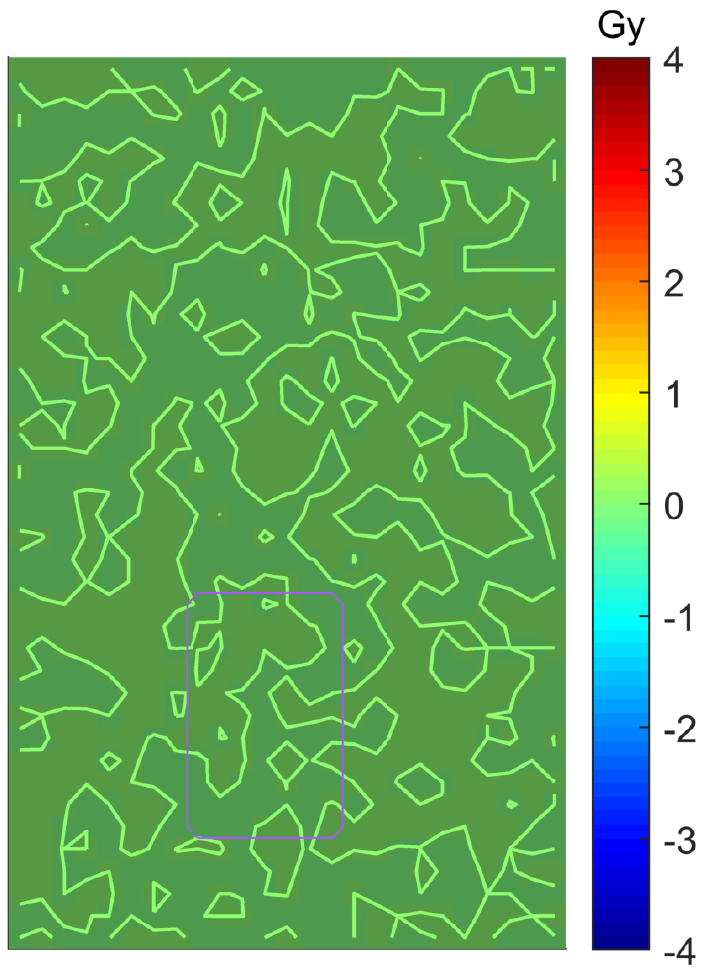}}\\
		
		$\boldsymbol{\sigma(d)}$&\raisebox{-.5\height}{\includegraphics[width=0.22\textwidth]{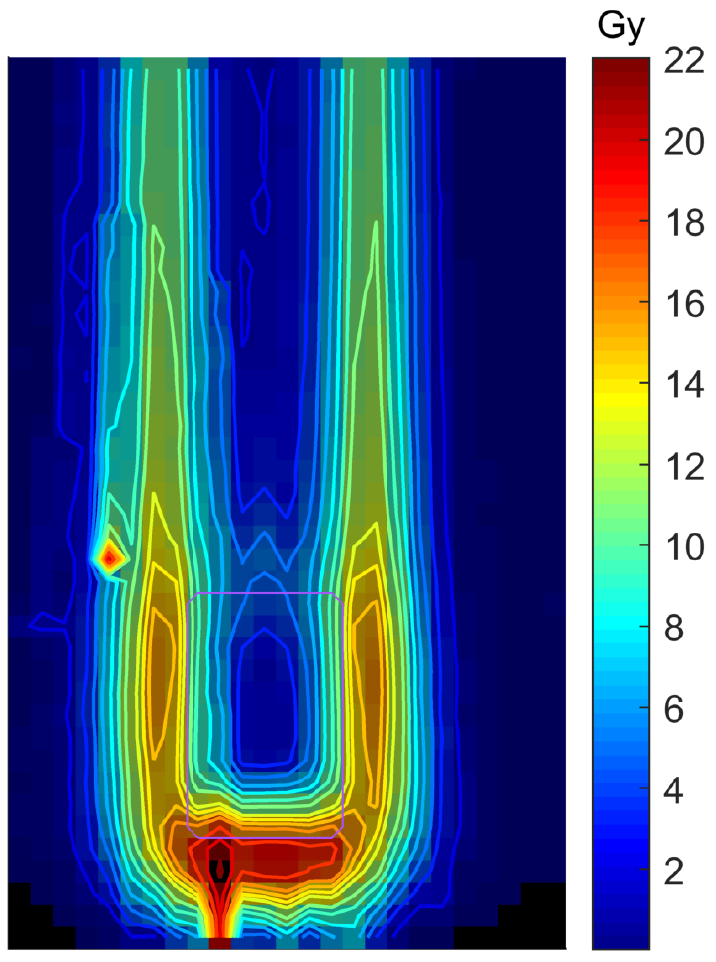}}&\raisebox{-.5\height}{\includegraphics[width=0.225\textwidth]{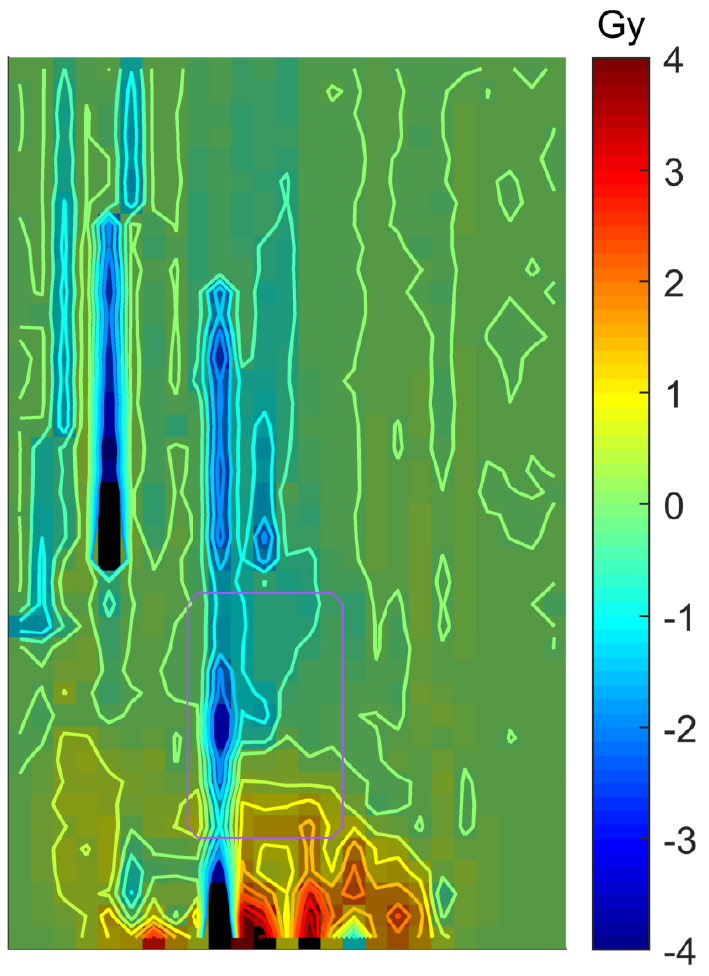}}&\raisebox{-.5\height}{\includegraphics[width=0.22\textwidth]{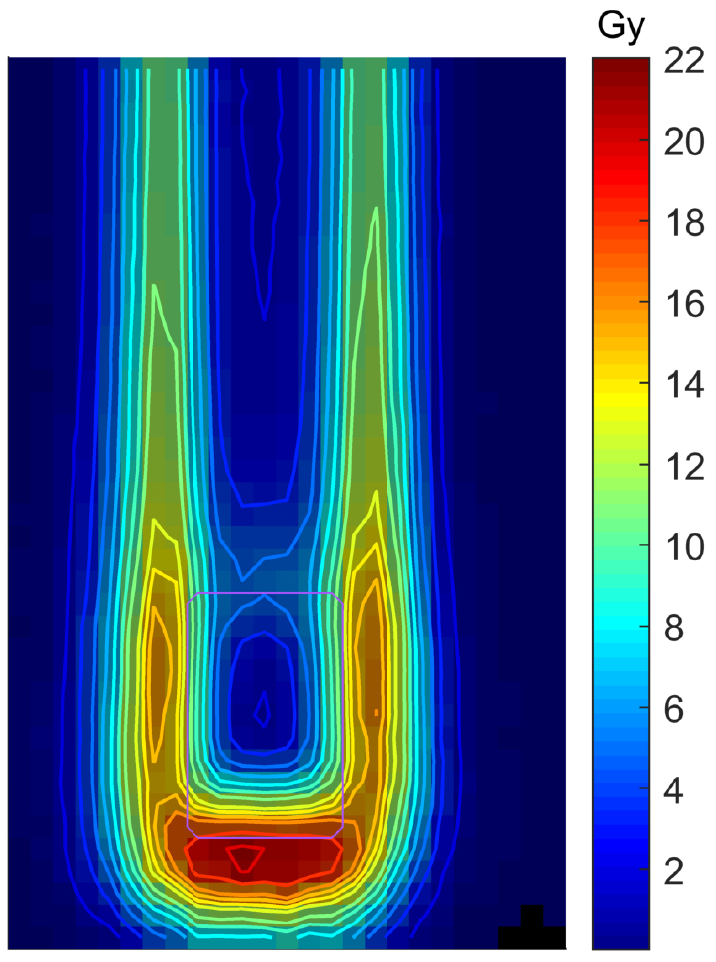}}&\raisebox{-.5\height}{\includegraphics[width=0.225\textwidth]{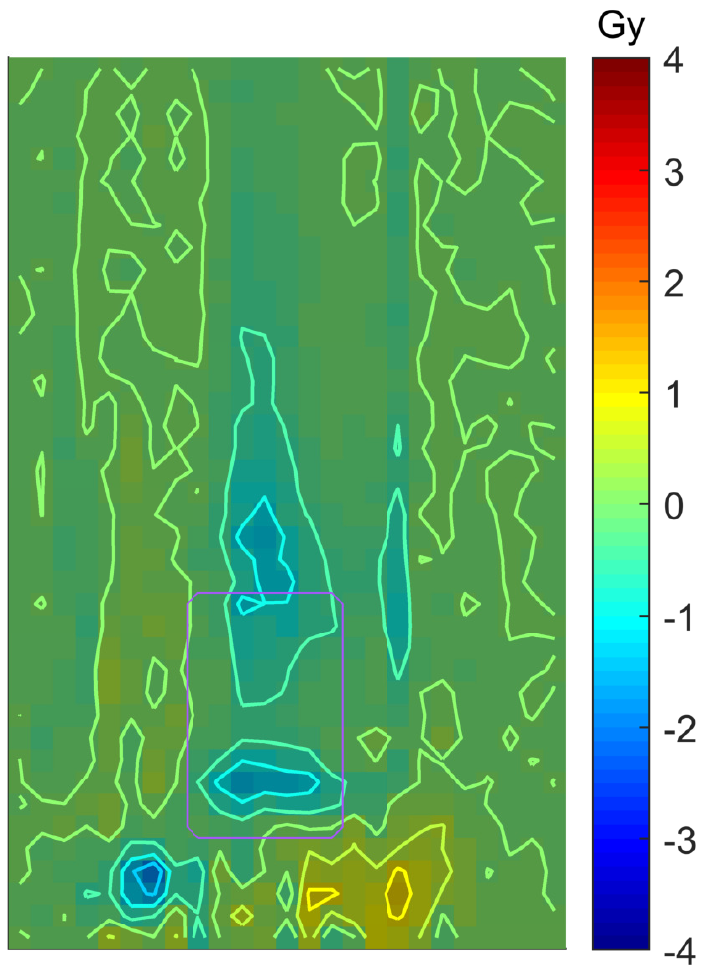}}\\
		& & & &\\[-1.5ex]
		& \multicolumn{4}{c}{(b) Range and set-up errors}\\
		
	\end{tabular}
	
	\caption{Expected dose $E[\boldsymbol{d}]$ and standard deviation $\boldsymbol{\sigma(d)}$ w.r.t.\ (a) range uncertainties with a $\SI{3}{\percent}$ standard error (b) and both range uncertainties as well as set-up errors with $\SI{3}{\milli\meter}$ standard deviation for a spread out Bragg peak in a water phantom. The left columns show the estimate computed with the proposed (re-)weighting approach, reconstructed either from the nominal distribution $\Phi_0$ or its convolution $\Psi$ witht the error kernel. The right columns show the difference to the corresponding references.}
	\label{fig:WB_Range}
\end{figure}

 In contrast to the set-up errors, for which dose estimates can also be shown to be mathematically accurate, range errors can only be modeled through an approximation introduced in \ref{sec:rangeError}. Figure \ref{fig:WB_Range} displays results for range errors as well as the combination of range and set-up errors in the water phantom.
 
The difference maps for both expected value and standard deviation show that the deviations when including range errors are expectedly higher. We observe a systematic bias primarily at the distal edge, where our method seems to consistently underestimate the variance. The standard deviation estimate using our importance weighting method also expresses strong local artifacts, as evident in the difference maps (compare figure \ref{fig:WB_Range}). This is an indicator of too little statistical mass, i.e., computed particle trajectories, in the original simulation. For more extreme error realizations, relatively high weights are assigned to a small number of particles, thereby amplifying single realizations or errors. Especially in case of a relatively small beam energy spread in the original simulation (here  $\SI{1}{\percent}$), compared to the range error of $\SI{3}{\percent}$, such artifacts are likely to appear. In order to prevent this, one could either compute a larger number of particle histories in the simulation or sample the particles from a different distribution which has more density mass in its outer regions or tails (compare \ref{sec:directExpDose}). 

To underline the explanation for the appearance of the artifacts above, we recomputed the estimates using the (re-)weighting method based on a direct computation of the expected value, which can be calculated using the convolution $\Psi$ of the Gaussian error kernel with the nominal phase space parameter distribution \ref{sec:directExpDose}. Figure \ref{fig:WB_Range} shows that this alleviates the discrepancy from the references, causing artifacts to disappear and also reducing the overall amount of deviation displayed in the difference maps.

\begin{figure}[H]
	\centering 
		\begin{tabular}{c@{\hspace{-0.1ex}} c@{\hspace{1ex}} c@{\hspace{1ex}} c@{\hspace{1ex}} c@{\hspace{1ex}}}
		& \textbf{Estimate ($\Phi_0$)}\hspace*{0.25cm} & \textbf{Difference}\hspace*{0.25cm}&  \textbf{Estimate ($\Psi$)}\hspace*{0.25cm} & \textbf{Difference}\hspace*{0.25cm} \\
		
		$E[\boldsymbol{d}]$&\raisebox{-.5\height}{\includegraphics[width=0.22\textwidth]{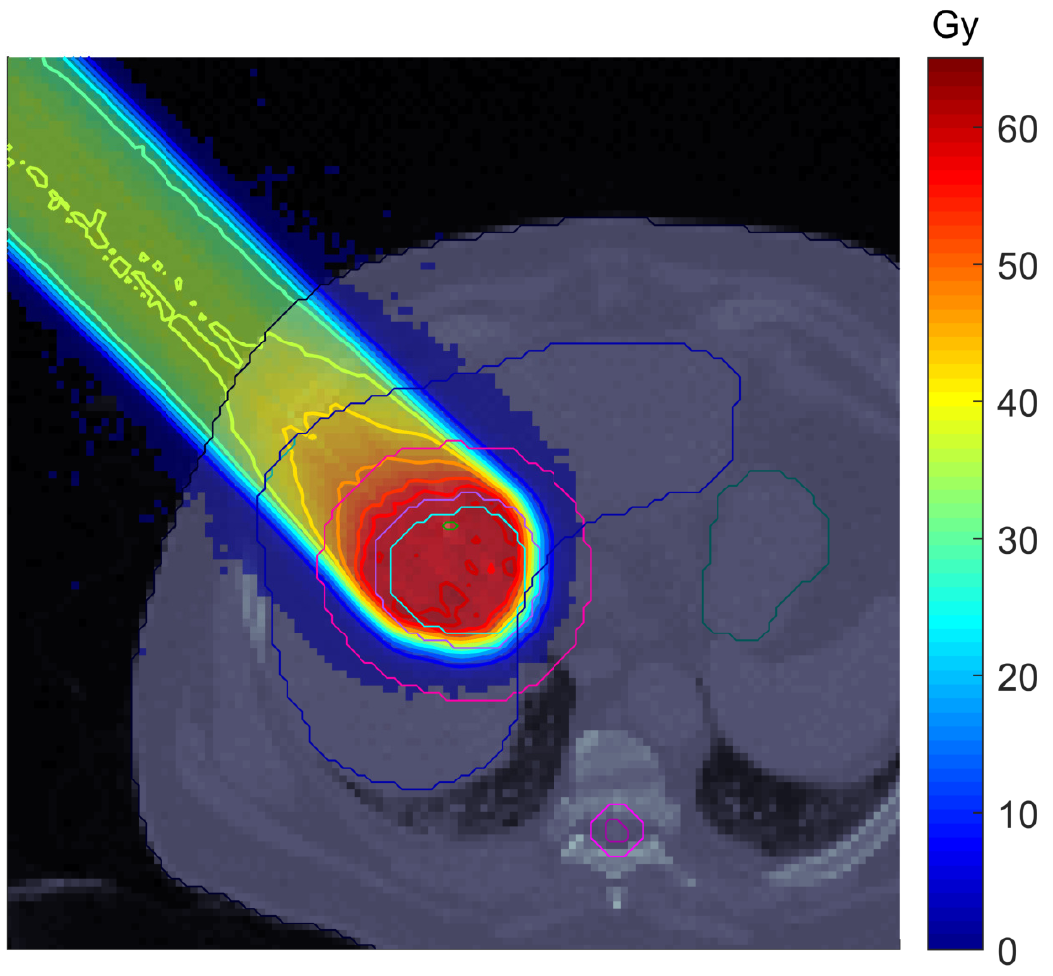}}&\raisebox{-.5\height}{\includegraphics[width=0.225\textwidth]{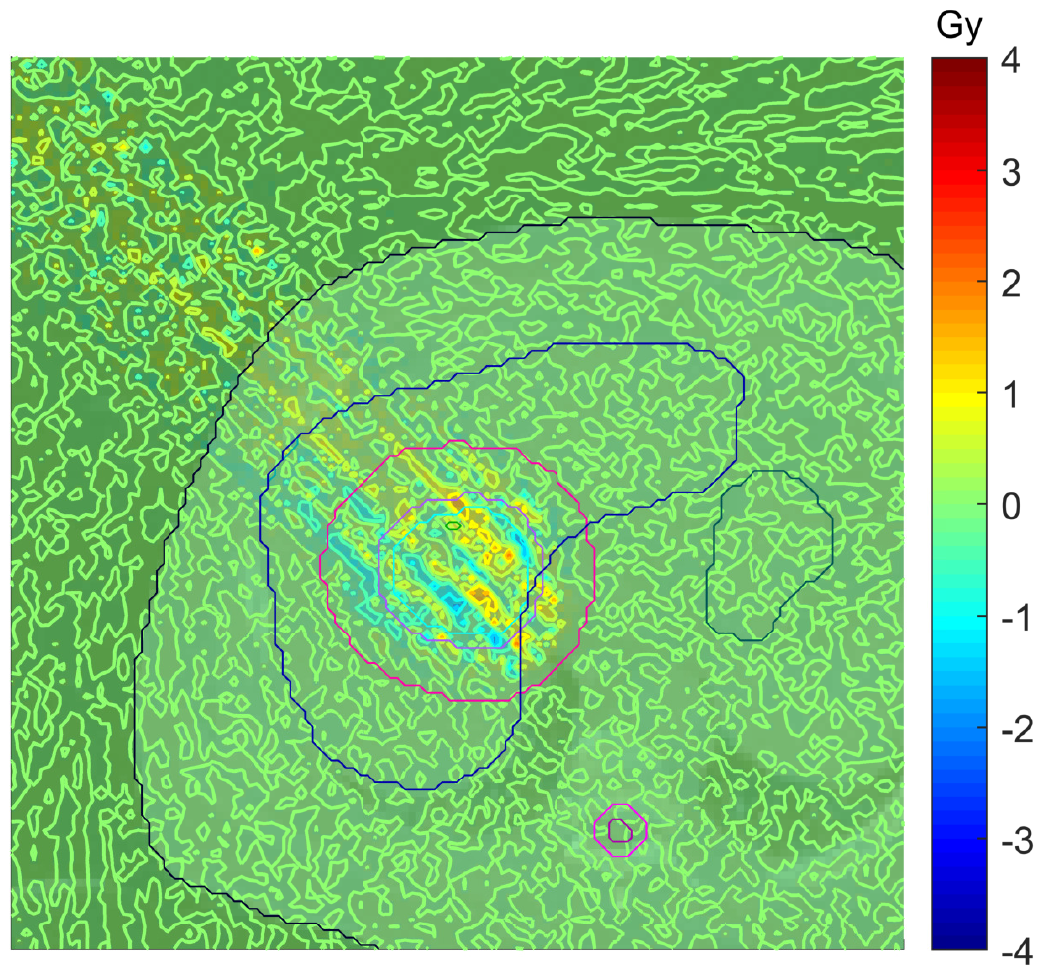}}&\raisebox{-.5\height}{\includegraphics[width=0.22\textwidth]{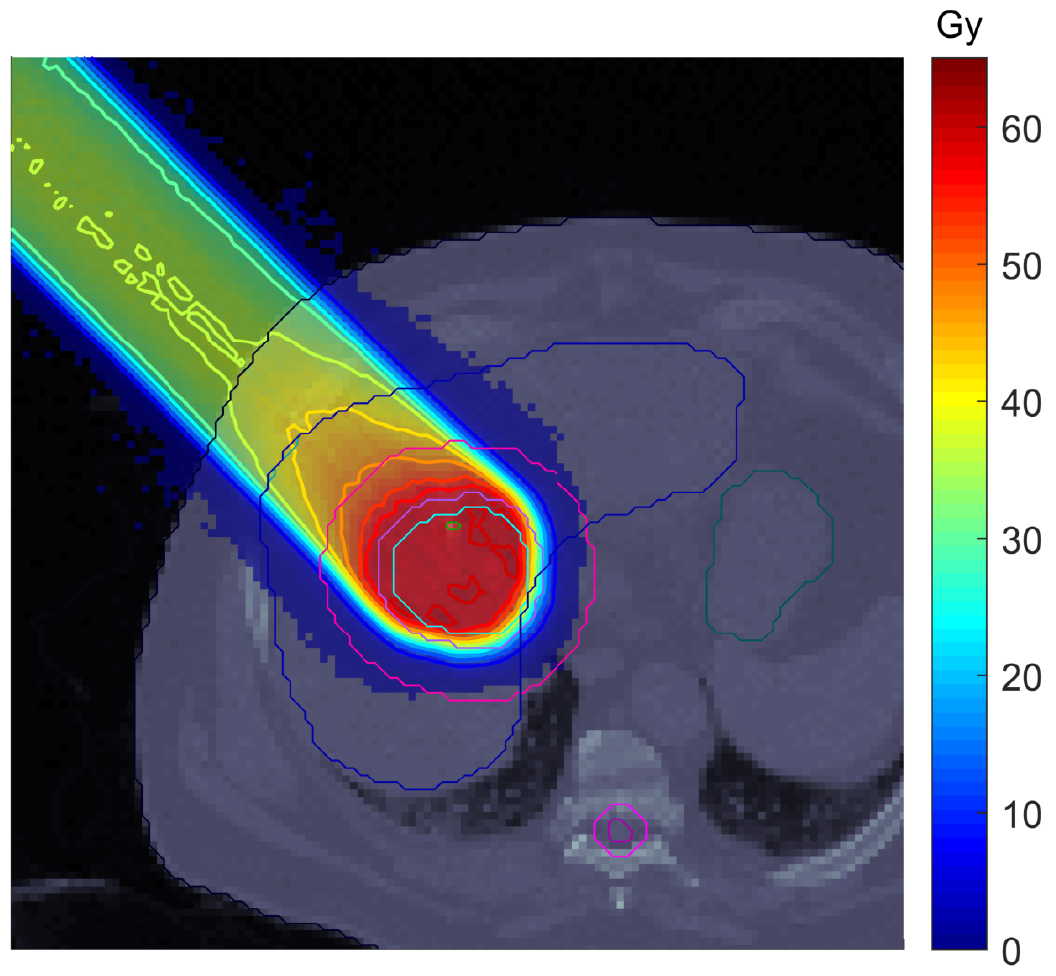}}&\raisebox{-.5\height}{\includegraphics[width=0.225\textwidth]{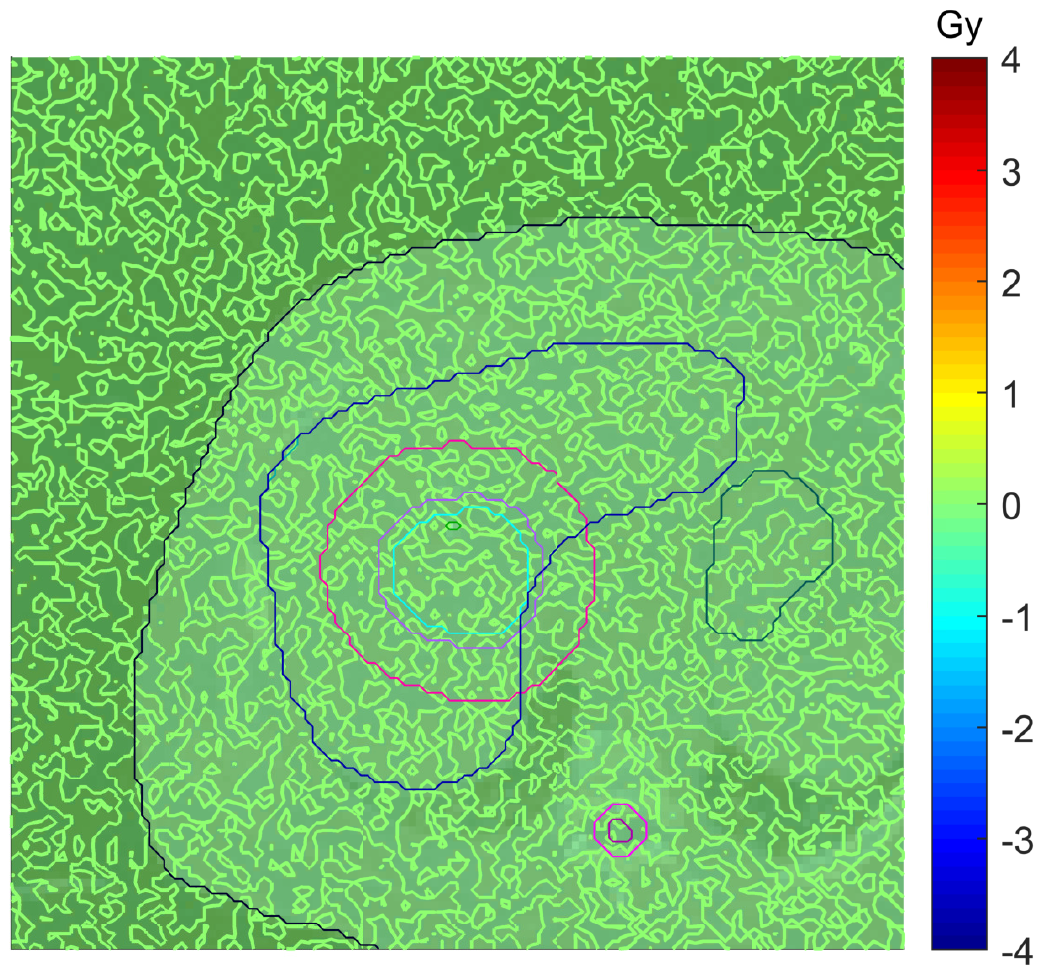}}\\
		
		$\boldsymbol{\sigma(d)}$&\raisebox{-.5\height}{\includegraphics[width=0.22\textwidth]{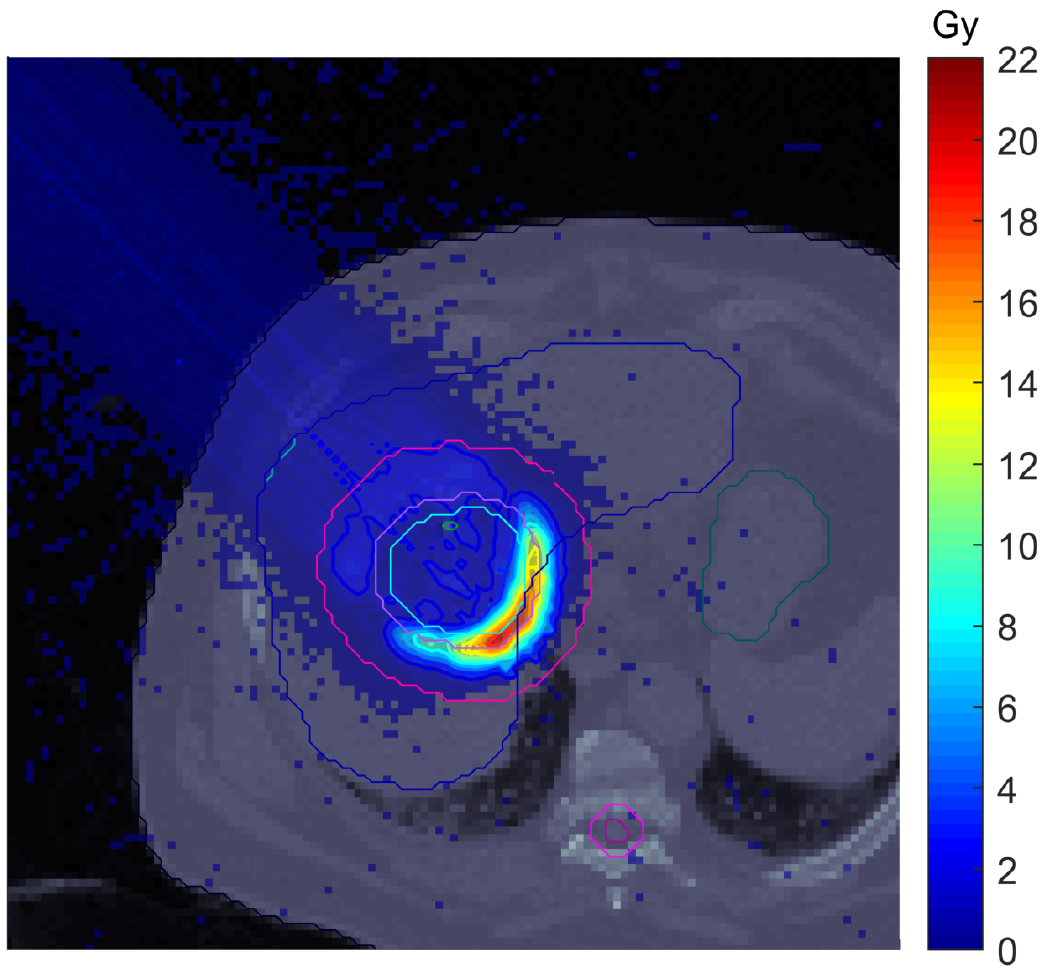}}&\raisebox{-.5\height}{\includegraphics[width=0.225\textwidth]{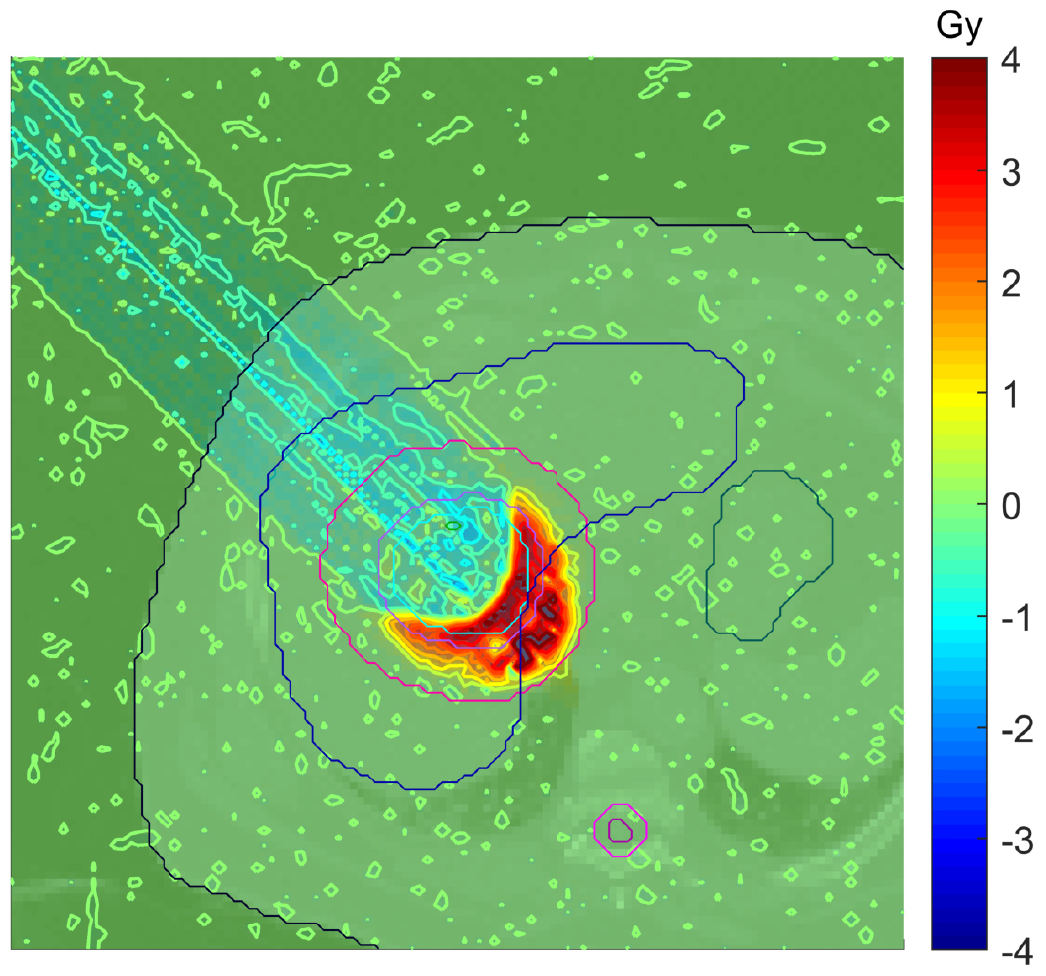}}&\raisebox{-.5\height}{\includegraphics[width=0.22\textwidth]{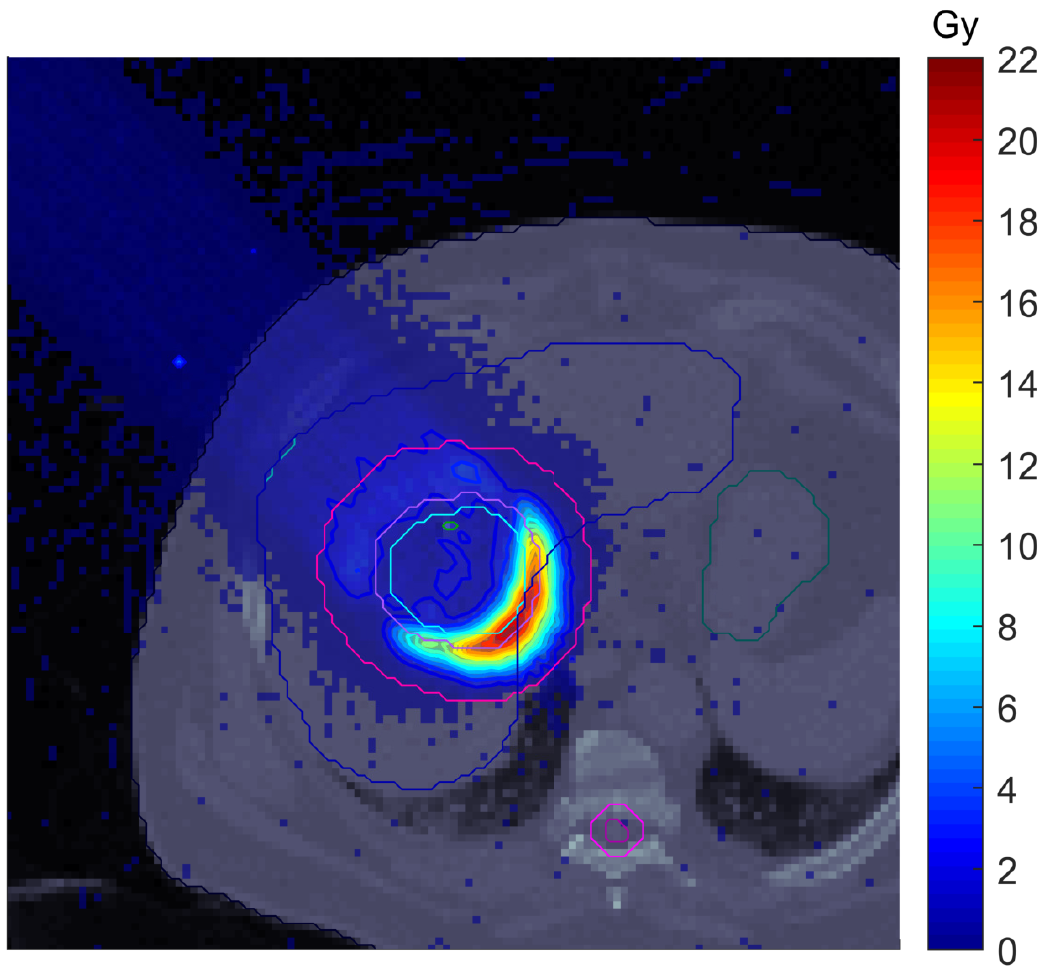}}&\raisebox{-.5\height}{\includegraphics[width=0.225\textwidth]{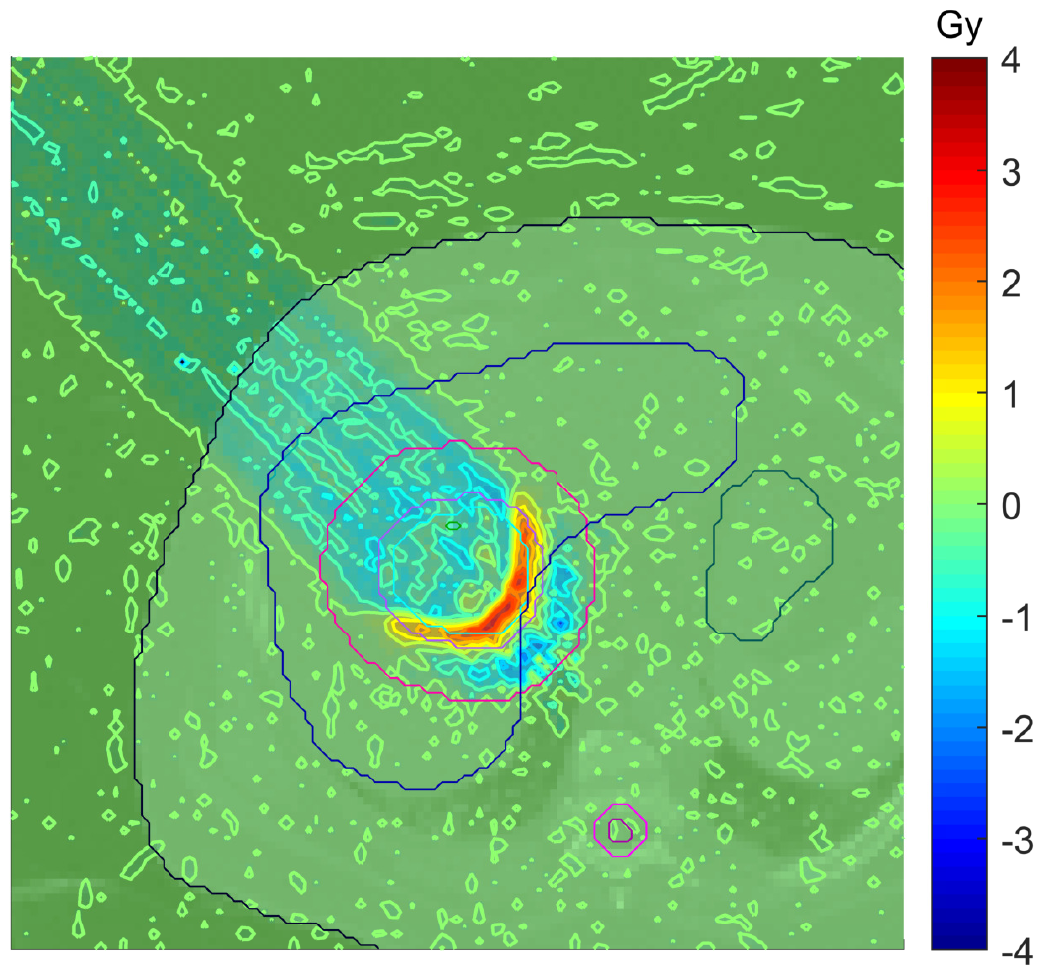}}\\
		& & & &\\[-1.5ex]
		& \multicolumn{4}{c}{(a) Range errors}\\
		
		& \textbf{Estimate ($\Phi_0$)}\hspace*{0.5cm} & \textbf{Difference}\hspace*{0.5cm}&  \textbf{Estimate ($\Psi$)}\hspace*{0.5cm} & \textbf{Difference}\hspace*{0.5cm} \\
		
		$E[\boldsymbol{d}]$&\raisebox{-.5\height}{\includegraphics[width=0.22\textwidth]{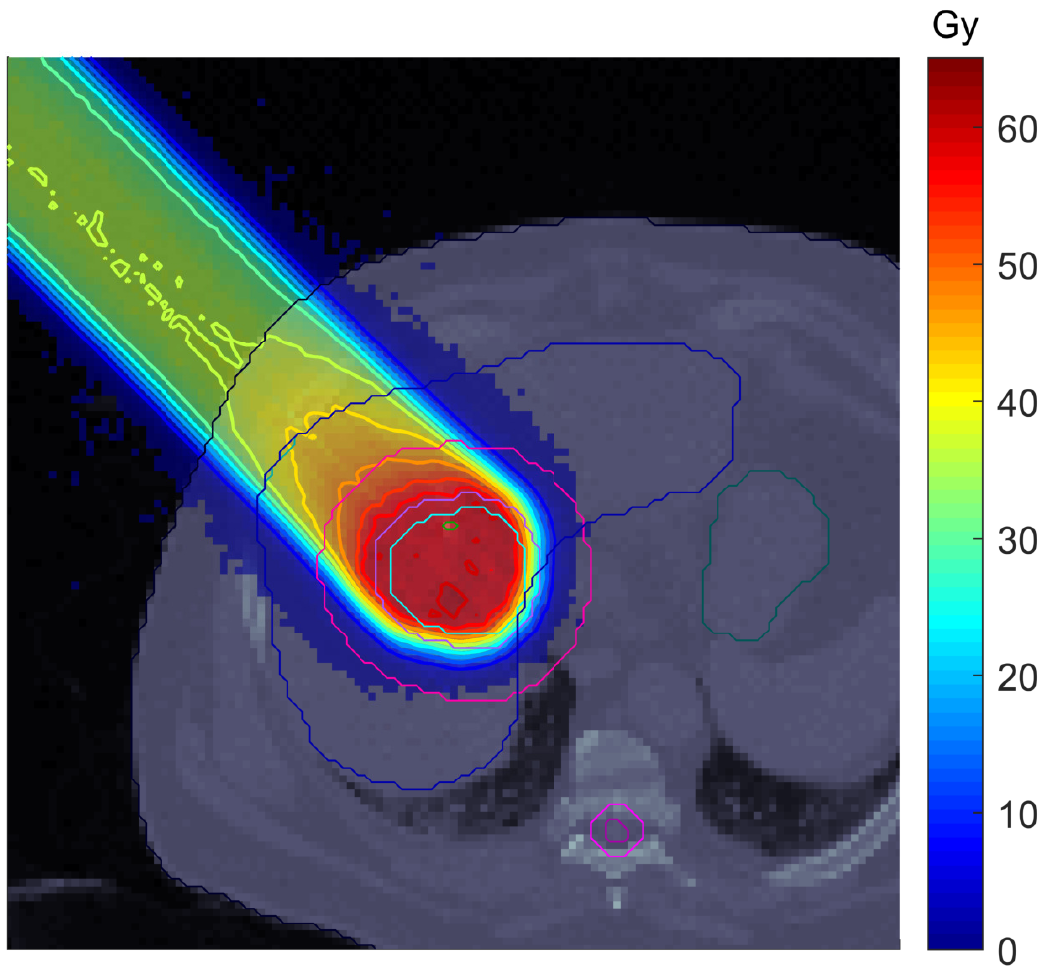}}&\raisebox{-.5\height}{\includegraphics[width=0.225\textwidth]{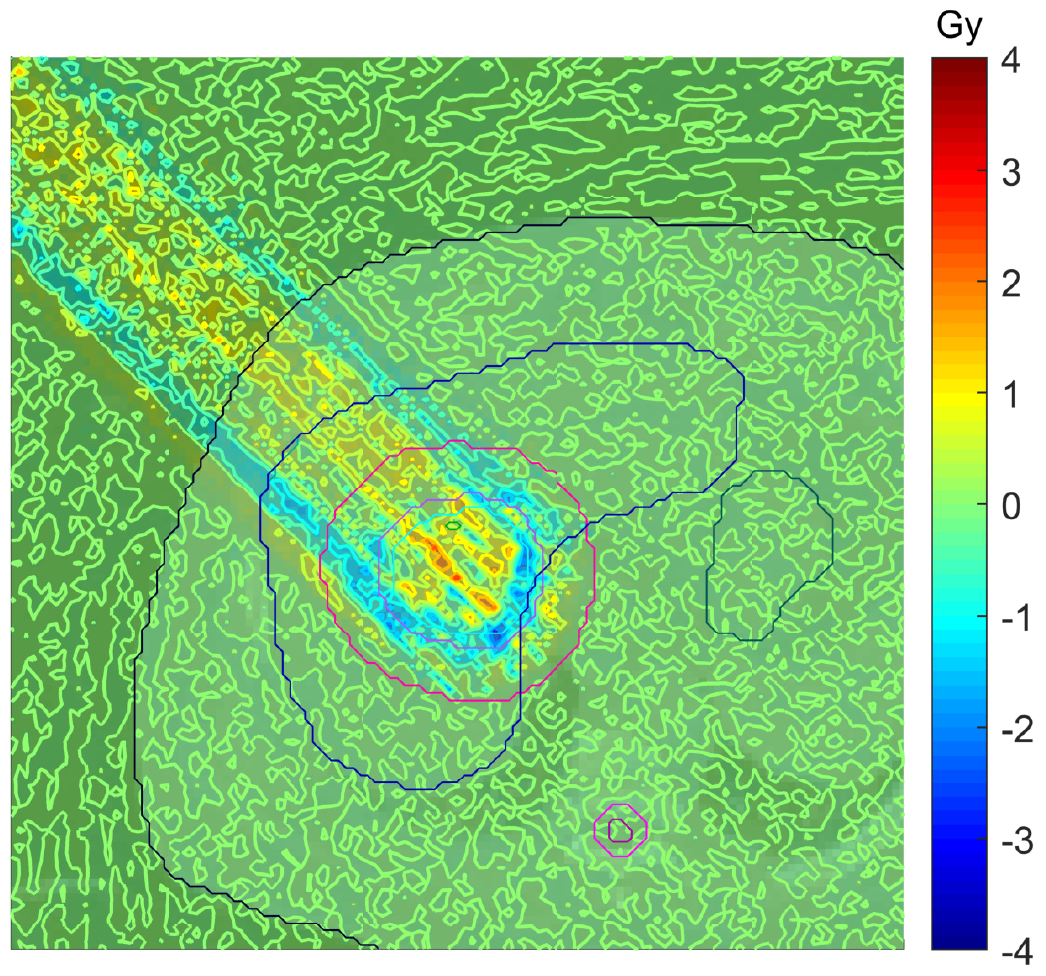}}&\raisebox{-.5\height}{\includegraphics[width=0.22\textwidth]{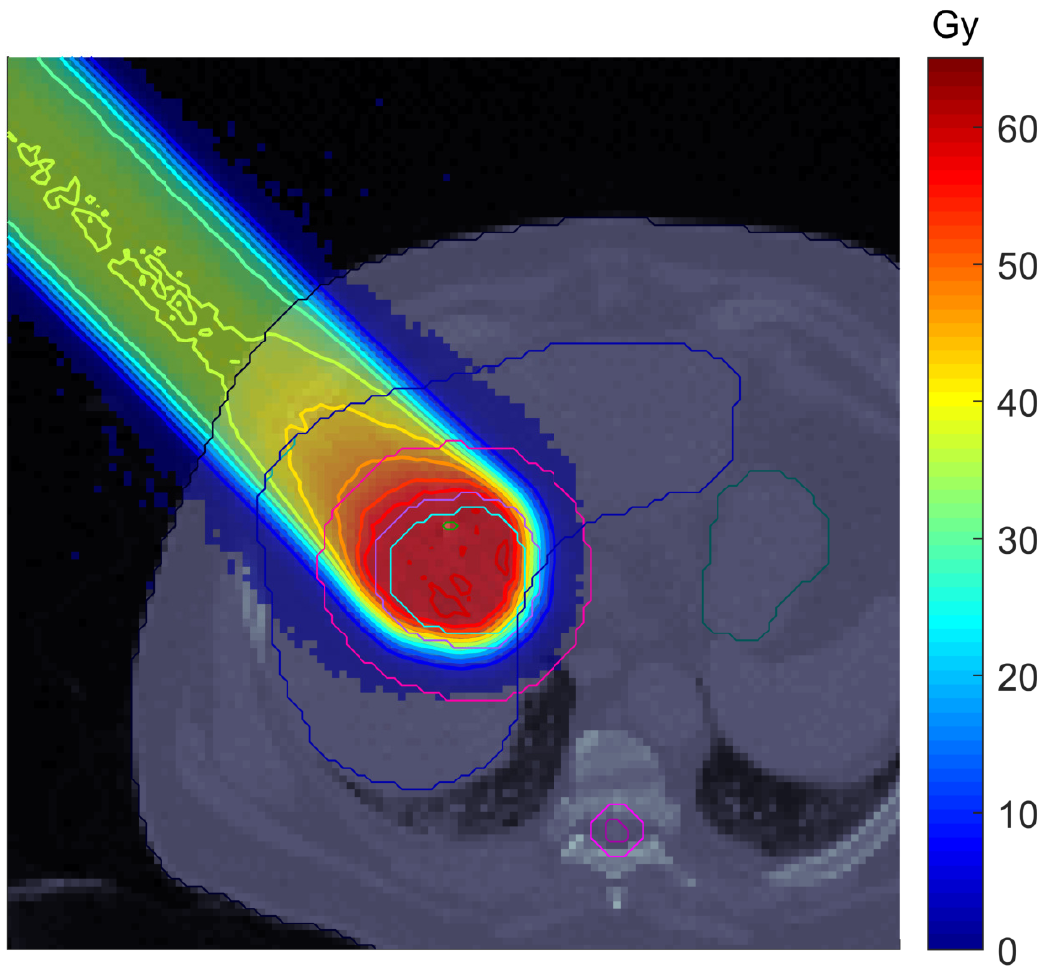}}&\raisebox{-.5\height}{\includegraphics[width=0.225\textwidth]{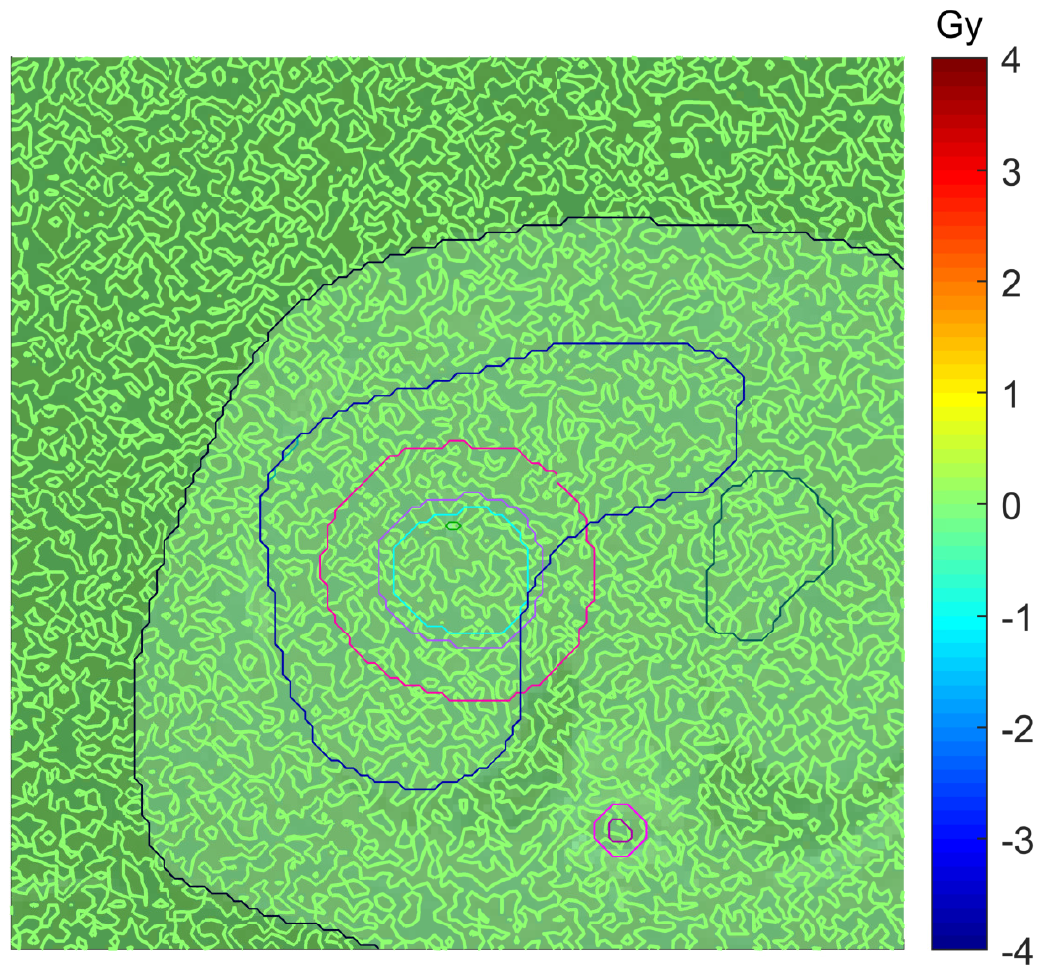}}\\
		
		$\boldsymbol{\sigma(d)}$&\raisebox{-.5\height}{\includegraphics[width=0.22\textwidth]{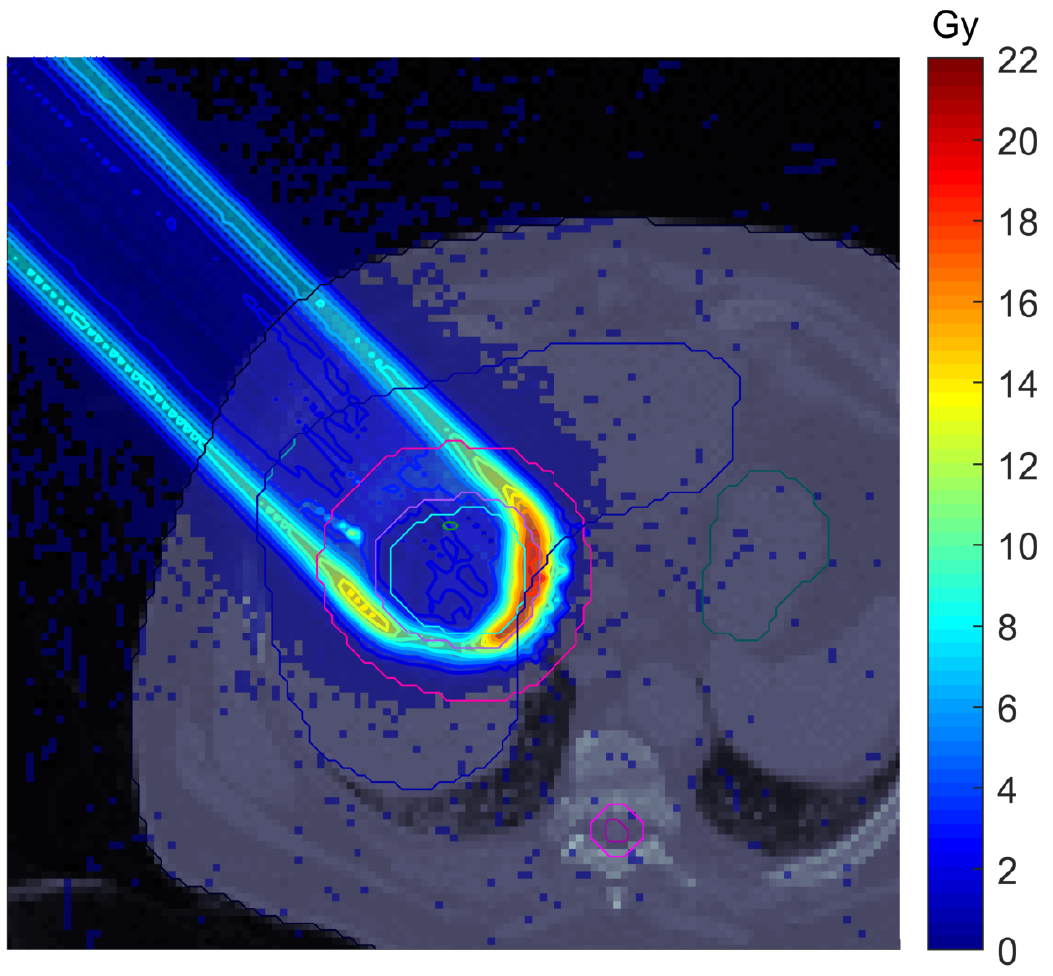}}&\raisebox{-.5\height}{\includegraphics[width=0.225\textwidth]{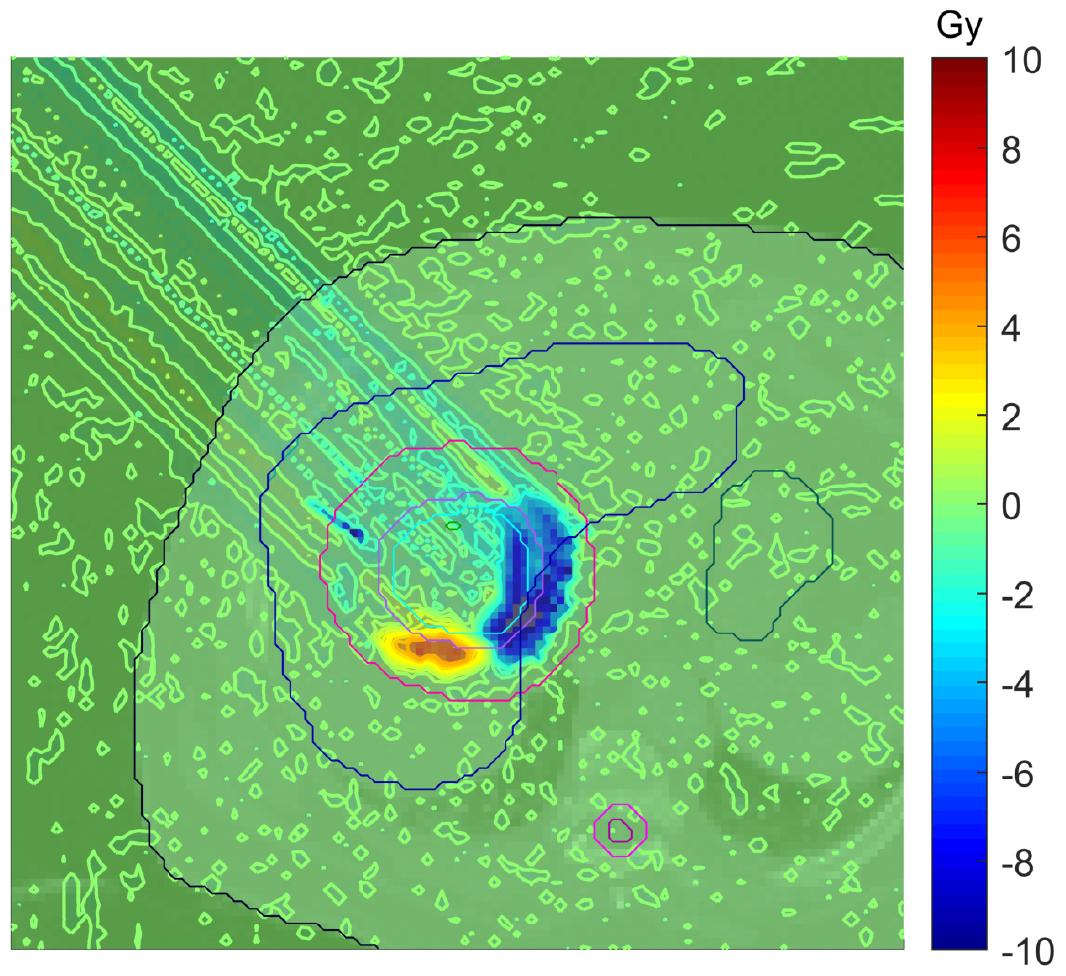}}&\raisebox{-.5\height}{\includegraphics[width=0.22\textwidth]{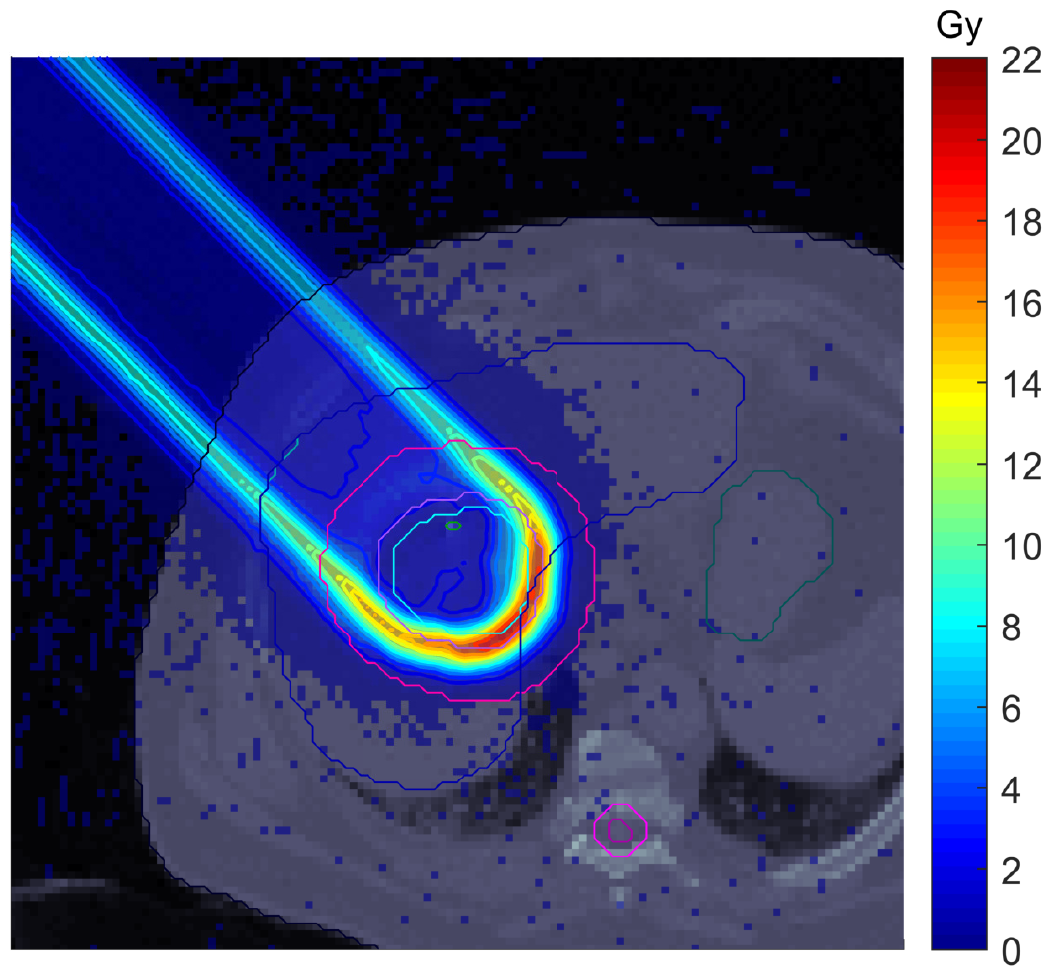}}&\raisebox{-.5\height}{\includegraphics[width=0.225\textwidth]{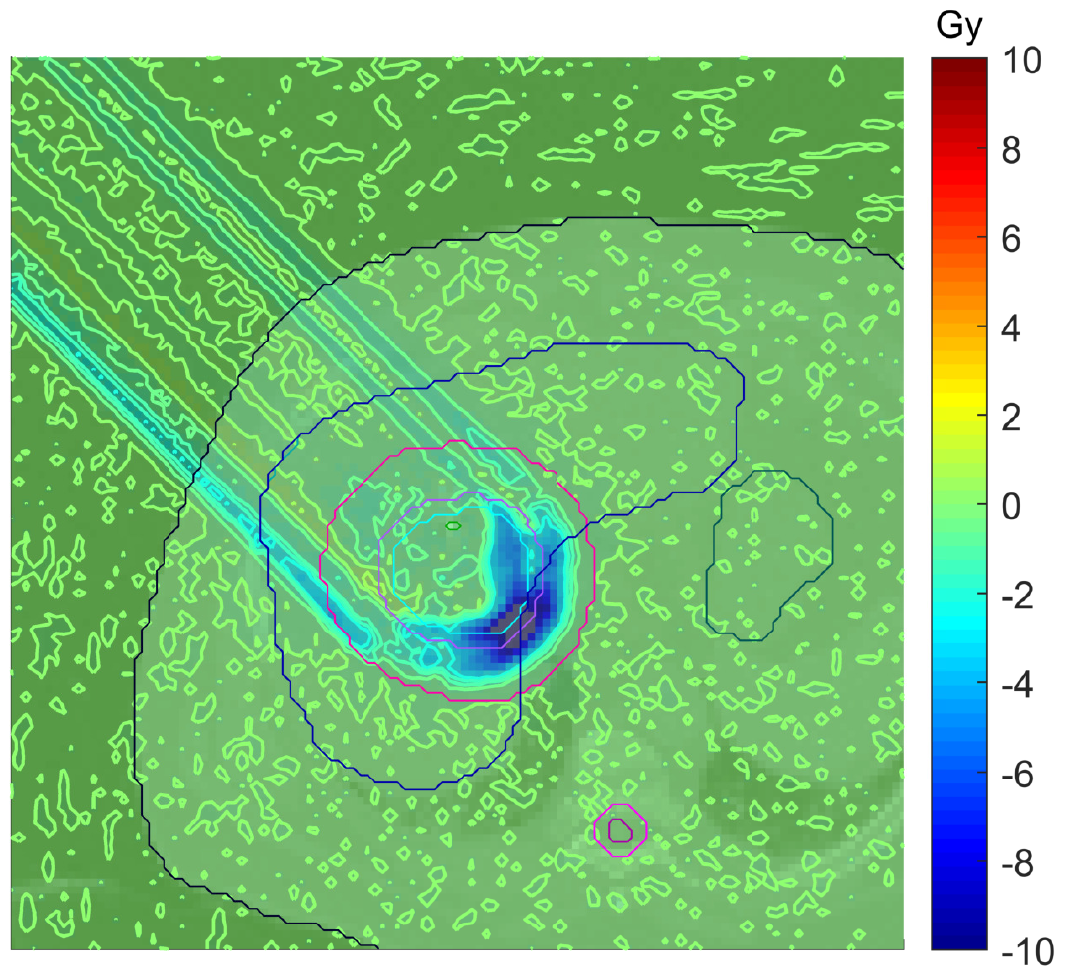}}\\
		& & & &\\[-1.5ex]
		& \multicolumn{4}{c}{(b) Range and set-up errors}\\
		
	\end{tabular}

	\caption{Expected dose $E[\boldsymbol{d}]$ and standard deviation $\boldsymbol{\sigma(d)}$ w.r.t.\ (a) range uncertainties with a $\SI{3}{\percent}$ standard error and (b) both range uncertainties as well as set-up errors with $\SI{3}{\milli\meter}$ standard deviation in a liver patient (couch angle \ang{0}, gantry angle \ang{315}). The left columns show the estimate computed with the proposed (re-)weighting approach, reconstructed either from the nominal distribution $\Phi_0$ or its convolution $\Psi$ with the error kernel. The right columns show the difference to the corresponding references.}
	\label{fig:Liver_Range}
\end{figure}
 Thereby we can conclude that the irregularities in the solution can be attributed to the lack of statistical support in certain areas. Contrary to this, parts of the systematic differences remain and are thus most likely a result of the model approximations.

Figure \ref{fig:Liver_Range} validates these observations for a liver patient. The difference maps for estimates computed based on the expected distribution $\Psi$, have less severe artifacts and systematic deviations. For both set-up and range errors, the $\gamma_{\SI{3}{\percent}}^{\SI{3}{\milli\meter}}$-pass rate also increases from $\SI{93.12}{\percent}$ to $\SI{98.19}{\percent}$ (table \ref{table:gammaLiver}). However, we do not observe such an increase in the case of only range uncertainties.

 Also, it has to be noted, that using $\Psi$ to sample the initial particles leads to an expected dose estimate which is exact up to machine precision, but a nominal dose estimate which now shows deviations from a nominal standard Monte Carlo reference computation in the order of magnitude that we could previously observe for the expected dose.

\subsection{Correlation models}

\begin{figure}[H]
	\centering 
	\begin{tabular}{c@{\hspace{1ex}} c@{\hspace{1ex}} c@{\hspace{1ex}} c@{\hspace{1ex}}}
		\includegraphics[width=0.24\textwidth]{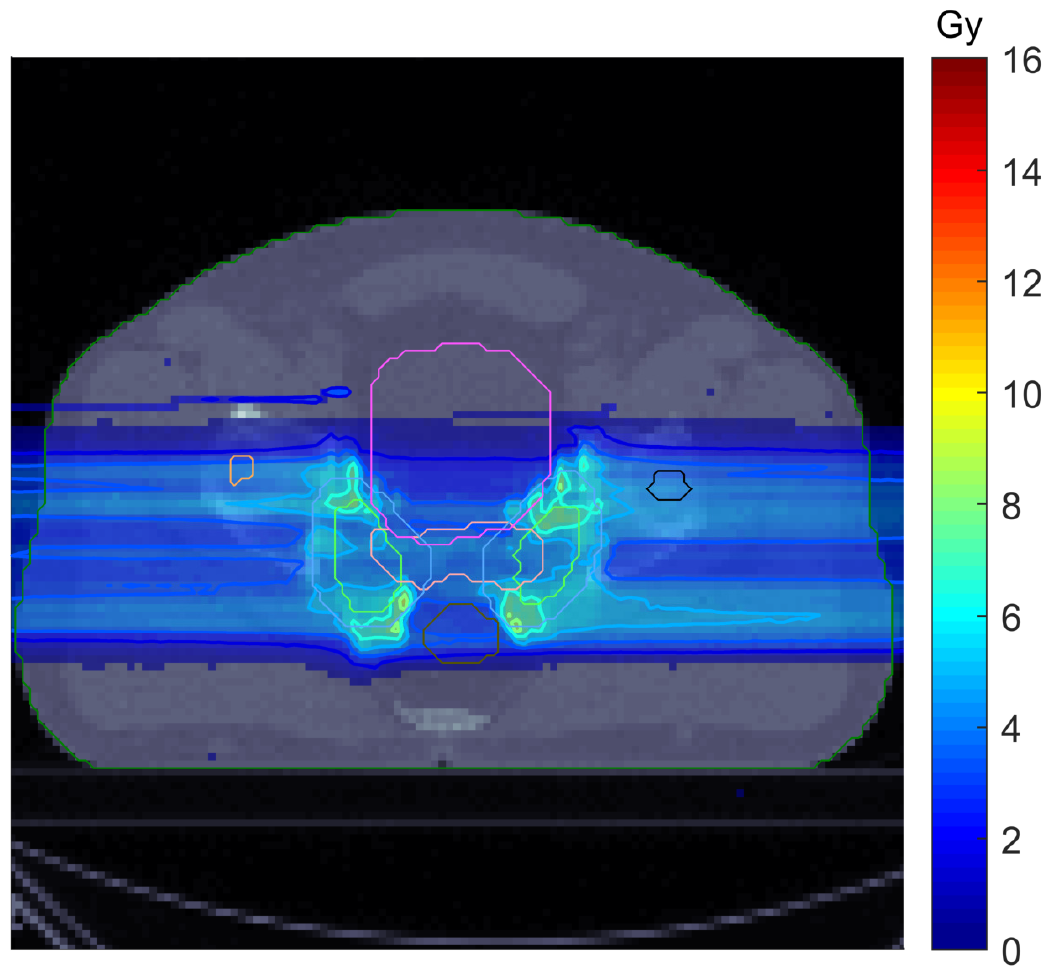} & \includegraphics[width=0.24\textwidth]{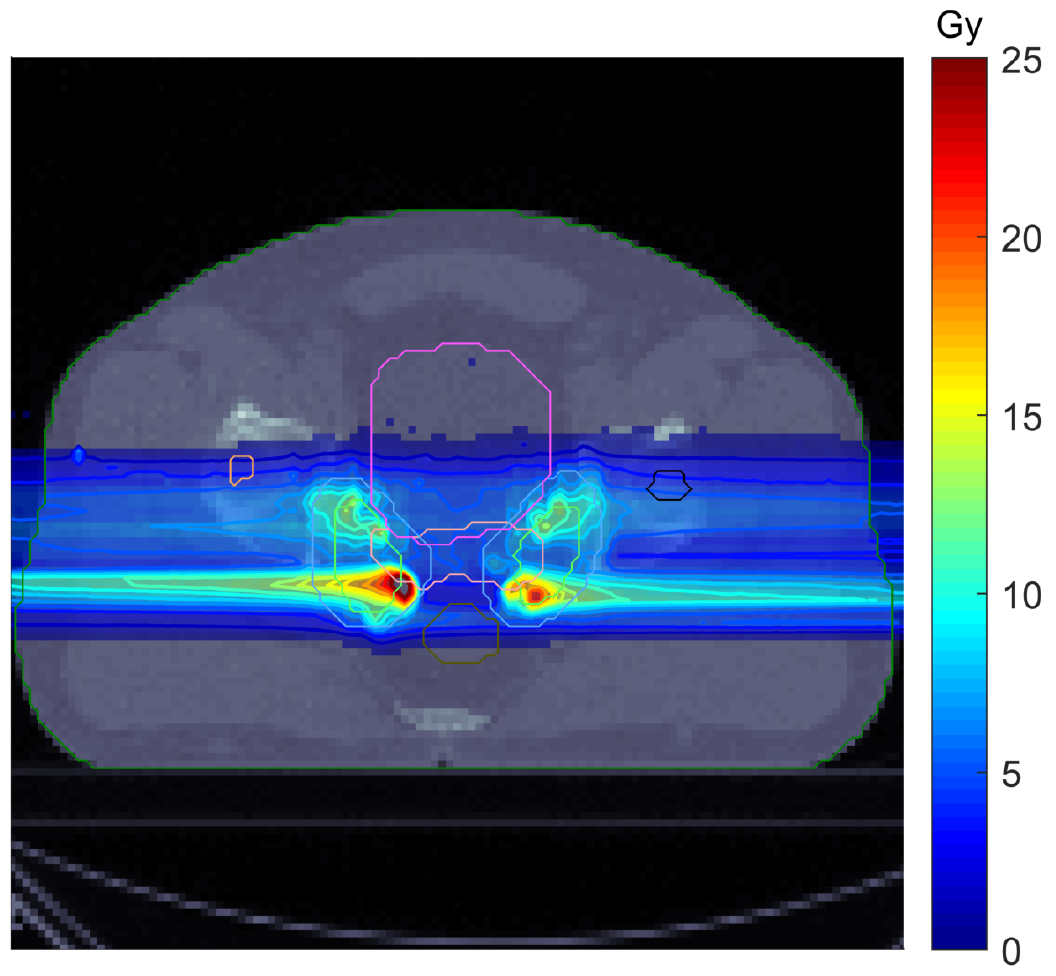} & \includegraphics[width=0.24\textwidth]{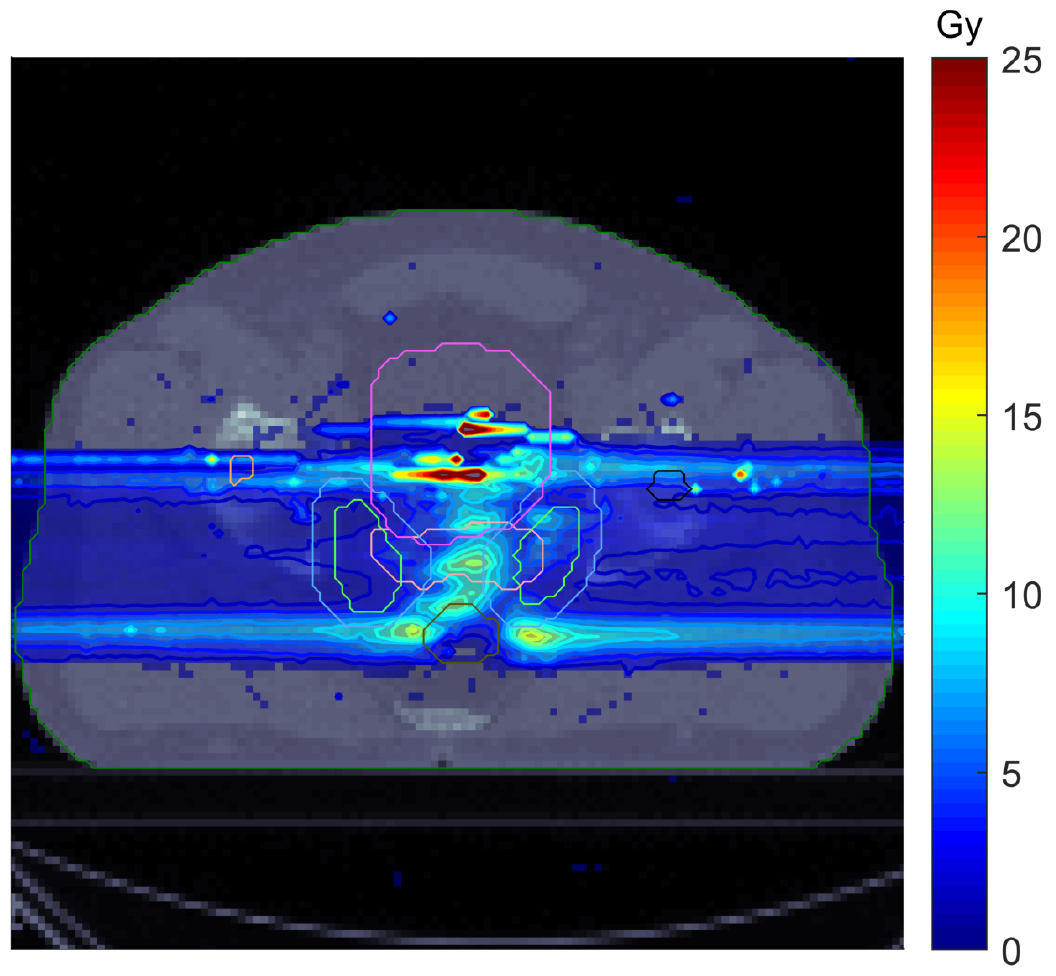}& \includegraphics[width=0.24\textwidth]{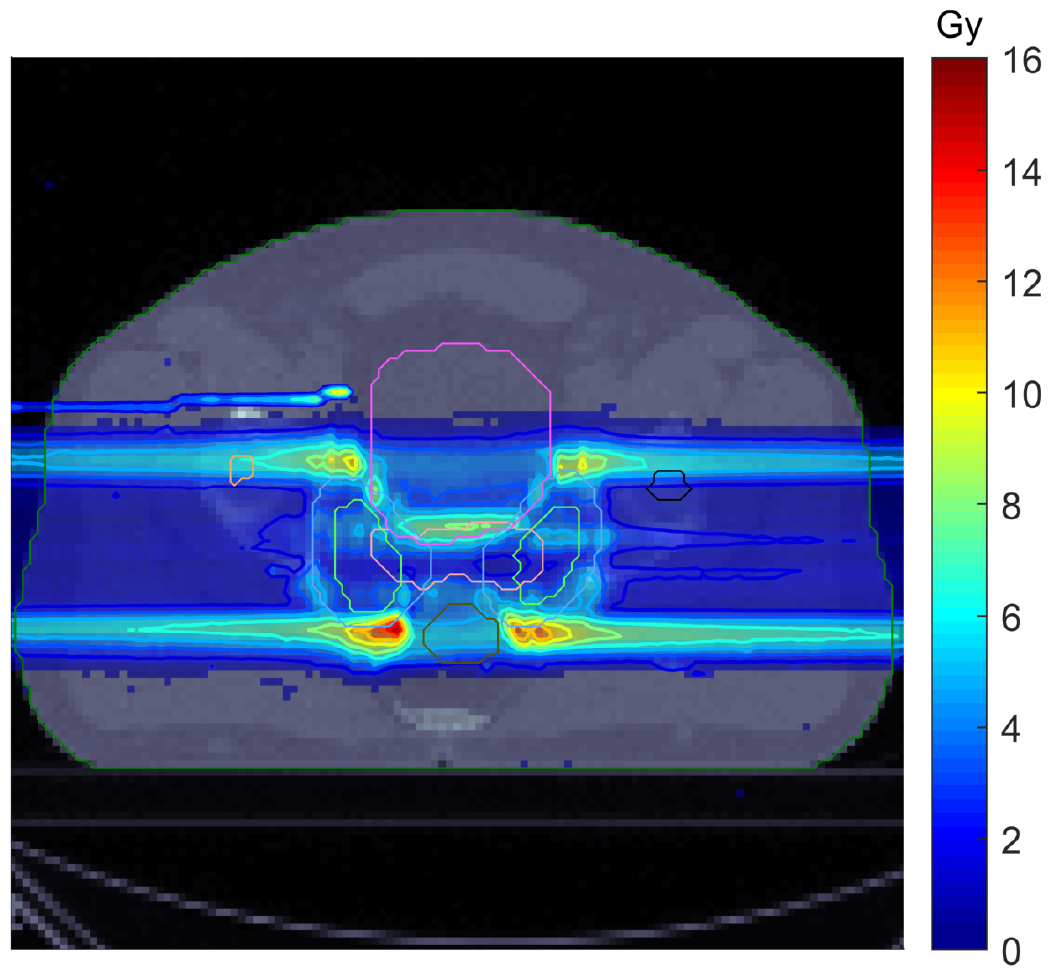} \\
		& & & \\[-1.5ex]
		(a)&(b) &(c)&(d) \\
	\end{tabular}
	\caption{Standard deviation of dose in a prostate patient for (a) no correlation (b) correlation between pencil beams in the same energy level (c) ray-wise correlation between pencil beams and (d) correlation between pencil beams with the same irradiation angle, w.r.t\ set-up errors and in case (c) also range errors.}
	\label{fig:Prostate_CorrModels}
\end{figure}

So far, we have only shown results for the case of fully correlated pencil beams, meaning one global shift of the patient position or scaling factor for the beam range. One of the advantages of the proposed method is, however, the high flexibility in changing the uncertainty model. In figure  \ref{fig:Prostate_CorrModels} we therefore present the standard deviation estimate for four examples of different error correlation models discussed in section \ref{sec:corrModels}.

The results indicate, that different correlation assumptions have a crucial impact on the standard deviation of dose distributions. While it is in principle possible to define arbitrary correlations within the proposed framework, estimates can be prone to artifacts due to a lack of statistical information, especially for the ray-wise correlation model. When sampling error realizations independently for smaller beam components, the reconstruction depends solely on the particle histories associated with these components. For rays with small weights, only very few histories are computed, therefore we observe similar artifacts as encountered in above range uncertainty computations (\ref{sec:resultsRangeError}).

\section{Accuracy and Convergence}
\label{sec:convergence}

Mathematically, it can be shown, that the expected and nominal dose estimates are unbiased. This also  holds for the doses corresponding to each individual error realization. While this does not generally apply for the variance, our results indicate that the bias does not have a significant impact on the quality of estimates.

\begin{figure}[H]
\centering
	\begin{subfigure}{0.45\textwidth}
		\includegraphics[width=\linewidth]{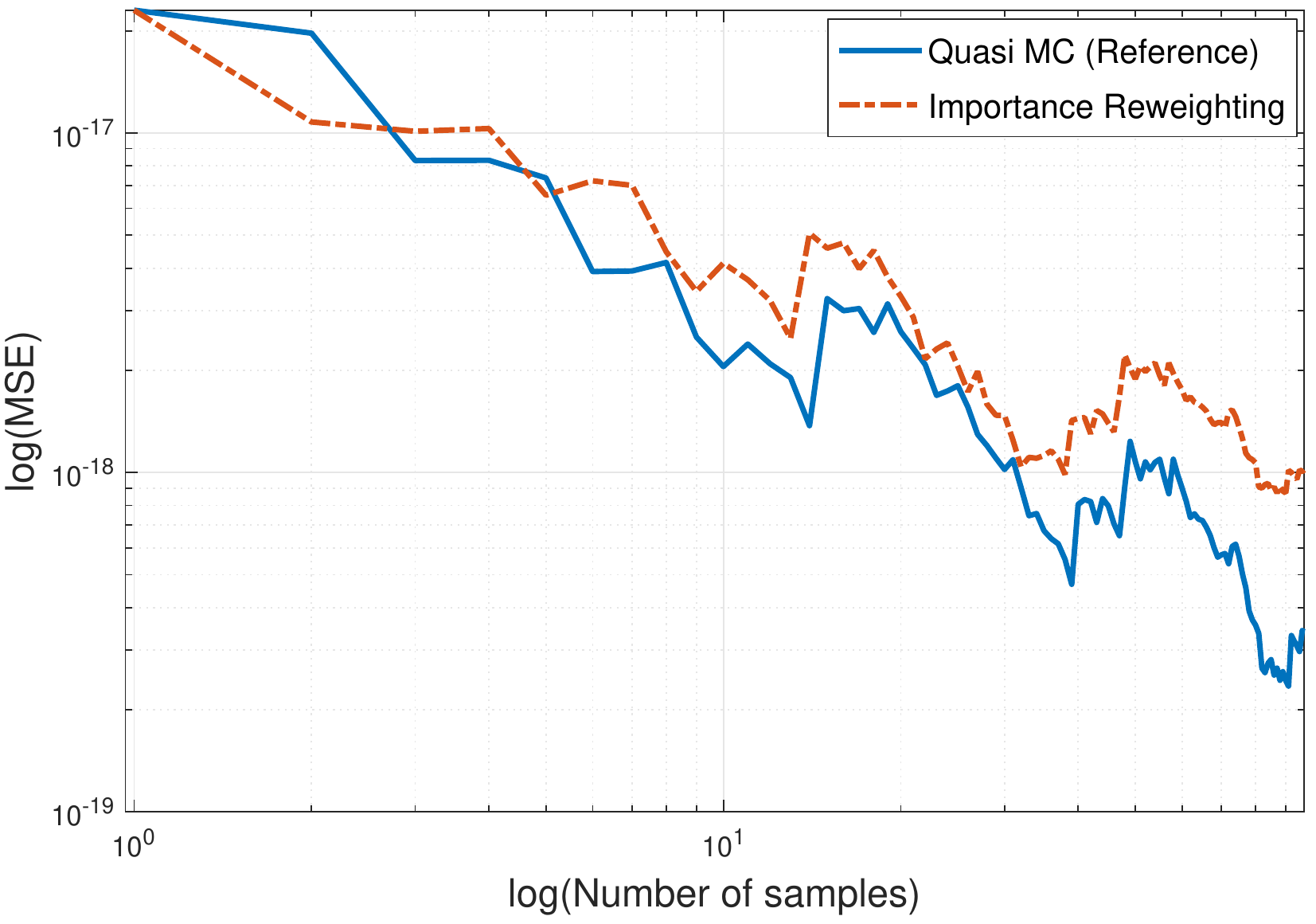}
		\caption*{(a)}
	\end{subfigure}
	\begin{subfigure}{0.45\textwidth}
		\includegraphics[width=\linewidth]{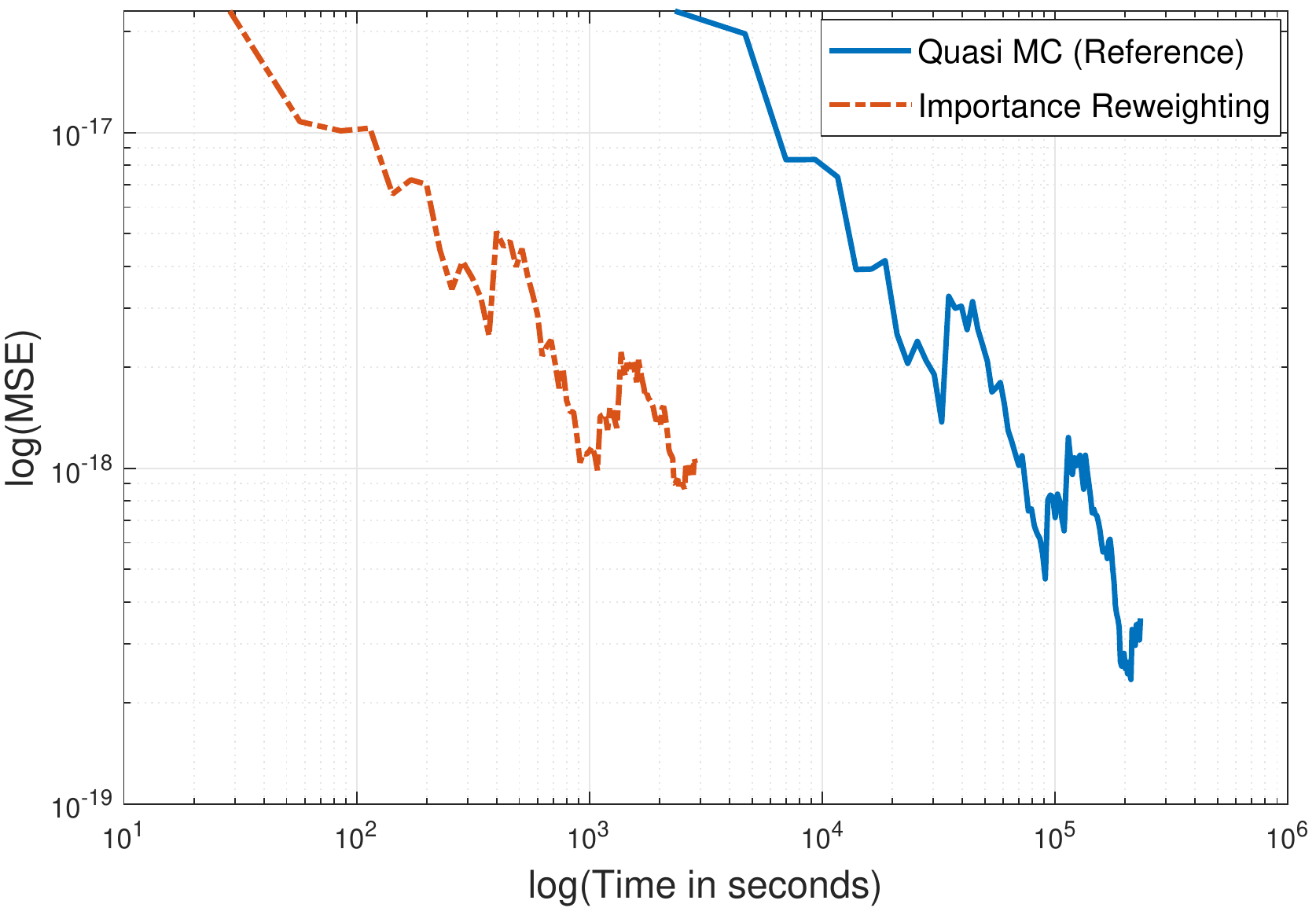}
		\caption*{(b)}
	\end{subfigure}
	\caption{Mean square error (MSE) of the dose standard deviation estimate for the water phantom, computed using a (quasi) Monte Carlo method and the importance (re-)weighting approach and compared for the error convergence (a) per number of samples and (b) per corresponding computation time.}
	\label{fig:convergencePlots}
\end{figure}

For quicker convergence we used quasi-random numbers throughout the whole comparisons, both for the reference computation and the importance (re-)weighting approach. Note, that the combination of importance sampling with quasi-Monte Carlo methods has been shown to be not only possible, but advantageous and preserves the convergence properties of quasi-Monte Carlo \citep{hormannQuasiImportanceSampling2005, oktenErrorReductionTechniques1999,caflischMonteCarloQuasiMonte1998,schurerAdaptiveQuasiMonteCarlo2004}. Since the procedure mimics a (quasi-) Monte Carlo method for uncertainty quantification, where the repeated simulation runs are replaced by (re-)weighting steps, the convergence of the variance per computed error realization is identical. However, due to the lower cost of the (re-)weighting steps, the convergence per time is much faster (see figure \ref{fig:convergencePlots}).  
\begin{table}[htb!]
	\centering
	\caption{CPU time comparison for the reference vs. (re-)weighting approach applied to different patients and computed on the same machine. All values are given in seconds. Note that the times for 100 realizations include the initialization times, while the time for a single realization only refers to the dose computation time.}
	\lineup
	\begin{tabular}{l l l l}
		\br
		 & & Reference& (Re-)weighting \\
		\mr
        \multirow{3}{*}{Water phantom}& Initialization & \,\,\0\0\0\0\0\0\02.35& \,\0\0\0\0\061.53 \\
                  				      & One realization &\,\,\0\0\02\,331.30  & \,\0\0\0\0\028.51\\
                   					  & 100 realizations & \,\0233\,126.93& \,\0\02\,912.53\\
                   					  &&&\\[-2ex]
        \multirow{3}{*}{Liver}		  & Initialization &\,\,\0\0\0\0\0\0\02.44 &\,\0\02\,038.75 \\
      								  & One realization &\,\,\0\039\,066.44  &\,\0\01\,198.74 \\
        							  & 100 realizations &3\,906\,650.90 &121\,912.75 \\
        							  &&&\\[-2ex]
       \multirow{3}{*}{ Prostate}	  & Initialization & \,\,\0\0\0\0\0\0\04.26& \,\0\04\,867.75\\
									  & One realization & \,\,\0\058\,762.40 & \,\0\02\,479.07\\
					  				  & 100 realizations & 5\,876\,253.86& 252\,774.75 \\
		\br
	\end{tabular}
	\label{table:CPUtimes}
\end{table}

For run-time comparisons, the reference computations using TOPAS and the (re-)weighting approach, implemented as post-processing in Matlab, were run on the same virtual machine\footnote{Virtual machine including 64 CPUs with 1.995 GHz and 200GB RAM}. We observe reduced CPU times by a factor of 80, 32 and 23 for the water phantom, liver and prostate patient, respectively (see table \ref{table:CPUtimes}).

\section{Discussion}
\label{sec:discussion}
In this paper, we introduce an efficient approach for uncertainty quantification in Monte Carlo dose calculations using history (re-)weighting. We demonstrate how particle histories from one simulation can be scored to construct estimates for error scenarios, the expected dose and standard deviation, for set-up and range errors in intensity modulated proton therapy. As demonstration example, Gaussian range and set-up uncertainties, with $\SI{3}{\percent}$ and  $\SI{3}{\mm}$ standard deviation respectively, were considered for a water phantom, a liver patient and a prostate patient. 

For set-up uncertainties, we observed good agreement of at least $\SI{99.99}{\percent}$ in the $\gamma_{\SI{3}{\mm}/\SI{3}{\percent}}$-criterion for all quantities of interest. Range error propagation could be approximated by transforming the assumed range uncertainty into energy uncertainty via the range-energy relationship. The error caused by this model approximation appears to be relatively minor for the expected dose. The standard deviation estimates are, however, sensitive to the number of histories and usage of the nominal Gaussian pencil beam width or the convolved distribution. Differences and visible artifacts in the standard deviation estimate can be partly eliminated by simulating the initial phase-space parameters using the convolved beam parameterization. While some systematic deviations remain, the order of magnitude as well as shape and extent of the dose standard deviation is sufficiently well-represented. However, this causes a reduction in accuracy for the nominal dose. Thus, improving the accuracy of the estimate for one quantity of interest comes at a cost for the accuracy of another and it remains up to the user and use-case to put the focus on either retaining accuracy in the nominal dose computation or trading it against better accuracy of the uncertainty estimate. 

We also demonstrate the use of different pencil beam correlation models within the framework. It is clear that the choice of correlation model has a significant impact on the standard deviation estimate. Therefore, it is particularly convenient that the (re-)weighting method allows for the definition of principally arbitrary correlation matrices to put into the underlying multivariate Gaussian error model. These could possibly be extended to simulate interplay effects or other dynamic influences in the context of 4D treatment planning. Since the applied correlation models are not only experimental but also difficult and time-consuming to evaluate in scenario sampling, we did not quantitatively compare them to reference computations. Further studies could explore whether they agree with other methods computing such correlations based on an analytical probabilistic dose engine \citep{bangertAnalyticalProbabilisticModeling2013,wahlEfficiencyAnalyticalSamplingbased2017,wieserImpactGaussianUncertainty2020}.

Compared to the reference scenario estimates, which rely on performing full MC dose calculations repeatedly, the CPU-time for standard deviation estimates could be reduced by more than an order of magnitude using our method in combination with a quasi-MC approach. This is achieved by reducing the costs of repeated expensive simulations to those of scoring based on matrix-vector multiplications. Consequently, it has to be noted, that the time reduction depends largely on the proportion of computational overhead of the initialization and simulation steps in the MC engine. Therefore the factor of speed increase varies strongly between different test-cases and, most likely, implementations. But even then, our method holds two performance advantages: First, it can directly compute the expected dose by using the convolved phase space parameterization (\ref{sec:directExpDose}) in \emph{one} standard simulation. Second, multiple uncertainty models with different correlation patterns and magnitude can be reconstructed from the same set of histories. This could for example be used to investigate the impact of fractionation effects, using the framework proposed by \citet{wahlAnalyticalIncorporationFractionation2018} or to consider a number of (worst case) scenarios besides the expected dose and standard deviation.
 
Further, we argue that computational performance can be further improved through a more efficient implementation and using better parallelization. Also, a combination of our approach with other efficient uncertainty quantification approaches, which rely on scenario computations could lead to run-time improvements. For instance a polynomial chaos expansion as introduced in \citet{perkoFastAccurateSensitivity2016} could be adjusted such that the evaluations are computed by (re-)weighting histories instead of the usual dose calculations.

When computing the estimates in post-processing, the regular Monte Carlo dose calculation is not slowed down perceptibly by storing particle histories for later reconstructions. The possible additional run-time is smaller than the variation between two runs of the same simulation and thereby barely detectable. An implementation as on-the-fly scoring is however also possible and might outperform post-processing in terms of overall run time for a single uncertainty model.

Last but not least, the method is not inherently limited to the discussed application in proton therapy; a calculation of uncertainty estimates using the (re-)weighting approach would also be feasible for other IMPT modalities like carbon ions but also photons. In its current description, it is however limited to uncertainties which can be modeled in terms of variations of phase space parameters with a prior probability distribution. Application to, for example, pre-simulated phase spaces might also be feasible using numerical convolution techniques. Also, a disproportionately high magnitude of uncertainties in relation to this probability distribution can compromise the accuracy of results. Furthermore, it needs to be mentioned, that the current computational speed, especially for the standard deviation, might still not be sufficient for optimization purposes, where a full dose influence matrix needs to be computed. Due to the simplicity of the process and the high flexibility in post-processing at virtually no cost for the original simulation, we are confident that the approach has the potential for further development and use.

\section{Conclusion}
\label{sec:conclusion}
Dose distributions in intensity modulated proton therapy are known to be sensitive to uncertainties. The computational efforts in estimating such uncertainties become particularly evident when Monte Carlo dose calculation is used. We showed how the concept of importance sampling can be adapted to estimate the expected dose and its variance using histories from only a single Monte Carlo simulation. Set-up uncertainties can be efficiently modeled and exhibit almost exact agreement with reference computations. The inclusion of range uncertainties, by modeling them as energy uncertainty via the range-energy relationship, yields less but sufficient accuracy for most application purposes. Further, the physical simulation of particles is completely decoupled from uncertainty quantification, thereby allowing for the incorporation of arbitrary correlation assumptions and the comparison of different scenarios, at no additional cost to the nominal dose calculation. Therefore, the presented approach has several benefits over classic non-intrusive methods and is a step towards reconciling efficient uncertainty quantification and, in the future, robust optimization based on Monte Carlo dose calculations. 

\ack
The present contribution is supported by the Helmholtz Association under the joint research school HIDSS4Health -- Helmholtz Information and Data Science School for Health.

\appendix
\section{Importance Sampling}

Importance sampling is a method most frequently used for variance reduction \citep{kahnRandomSamplingMonte1950, hastingsMonteCarloSampling1970}. Assume the integral  $ I(g) = \int g(x) p(x)dx$ is to be computed for $x \sim p(x)$ without directly sampling from $p(x)$. The importance sampling estimate is defined as

\begin{eqnarray}
\label{importanceSampling}
I(g) &= \int g(x)p(x) dx = \int g(x) \frac{p(x)}{q(x)} q(x) dx \nonumber \\
&\approx \sum_{i=1}^{N} g(X_i)  \frac{p(X_i)}{q(X_i)} \;, \;\; X_i \sim q(x)
\end{eqnarray}

Thus, the integral of $g(x)$ with the probability distribution $x\sim p(x)$ can be reconstructed from samples of a suitable density function $q(x)$.

\section{Full phase space model}

The simplified model used in the main paper can be extended to include distributions in the momentum direction $\boldsymbol{\varphi}=(\varphi_x, \varphi_y, \varphi_z)\neq 0$. The additional variables are included in the Gaussian mixture model:
\begin{equation}
\Phi_0(\boldsymbol{\xi})=\sum_{b=1}^B w_b \Phi_0^b(\boldsymbol{\xi})\;, \Phi_0^b(\boldsymbol{\xi})=\Phi_0^b(\boldsymbol{r},\boldsymbol{\varphi},E)=\mathcal{N}(\boldsymbol{\mu}_{\xi}^b,\boldsymbol{\Sigma}_{\xi}^b)
\end{equation}

where the respective entries in the covariance matrix $\Sigma_{\xi}^b$ can be chosen $\neq 0$ to model randomness in the momentum direction as well as correlations of the momentum directions with primary particle positions.

\begin{table}[htb!]
	\centering
	\caption{$\gamma^{\SI{3}{\milli\meter}}_{\SI{3}{\percent}}$-pass rates in volumes of interest (VOI) of the liver patient computed using the full phase space parameterizations.}
	\label{table:gammaLiverFullPhaseSpace}
	\lineup
	\begin{tabular}{l  lll lll}
		\br
		& \multicolumn{3}{c}{From $\Phi_0$} & \multicolumn{3}{c}{From $\Psi$}  \\
		\mr
		Error type &	\multicolumn{6}{c}{Set-up} \\
		\mr
		Liver  & $\boldsymbol{d}$ & $E[\boldsymbol{d}]$ &$\boldsymbol{\sigma(d)}$ & $\boldsymbol{d}$ & $E[\boldsymbol{d}]$ &$\boldsymbol{\sigma(d)}$\\
		\mr
		\textbf{Overall} &100& \099.99 & \099.77 & \099.99& 100&\099.74\\
		GTV &100& 100&100 &100 &100&100\\
		Liver &100& 100& \099.78&100 &100&\099.79\\
		Heart &100& 100& \099.83&100 &100&\099.53\\
		CTV &100& 100&\099.64 &100 &100&\099.51\\
		Contour &100&100 & \099.81 &100 &100&\099.70\\
		PTV &100&100 &\099.54 &100 &100&\098.72\\
		\br
	\end{tabular}
\end{table}

\begin{figure}[H]
	\centering 
	\begin{minipage}{0.05\textwidth}
		\vspace*{\fill}
		$\boldsymbol{d}$
		\vspace*{\fill}
	\end{minipage}\hfil
	\begin{minipage}{0.8\textwidth}
		\begin{subfigure}{0.31\textwidth}
			\caption*{\textbf{Estimate ($\Phi_0$)}}
			\includegraphics[width=\linewidth]{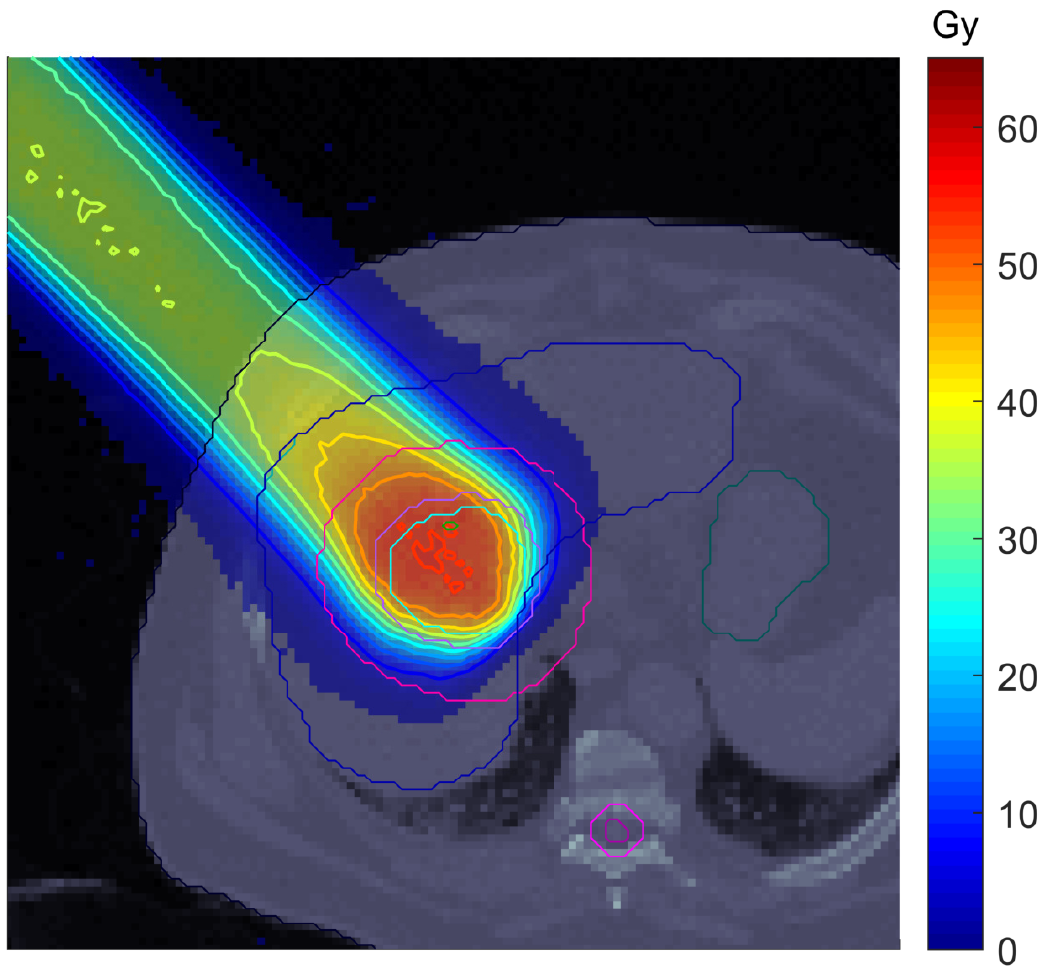}
		\end{subfigure} 
		\begin{subfigure}{0.31\textwidth}
			\caption*{\textbf{Estimate ($\Psi$)}}
			\includegraphics[width=\linewidth]{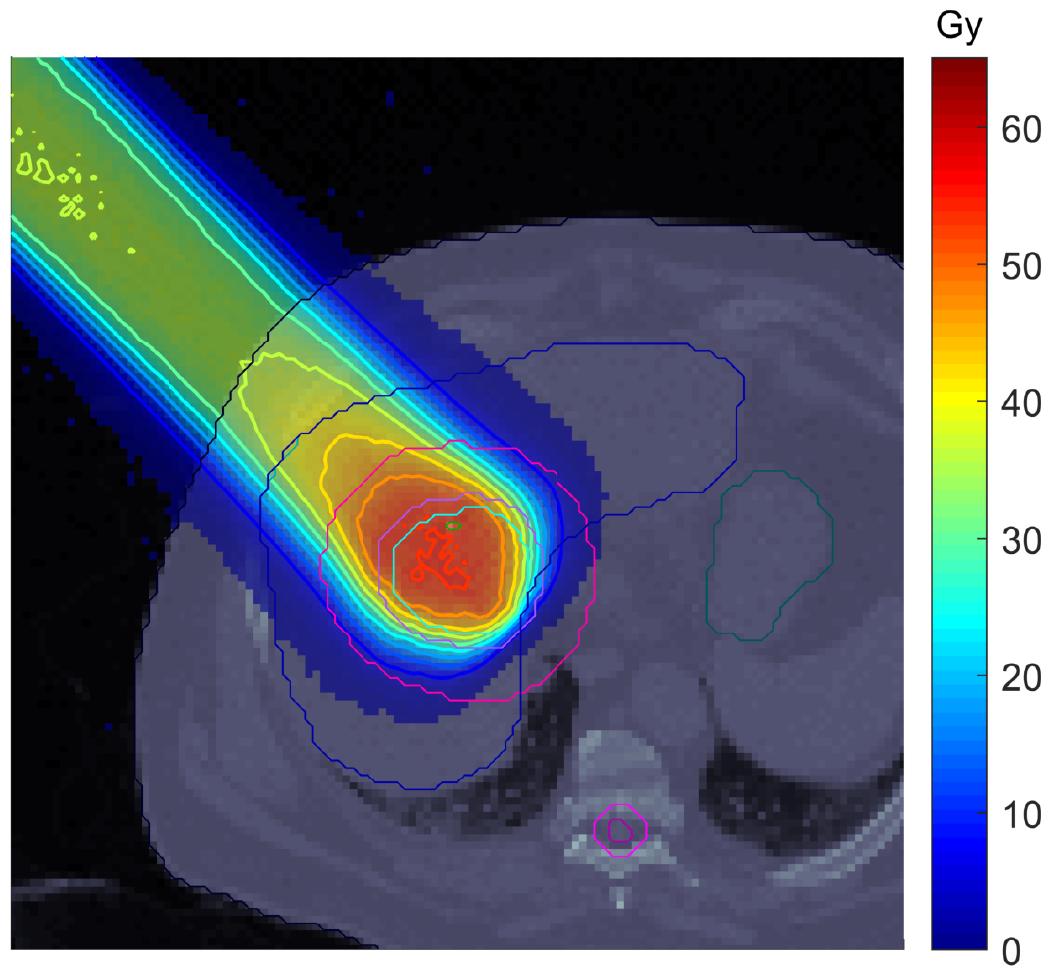}
		\end{subfigure}
		\begin{subfigure}{0.31\textwidth}
			\caption*{\textbf{Reference}}
			\includegraphics[width=\linewidth]{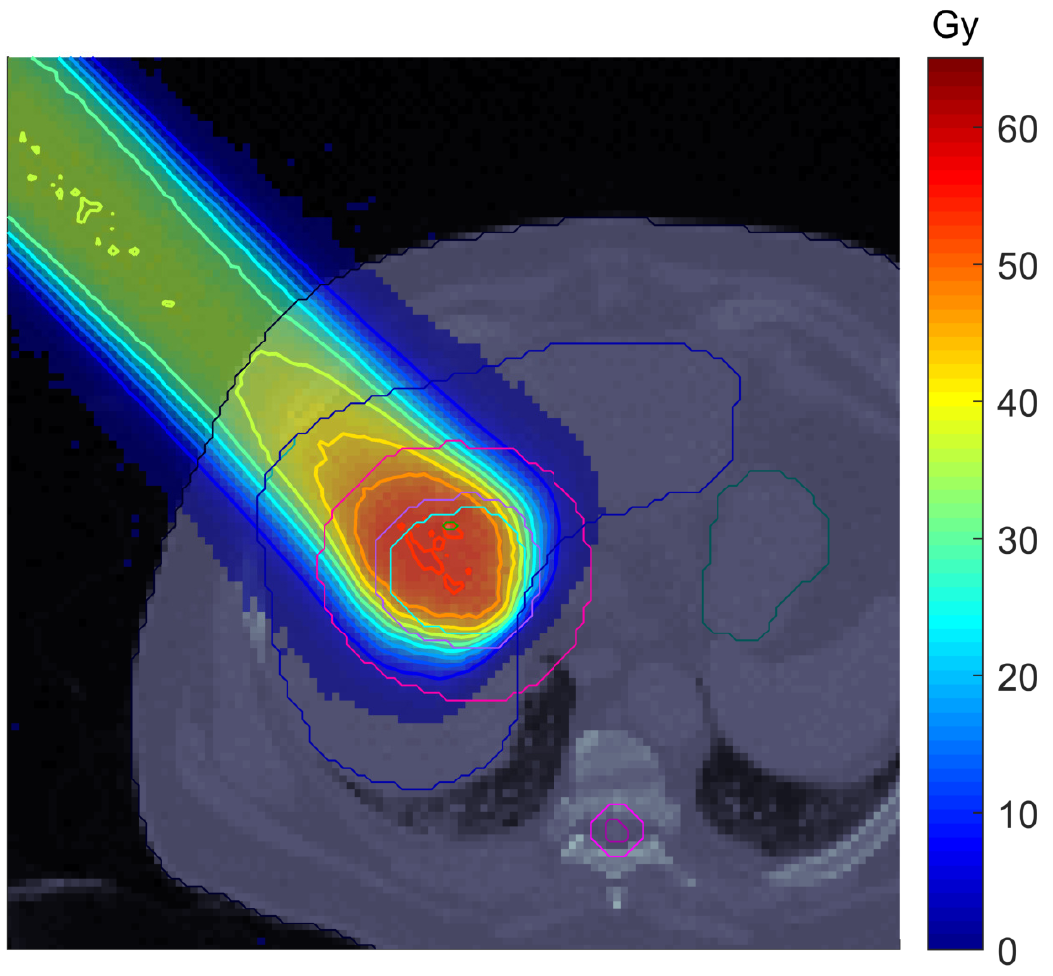}
		\end{subfigure}
	\end{minipage} 
	
	\begin{minipage}{0.05\textwidth}
		\vspace*{\fill}
		$E[\boldsymbol{d}]$
		\vspace*{\fill}
	\end{minipage}\hfil
	\begin{minipage}{0.8\textwidth}
		\begin{subfigure}{0.31\textwidth}
			\includegraphics[width=\linewidth]{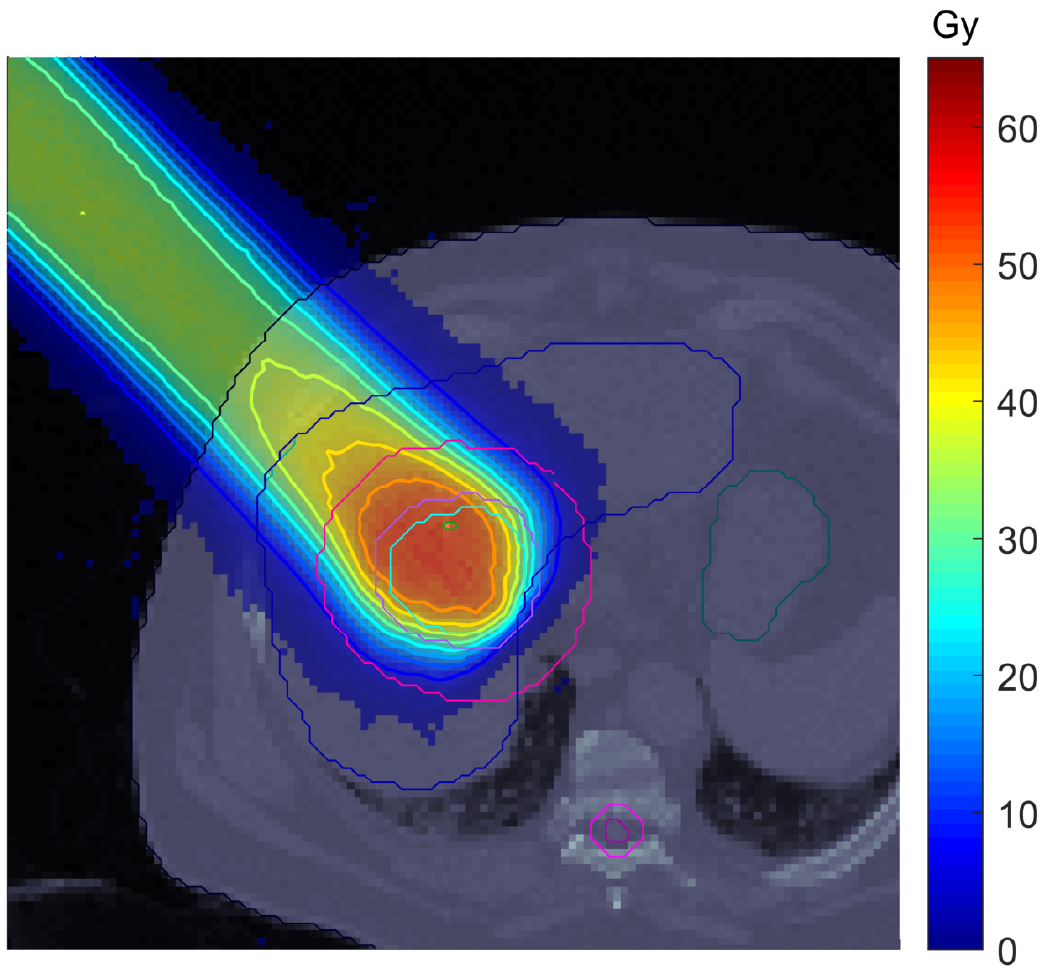}
		\end{subfigure}
		\begin{subfigure}{0.31\textwidth}
			\includegraphics[width=\linewidth]{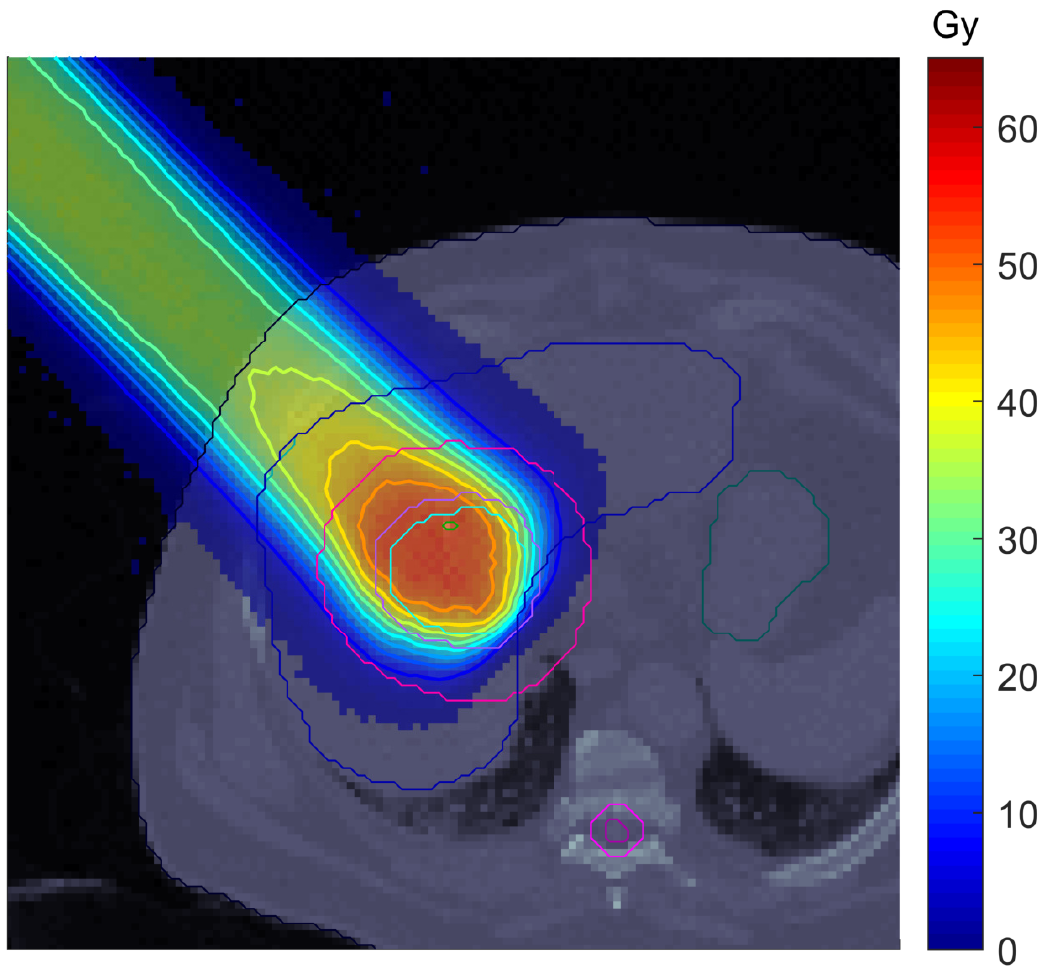}
		\end{subfigure}
		\begin{subfigure}{0.31\textwidth}
			\includegraphics[width=\linewidth]{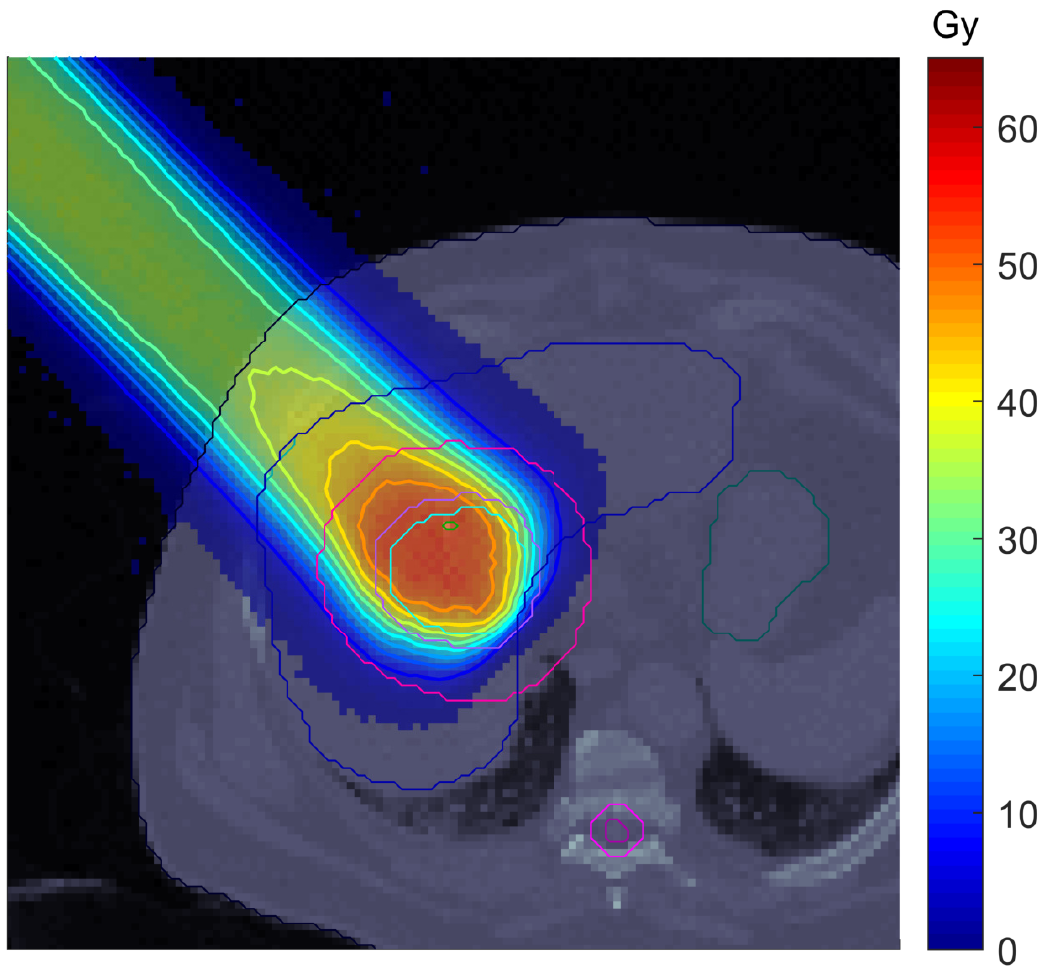}	
		\end{subfigure}
	\end{minipage}
	
	\begin{minipage}{0.05\textwidth}
		\vspace*{\fill}
		$\boldsymbol{\sigma}$
		\vspace*{\fill}
	\end{minipage}\hfil
	\begin{minipage}{0.8\textwidth}
		\begin{subfigure}{0.31\textwidth}
			\includegraphics[width=\linewidth]{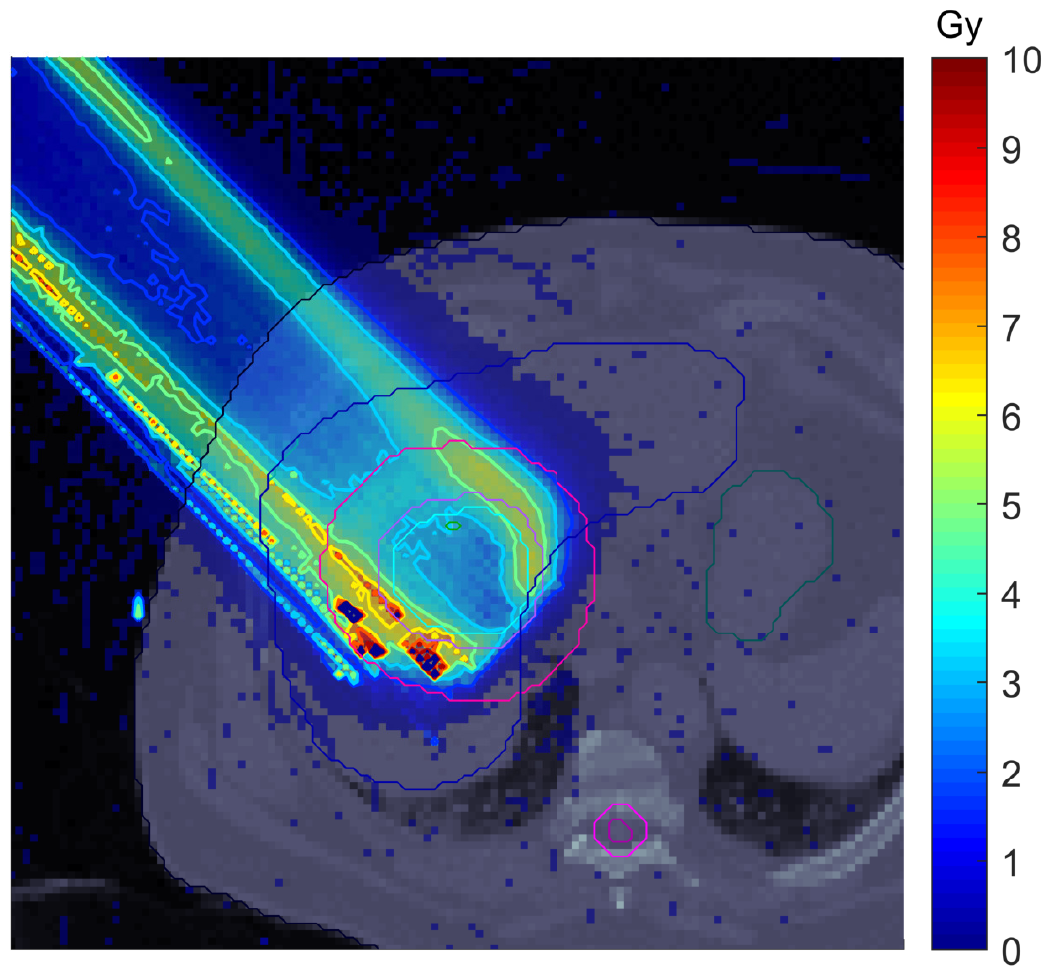}
		\end{subfigure} 
		\begin{subfigure}{0.31\textwidth}
			\includegraphics[width=\linewidth]{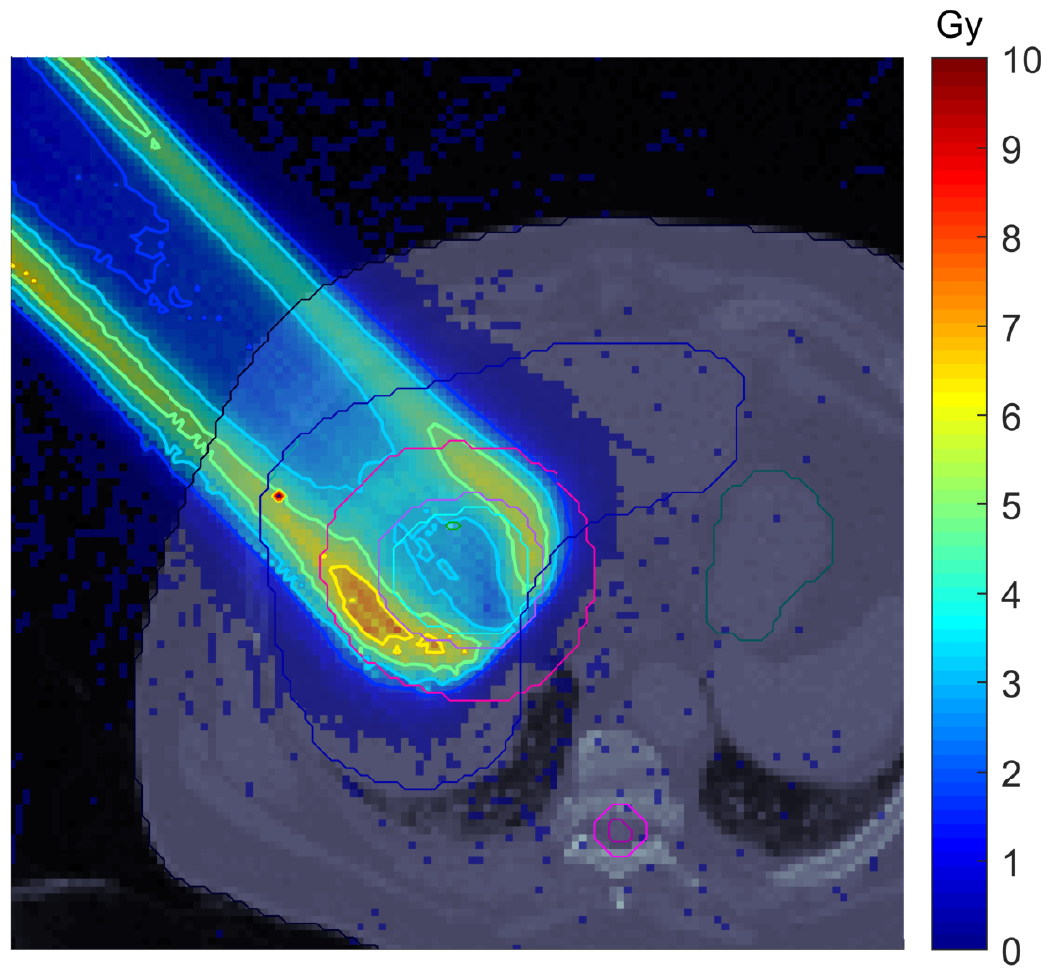}	
		\end{subfigure}
		\begin{subfigure}{0.31\textwidth}
			\includegraphics[width=\linewidth]{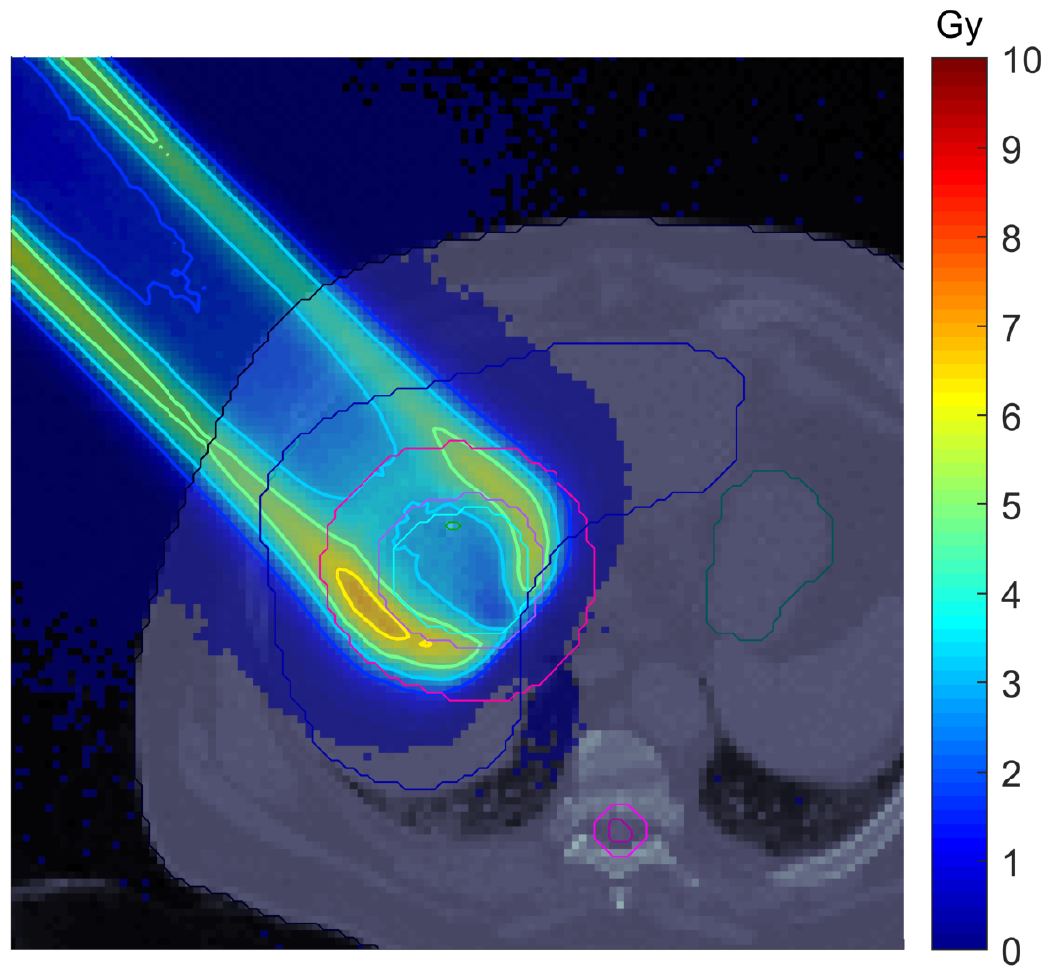}
		\end{subfigure}
	\end{minipage}
	\caption{Nominal dose, expected dose and standard deviation w.r.t.\ set-up uncertainties with $\SI{3}{\milli\meter}$ standard deviation for one beam (couch angle $\ang{0}$, gantry angle $\ang{315}$), computed using the full phase space parameterizations.}
	\label{fig:BeamDivergence}
\end{figure}
Figure \ref{fig:BeamDivergence} presents results for the nominal dose, expected dose and standard deviation in a liver patient, for set-up uncertainties with  \SI{3}{\milli\meter} standard deviation, $0.2$ standard deviation in the momentum direction and $0.3$ correlation between $\varphi_v$ and $r_v, \; v \in \{x,y\}$. Estimates were computed based on the convolution function $\Psi$ of the error and beam parameter densities, as well as the nominal parameter density $\Phi_0$. The corresponding global $\gamma$-analysis pass rates can be found in table \ref{table:gammaLiverFullPhaseSpace}.

\end{document}